# Experimental ten-photon entanglement

Xi-Lin Wang, Luo-Kan Chen, Wei Li, He-Liang Huang, Chang Liu, Chao Chen, Yi-Han Luo, Zu-En Su, Dian Wu, Zheng-Da Li, He Lu, Yi Hu, Xiao Jiang, Cheng-Zhi Peng, Li Li, Nai-Le Liu, Yu-Ao Chen, Chao-Yang Lu, and Jian-Wei Pan

*Hefei National Laboratory for Physical Sciences at Microscale and Department of Modern Physics, University of Science and Technology of China,*
*& CAS Centre for Excellence and Synergetic Innovation Centre in Quantum Information and Quantum Physics, Hefei, Anhui 230026, China,*
*& CAS-Alibaba Quantum Computing Laboratory, Shanghai, 201315, China*

**Abstract:**

**We report the first experimental demonstration of quantum entanglement of ten spatially separated single photons. A near-optimal entangled photon pair source is developed with simultaneously a source brightness of ~12 MHz/W, a collection efficiency of ~70% and an indistinguishability of ~91% between independent photons, and is used for a step-by-step engineering of multi-photon entanglement. Further, our experiment is able to identify two regimes, under different pump power, where the ten photons are in genuine entanglement or distillable entanglement. Our work creates a state-of-the-art platform for multi-photon experiments, and provide enabling technologies for many challenging optical quantum information tasks, such as the realization of Shor's error correction code, teleportation of three degrees of freedom in a single photon, and scattershot boson sampling.**

Quantum entanglement[1-3] among multiple spatially separated particles is of fundamental interest, and can serve as central resources for studies in quantum nonlocality, quantum-to-classical transition[4], quantum error correction[5], and quantum simulation[6]. The ability of generating an increasing number of entangled particles is an important benchmark for quantum information processing[7]. The largest entangled states were previously created with fourteen trapped ions[8], eight photons[9], and five superconducting qubits[10].

Despite fast progress in linear optics quantum computing[11-13] in the last decades, the limiting bottleneck remains the number of experimentally controlled single photons. Even for non-universal quantum computation models such as boson sampling[12] that demands minimal physical resources, 20-30 single photons are required to reach quantum supremacy. The previous demonstrated eight-photon entangled state had a coincidence count rate of ~9 per hour[9]. Based on the same count rate, ten-photon count rate would be ~170 per year, which was a formidable challenge.

For multi-photon experiments with spontaneous parametric down-conversion[13] (SPDC), the probability ($p$) of generating a single photon pair per pump pulse should be kept small (typically <0.1), to suppress the undesired noise contribution from double-pair emission rate (~$p^2$). To increase the count rate of the entangled photons without compromising the single photon purity and indistinguishability, it is most crucial to engineer the spatial and spectral properties of the pulsed SPDC to enhance the photon collection efficiency into a single spatial mode. In previous multi-photon entanglement experiments[9,14], however, the collection efficiency of the SPDC photons was not sufficient to demonstrate ten-photon entanglement.

The key to our ten-photon entanglement experiment is a pulsed SPDC photon pair source with simultaneously high brightness ( $\sim 1.2 \times 10^7$ photon pairs generated per one Walt pump power), high collection efficiency (~70%) and high photon indistinguishability (~91%), as shown in Fig.1. To achieve a high coupling efficiency, we adopted beam-like type-II

SPDC[14-19] where the signal-idler photon pairs are emitted as two separate circular beams in a Gaussian-like intensity distribution, confirmed by the recorded image displayed in the right panel of Fig.1a. This is favorable for collecting the produced fluorescence into a single-mode fibre[17], a significant advantage compared to the non-collinear SPDC where the photon pairs are collected from two intersections of the down-converted rings[20]. In our experiment, using a single 2-mm-thick beta-barium borate (BBO) crystal, the collection efficiency was measured to be 88% without narrowband filtering.

To generate polarization-entangled photon pairs, we adopted a sandwich-like geometry[14,18,19] where a half-wave plate (HWP) is sandwiched between two 2-mm-thick, identically cut BBOs (see *Supplemental Materials*). The polarization state of the correlated photon pairs generated from the left BBO can be written as $\left|V_{1,o}\right\rangle\left|H_{2,e}\right\rangle$, where $H$ ($V$) denotes horizontal (vertical) polarization and the subscripts $1,o$ ($2,e$) indicates the single photon being the down-converted ordinary (extraordinary) light in spatial mode 1 (2). The photons pass through the HWP and are rotated to its orthogonal state $\left|H_{1,o}\right\rangle\left|V_{2,e}\right\rangle$, which are then coherently superposed with the photon pair $\left|V_{1,o}\right\rangle\left|H_{2,e}\right\rangle$ generated from the right BBO. The sandwich-like structure engineers the $e$ and $o$ photon emission to be spatially separated, allowing efficient narrowband filtering for the $e$ and $o$ photons[21]. By careful spatial and temporal compensations (see *Supplemental Materials*), entangled photons in the state $(\left|H_{1,o}\right\rangle\left|V_{2,e}\right\rangle+\left|V_{1,o}\right\rangle\left|H_{2,e}\right\rangle)/\sqrt{2}$ were prepared.

The observed count rate and fidelity of the two-photon entanglement as a function of pump power are summarized in Fig.1b and 1c. At a pump power of 0.57 W, with (3+8) nm filters and without, 1.2 and 3 million two-photon counts per second were detected, respectively, with a state fidelity (the overlap with an ideal two-photon entangled state) of 0.97 and 0.93. Thanks to the new source design with a record high collection efficiency of 70%, we obtained ~4 times brighter entangled photon source than the previous result[9], by using only 65% pump power. This increases the ten-photon coincidence count rate by 2-3 orders of magnitude and thus makes the experimental observation possible.

For entangling independent single photons, a crucial requirement is that the photons should possess a high indistinguishability. For example, any spectral or temporal information that labels an individual photon could reveal which way it comes from, thus eliminating the quantum interference. We measure the joint spectrum of the two photons to quantify their frequency correlation, as shown in Fig.1d. We observe a tilted elliptical joint spectral intensity distribution. Its long axis is about $25°$ with respective to the vertical direction, indicating that the $o$ and $e$ photons remain frequency correlated[22]. With 3 nm and 8 nm filtering for the $e$ and $o$ photon respectively, we can expect a photon purity of 0.93. The photon indistinguishability is experimentally tested through quantum interference between two independent photons (see Fig.1e and *Supplemental Materials*). We observe a raw interference visibility of 0.91(1) under a low pump power of ~44 mW where the higher-order emission rate is kept small, thus this 0.91 visibility indicates a lower limit of photon indistinguishability. Under a high pump power of 1 W, the visibility is reduced to 0.76(1) which as expected is due to the noise contributions from double-pair emission of SPDC.

The experimental setup for entangling five successively generated photon pairs from SPDC into a ten-photon Schrödinger cat-like state is illustrated in Fig.2. Five $e$ photons, each from an entangled pair, are combined on a linear optical network consisting of four polarizing beam splitters (PBSs). The time delays between different paths were finely adjusted to ensure that the photons arrived at the PBSs simultaneously. After the PBSs, the single photons are coupled into single-mode fibers and detected by silicon single-photon detectors.

As the PBSs transmit $H$ and reflect $V$ polarizations, it is straightforward to check that after the four PBSs, if we detect one and only one single photon in each output port (i.e., a ten-photon coincidence), the ten photons are projected into the Schrödinger cat like state[2], also known as a Greenberger-Horne-Zeilinger (GHZ) state[3]:

$$\left| \psi^{10} \right\rangle = (\left| H \right\rangle^{\otimes 10} + \left| V \right\rangle^{\otimes 10}) / \sqrt{2} \ .$$

This is because the ten-photon coincidence events can only originate from either the case that all the photons are $H$ polarized or all are $V$ polarized—two possibilities quantum mechanically indistinguishable. The entangled states and their creation process can be intuitively illustrated in Fig.2b using graph-state presentation[23], where each vertex represents a qubit and each edge indicates a controlled logic gate having been applied between the two connected qubits. The combination of two single photons at a PBS can be described by the parity-check operator: $\left| H \right\rangle \left| H \right\rangle \left\langle H \right| \left\langle H \right| + \left| V \right\rangle \left| V \right\rangle \left\langle V \right| \left\langle V \right|$, which leads to the "fusion" of separate two-qubit graph into a larger star-shaped graph state. Other $N$-photon GHZ states (e.g. $N = 2,4,6,8$) in the form $\left| \psi^{N} \right\rangle = (\left| H \right\rangle^{\otimes N} + \left| V \right\rangle^{\otimes N}) / \sqrt{2}$ can be engineered step by step using similar method in the process of generating the ten-photon GHZ state.

To demonstrate the $N$-photon coherence of the generated GHZ states, we measure the $N$ single photons individually along the basis of $(\left| H \right\rangle \pm e^{i\theta} \left| V \right\rangle) / \sqrt{2}$ and detect all the $2^{N}$ combinations of $N$-photon output. From these measurements, we obtain the experimentally estimated expectation values of the observable $M_{\theta}^{\otimes N} = (\cos \theta \sigma_{x} + \sin \theta \sigma_{y})^{\otimes N}$. The coherence of the $N$-qubit GHZ state, which is defined by the off-diagonal element of its density matrix and reflects the coherent superposition between the $\left| H \right\rangle^{\otimes N}$ and $\left| V \right\rangle^{\otimes N}$ component of the GHZ state, can be calculated by[24]

$$C^{N} = \frac{1}{N} \sum_{k=0}^{N-1} (-1)^{N} \left\langle M_{(k\pi/N)}^{\otimes N} \right\rangle.$$

For the 2-, 4-, 6-, 8-, and 10-photon GHZ states, the coherence is calculated to be 0.9305(3), 0.750(6), 0.612(28), 0.538(29), and 0.438(27) respectively, from measurements at certain angles $\theta = k\pi / N$, $k = 0, 1, ..., N-1$.

Figure 3 shows the experimentally obtained expectation values of $\left\langle M_\theta^{\otimes N} \right\rangle$ ($N$=1, 2, 4, 6, 8) with $\theta$ ramping continuously from 0 to $\pi$. The data are fitted to sinusoidal fringes that show an $N$-times increase of the oscillatory frequencies for the $N$-photon GHZ states, with a gradual reduction of fringe visibility which corresponds to the coherence of the GHZ states as mentioned above. Ideally, for $N = 1$, i.e., a single-photon state $(|H\rangle + |V\rangle) / \sqrt{2}$, the expectation value is $\propto \cos \theta$. For $N$-photon GHZ state $|\psi^N\rangle = (|H\rangle^{\otimes N} + |V\rangle^{\otimes N}) / \sqrt{2}$ where all the entangled photons collectively responds to the phase change, the expectation value is $\propto \cos N\theta$. This observed $N\theta$ oscillation behaviour highlights the potential of the GHZ states for entanglement-enhanced super-resolving phase measurements[25].

For a more complete characterization of the $N$-photon states, we measure their state fidelities, that is, the overlap of the experimentally produced state with the ideal one: $F(\psi^N) = \left\langle \psi^N \middle| \rho_{\text{exp}} \middle| \psi^N \right\rangle = Tr(\rho_{\text{ideal}} \rho_{\text{exp}})$ and $\rho_{\text{ideal}} = |\psi^N\rangle\langle\psi^N| = (P^N + C^N) / 2$, where $P^N = (|H\rangle\langle H|)^{\otimes N} + (|V\rangle\langle V|)^{\otimes N}$ denotes the population of the $|H\rangle^{\otimes N}$ and $|V\rangle^{\otimes N}$ components of the GHZ state. The fidelity can be calculated by the expectation values of the average of the population and coherence $F(\psi^N) = (\left\langle P^N \right\rangle + \left\langle C^N \right\rangle) / 2$. The experimental data for the ten-photon state is shown in Fig. 4. Under a laser pump power of 0.57 W, we achieve a ten-photon count rate of ~4 per hour, and a state fidelity of 0.573(23). For the GHZ-type entangled states, if the measured fidelity exceeds the threshold of 0.5, we can establish the presence of genuine multipartite entanglement[26], excluding the produced state from any incompletely entangled (e.g. biseparable) state. Meanwhile, we systematically measured the state fidelities of 2-, 4-, 6-, and 8-photon GHZ states together with their count rate following the same method, which are listed in Table I.

The main source of noise causing the imperfection of the $N$-photon states are from the double pair emission of SPDC. Increasing the laser pump power can boost the photon pair generation probability but also the double pair emission rate. As an example, we further tested with a laser pump power of 0.7 W where we detected two-photon count rate of 1.5 million per second. Under this condition, we detected a higher ten-photon count rate of ~11 per hour, but a reduced state fidelity of 0.429(21) (see Supplementary Fig.S1 for the data), demonstrating the detrimental effect of high-order SPDC emission. This fidelity is below the threshold of 0.5 for proving the genuine ten-particle entanglement. A less stringent criteria is called distillable entanglement[27], meaning that out of many copies of the imperfectly created states, ten-photon entanglement can be produced by local operations and classical communications. Using this criteria, our data proves ten-photon distillable entanglement with a statistical significance of 7.2 standard deviations (see Supplementary Information). Using high-efficiency photon-number-discriminating single-photon detectors, the noise from double pair emission of SPDC can be eliminated.

In summary, we have demonstrated the first ten-photon entanglement in experiment. The ability to control ten single photons will enable many previously challenging experiments such as realization of universal quantum error correction code[28], teleportation of three degrees of freedom in a single photon[29] and ground-to-satellite teleportation overcoming high channel loss[30,31]. The combination of a high collection efficiency and a high indistinguishability in SPDC is a crucial for scalable photonic quantum technologies[11-13]. The methods developed here are of particular interest for scattershot boson sampling[32] to significantly enhance $N$-photon count rate, which may provide a promising route to demonstrate quantum supremacy using photonic quantum simulators in the near term.

## References

1. Einstein, A., Podolsky, B. & Rosen, N. Can quantum-mechanical description of physical reality be considered complete? *Phys. Rev.* **47**, 777–780 (1935).

2. Schrödinger, E. Die Gegenwartige Situation in der Quantenmechanik. *Naturwissenschaften* **23**, 807–812, 823–828, 844–849 (1935).

3. Greenberger, D. M., Horne, M., Shimony, A. & Zeilinger, A. Bell's theorem without inequalities. *Am. J. Phys.* **58**, 1131–11 43 (1990).

4. Leggett, A. J. Realism and the physical world. *Rep. Prog. Phys.* **71**, 02200 1 (2008).

5. Shor, P. W. in *Proceedings of the 35th Annual Symposium on Foundations of Computer Science* (ed. Goldwasser, S.) 124–134 (IEEE Computer Society Press, 1994).

6. Lloyd, S. Universal quantum simulators. *Science* **273**, 1073–1078 (1996).

7. Zoller, P. *et al*. Quantum information processing and communication. *Eur. Phys. J. D* **36**, 203–228 (2005).

8. Monz, T. *et al*. 14-qubit entanglement: creation and coherence. *Phys. Rev. Lett.* **106**, 130506 (2011).

9. Yao, X.- C. *et al*. Observation of eight-photon entanglement. *Nature Photon*. **6**, 225-228 (2012).

10. Barends, R. *et al*. Superconducting quantum circuits at the surface code threshold for fault tolerance. *Nature* **508**, 500–503 (2014).

11. Kok, P. *et al.* Linear optical quantum computing with photonic qubits. *Rev. Mod. Phys.* **79**, 135-174 (2007).

12. Aaronson, S. & Arkhipov, A. *in Proceedings of the 43rd Annual ACM Symposium on Theory of Computing, San Jose* 333–342 (ACM Press, 2011).

13. Pan, J.-W. *et al*. Multiphoton entanglement and interferometry. *Rev. Mod. Phys*. **84**, 777–838 (2012).

14. Zhang, C. *et al*. Experimental Greenberger-Horne-Zeilinger-type six-photon quantum nonlocality. *Phys. Rev. Lett.* **115**, 260402 (2015).

15. Takeuchi, S. Beamlike twin-photon generation by use of type II parametric downconversion. *Opt. Lett.* **26**, 843-845 (2001).

16. Kurtsiefer, C., Oberparleiter, M. & Weinfurter, H. Generation of correlated photon pairs in type-II parametric downconversion – revisited. *J. Mod. Opt.* **48**, 1997-2007 (2001).

17. Kwon, O., Cho, Y.-W. & Kim, Y.-H. Single-mode coupling efficiencies of type-II spontaneous parametric down-conversion: Collinear, noncollinear, and beamlike phase matching. *Phys. Rev. A* **78**, 053825 (2008).

18. Kim, Y.-H. Quantum interference with beamlike type-II spontaneous parametric down-conversion. *Phys. Rev. A* **68**, 013804 (2003).

19. Niu, X.-L. *et al*. Beamlike high-brightness source of polarization-entangled photon pairs. *Opt. Lett.* **33**, 968-970 (2008).

20. Kwiat, P. G. et al. New high-intensity source of polarization-entangled photon pairs. *Phys. Rev. Lett*. **75**, 4337–4341 (1995).

21. Kim, Y.-H., Kulik, S. P., Chekhova, M. V., Grice, W. P. & Shih, Y. Experimental entanglement concentration and universal Bell-state synthesizer. *Phys. Rev. A* **67**, 010301 (2003).

22. Mosley, P. J. *et al.* Heralded generation of ultrafast single photons in pure quantum states. *Phys. Rev. Lett.* **100**, 133601 (2008).

23. Hein, M., Eisert, J. & Briegel, H.-J. Multiparty entanglement in graph states. *Phys. Rev. A* **69**, 062311 (2004).

24. Sackett, C. A. *et al.* Experimental entanglement of four particles. *Nature* **404**, 256–259 (2000).

25. Giovannetti, V., Lloyd, S. & Maccone, L., Advances in quantum metrology. *Nature Photonics* **5**, 222–229 (2011)

26. Gühne, O. & Toth, G. Entanglement detection. *Phys. Rep.* **474**, 1–75 (2009).

27. Dür, W. & Cirac, J. I. Multiparticle entanglement and its experimental detection. *J. Phys. A: Math. Gen.* **34** 6837–6850 (2001).

28. P. W. Shor. Scheme for reducing decoherence in quantum computer memory. *Phys. Rev. A.* **52**, R2493 (1995).

29. Wang X.-L. *et al.* Quantum teleportation of multiple degrees of freedom in a single photon. *Nature* **518**, 516-519 (2015).

30. Yin, J. *et al*. Quantum teleportation and entanglement distribution over 100-kilometre free-space channels. *Nature* **488**, 185-188 (2012).

31. Ma, X.-S. *et al*. Quantum teleportation over 143 kilometres using active feed-forward. *Nature* **489**, 269-273 (2012).

32. Lund, A. P. *et al.* Boson sampling from a Gaussian state. *Phys. Rev. Lett.* **113**, 100502 (2014).

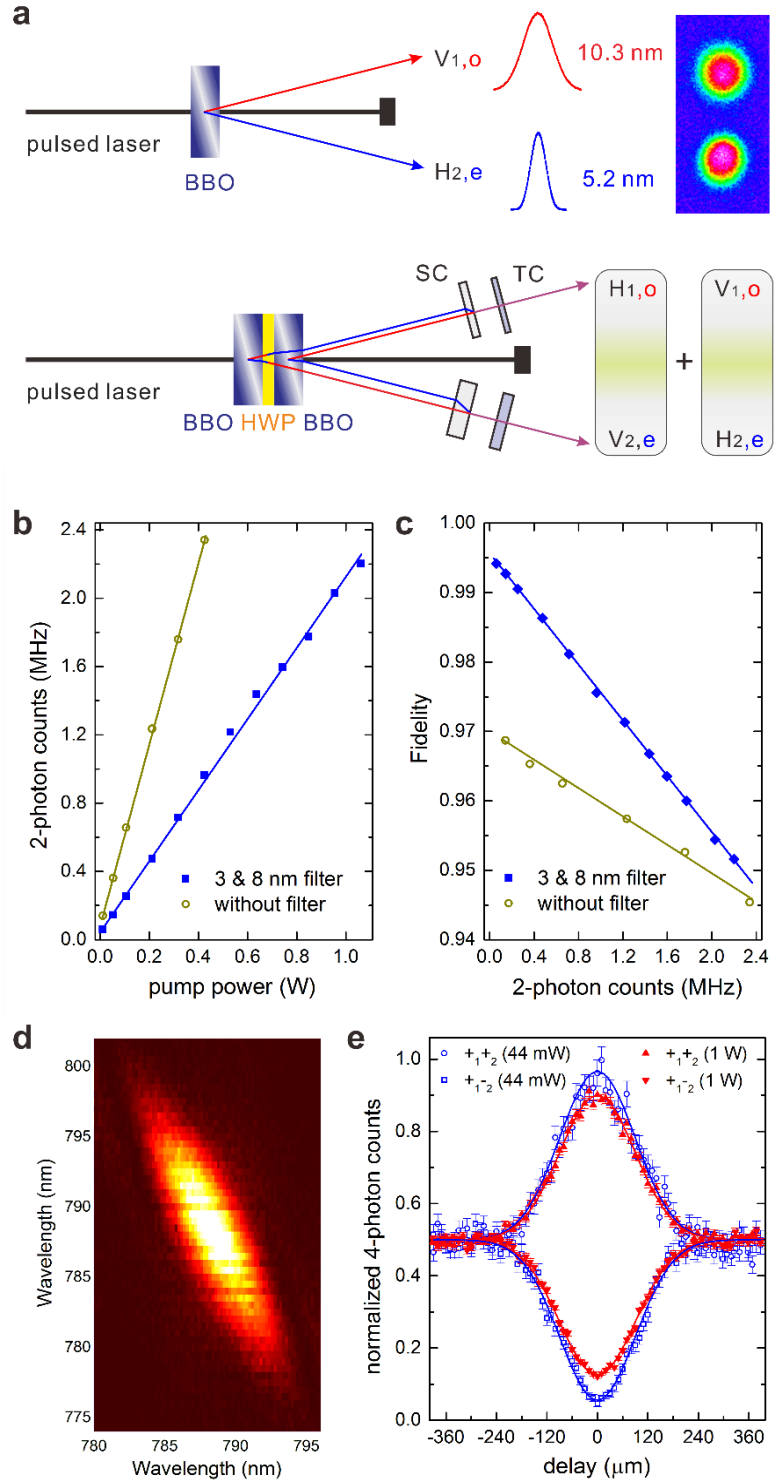

**Figure 1 | Near-optimal entangled photons from SPDC with high collection efficiency and photon indistinguishability. a,** Beamlike type-II SPDC. The top panel shows a single BBO setup for generating signal-idler photon pairs in two separate circular beams, evident by the image of the photo emission profile taken using an EMCCD. The spectra for the *e* and *o* photons are measured using a spectrometer, showing a different bandwidth of 5.2 nm and 10.3

nm, respectively. The bottom panel shows a sandwich-like BBO+HWP+BBO geometry for generating entangled photons, after careful birefringent compensations. **b,** Experimentally detected two-photon count rate as a function of laser pump power, with and without narrowband filtering. **c,** The entangled state fidelity at different two-photon count rate. **d,** Measurement of the joint spectrum of the photon pair. **e,** Test of the indistinguishability between the single photons created from independent SPDC through PBS-based two-photon interference experiments at different pump power (see *Supplemental Materials*).

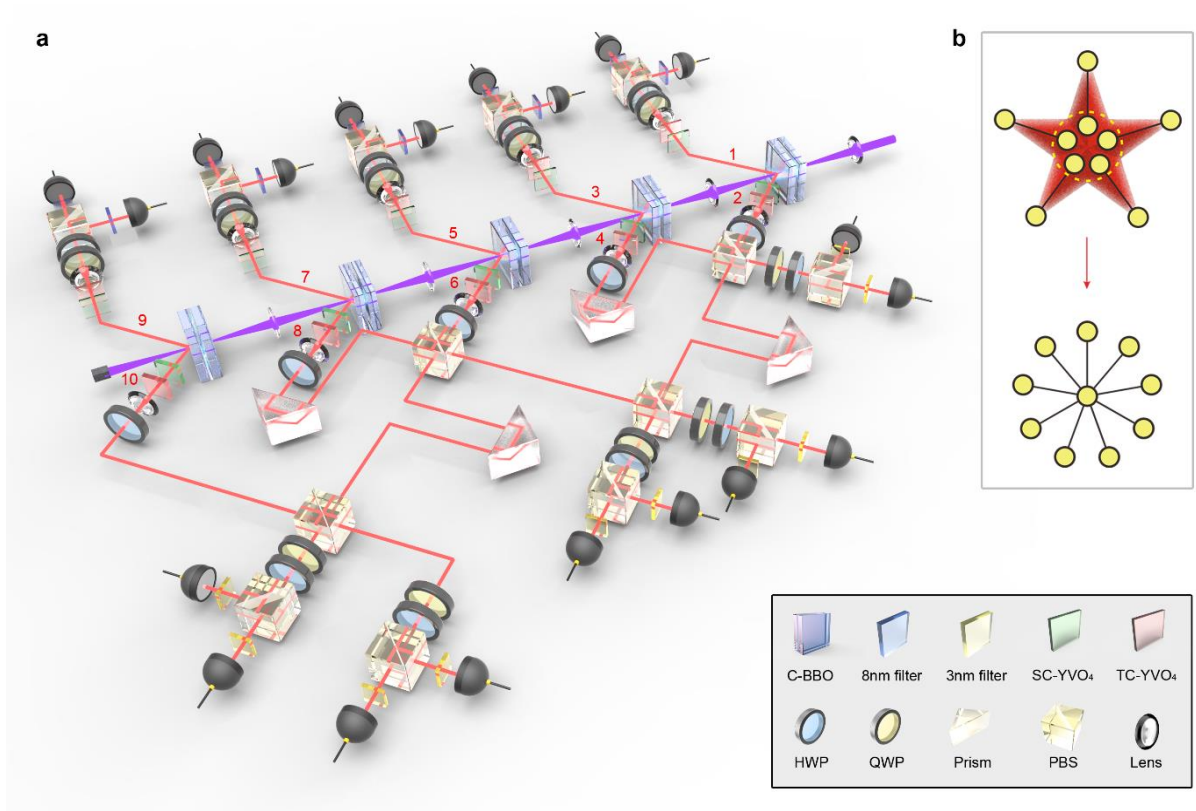

**Figure 2 | Experiment setup for generating ten-photon polarization-entangled GHZ state**. **a,** A pulsed ultraviolet laser was focused on five sandwich-like bulks, consisted of two beam-like type II β-barium borate (BBO) crystals and one HWP, to produce five entangled photon pairs in spatial modes 1-2, 3-4, 5-6, 7-8 and 9-10. All the *o* photons propagated in the modes 1, 3, 5, 7 and 9, while the *e* photons propagated in the modes 2, 4, 6, 8 and 10 and were combined on four PBSs. Four prisms were utilized to adjust delay to ensure that the incoming

photons simultaneously arrive at the PBSs. All the *e* photons were spectrally filtered by 3 nm filters, and all the *o* photons with 8 nm filters. The outputs were detected by 20 single-photon detectors where all the 1024 ten-photon coincidence events were simultaneously recorded by a homemade coincidence count system. **b,** Graph state representation of the process to combine the five separate photon pairs into the ten-photon entangled GHZ state. The graph states can be associated with graphs where each vertex represents a qubit prepared in the state $(|H\rangle + |V\rangle)/\sqrt{2}$ and each edge represents a controlled phase gate having been applied between the two connected qubits.

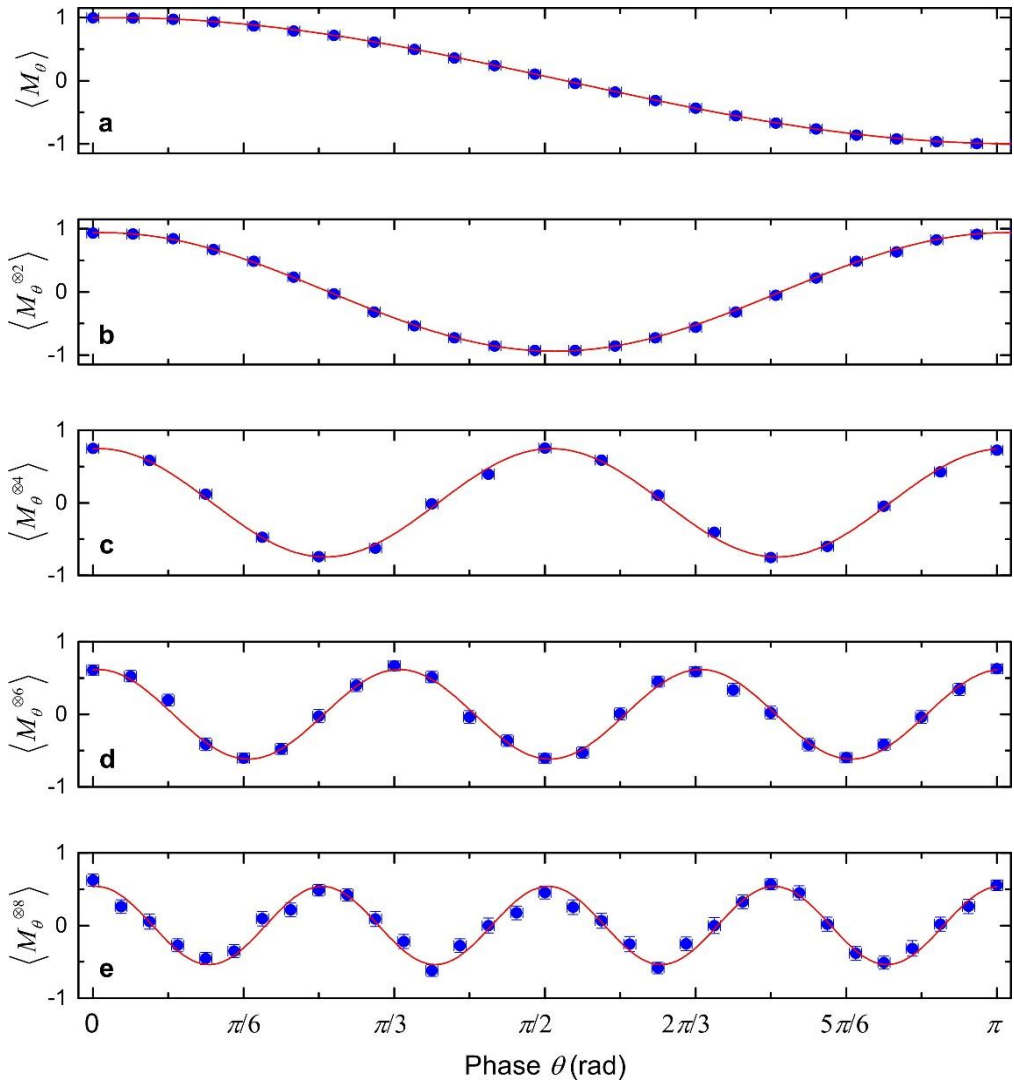

**Figure 3 | *N*-photon coherence.** The *N* single photons were individually measured in the basis of $(|H\rangle \pm e^{i\theta}|V\rangle)/\sqrt{2}$, which were eventually measured by $2N$ single-photon detectors where all the $2^N$ *N*-photon coincidence events were simultaneously recorded. **a**, *N*=1. **b**, *N*=2. **c**, *N*=4. **d**, *N*=6. **e**, *N*=8. Each of the $2^N$ *N*-photon events signals the observation of an eigenstate of the observable $M_\theta^{\otimes N}$ with corresponding eigenvalue of $v_j = +1$ or $v_j = -1$. From the relative probabilities of the *N*-photon detection events $p_j$, $j = 1,...,2^N$, we can then compute the expectation values of the observables by $\left\langle M_\theta^{\otimes N} \right\rangle = \sum_{j=1}^{2^N} p_j v_j$. The x axis is the phase shift $\theta$ between *H* and *V*. The y axis is the experimentally obtained $\left\langle M_\theta^{\otimes N} \right\rangle$. The error bars represent one standard deviation (in **a**, **b** and **c**, the error bars are smaller than the data points), calculated from Poissonian counting statistics of the raw detection events.

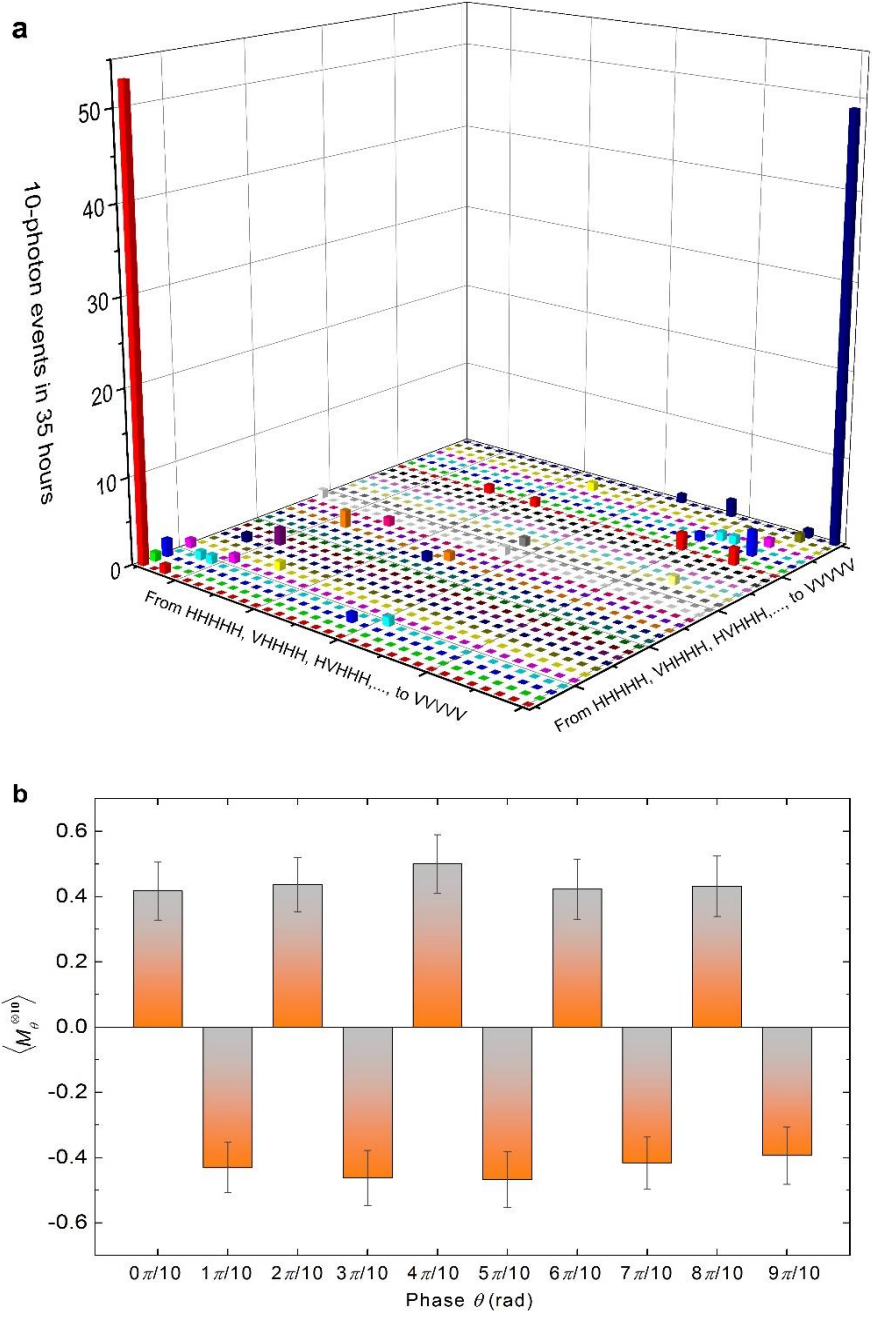

**Figure 4 | Experiment results for ten-photon genuine entanglement at pump power of 0.57 W. a**, Ten-photon coincidence counts in $|H\rangle/|V\rangle$ basis accumulated for 35 hours. **b**, Expectation values of $M_{k\pi/10}^{\otimes 10} = \cos(k\pi/10)\sigma_x + \sin(k\pi/10)\sigma_y$ $(k=0,1,...9)$ obtained by the measurement in the basis of $(|H\rangle \pm e^{ik\pi/10}|V\rangle)/\sqrt{2}$. Each setting is measured in ~26 hours to accumulate ~100 events. The error bars represent one standard deviation, calculated from Poissonian counting statistics of the raw detection events.

| $N$-photon GHZ state | 2 (*) | 4 (*) | 6 (*) | 8 (*) | 10 (*) | 10 (**) |
|---|---|---|---|---|---|---|
| Fidelity | 0.9720(1) | 0.833(4) | 0.710(16) | 0.644(22) | 0.573(23) | 0.429(21) |
| Distillable entanglement, $\sigma$ | 5215 | 122.3 | 20.8 | 17.7 | 15.2 | 7.2 |
| Genuine entanglement, $\sigma$ | 4334 | 84.6 | 13.3 | 6.5 | 3.1 | -3.4 |
| Count rate (Hz) | 1200000 | 6000 | 39 | 0.2 | 0.0011 | 0.0031 |

\* under a laser pump power of 0.57 W.          \*\* under a laser pump power of 0.7 W.

**Table I | Summary of the measured $N$-photon count rate, state fidelities and tests of $N$-qubit entanglement.** The errors in brackets represent one standard deviation, deduced from propagated Poissonian counting statistics of the raw detection events. Both distillable and genuine entanglement criteria were supported by $\sigma$ standard deviations.

# Experimental ten-photon entanglement

Xi-Lin Wang, Luo-Kan Chen, Wei Li, He-Liang Huang, Chang Liu, Chao Chen, Yi-Han Luo, Zu-En Su, Dian Wu, Zheng-Da Li, He Lu, Yi Hu, Xiao Jiang, Cheng-Zhi Peng, Li Li, Nai-Le Liu, Yu-Ao Chen, Chao-Yang Lu, and Jian-Wei Pan

*Hefei National Laboratory for Physical Sciences at Microscale and Department of Modern Physics, University of Science and Technology of China,*
*& CAS Centre for Excellence and Synergetic Innovation Centre in Quantum Information and Quantum Physics, Hefei, Anhui 230026, China,*
*& CAS-Alibaba Quantum Computing Laboratory, Shanghai, 201315, China*

## S1. Accurate assembling of double BBOs

We use a sandwich-like BBO+HWP+BBO geometry to generate entangled photons. To minimize the separation of the two BBOs, a 44-μm-thick true-zero-order HWP is utilized. As the propagation direction of down-converted photons are very sensitive to the angle between the optic axis of BBO and the incident direction of the pump beam, the two optical axes of left and right BBOs must be placed parallel to each other (numerical simulation shows that a twist within 0.7 mrad is necessary). In this work, we assembled the BBOs and HWP as a bulk fixed in an adjustable mount, where the two BBOs are accurately cut and assembled. This achieves a high stability and robustness required for the multi-photon measurements.

## S2. Compact Bell-state synthesizer

In type-II SPDC with ultrafast laser pulse as the pump source, the emitted $e$ and $o$ light exhibits different spectral linewidth, whose full width at maximum (FWHM) was measured to be 5.2 nm and 10.3 nm (see spectra in Fig.1a), respectively. The BBO+HWP+BBO structure enables the $o$ and $e$ light are emitted into two different spatial modes 1 and 2, respectively. Thus, in a later stage of narrowband filtering, we can choose an independent, more efficient scheme: 3 nm filtering for the $e$ photon and 8 nm filtering for the $o$ photon. This is in similar spirit as the interferometric Bell-state synthesizer which had been used in the previous eight-photon experiment [S1] to increase the photon collection efficiency by ~60% compared to the $e$ and $o$ spatially mixed SPDC photon pair sources [S2]. In addition, the sandwich-like geometry doesn't require interferometric setup as in ref. S1 and is more compact and stable.

### S3. Spatial and temporal walk-offs compensations

As shown in the bottom panel of Fig. 1a, polarization-entangled photon pairs in the state $(\left|H_{1,o}\right\rangle\left|V_{2,e}\right\rangle+\left|V_{1,o}\right\rangle\left|H_{2,e}\right\rangle)/\sqrt{2}$ were generated from two BBOs sandwiching a HWP. The spatial walk-off were resulted from two aspects: the different pumping positions in the two BBOs, and the different propagation directions for $e$ and $o$ photons. We note that the spatial and temporal work-off compensations in our experiment were crucial for both achieving the record-high collection efficiency into a single-mode fibre (70%) and near-unity two-photon fidelity (99.5% at low power). We utilized a birefringent YVO$_4$ crystal with a thickness of 2.3 (0.7) mm cut at $45°$ with respective to the propagation direction in the horizontal plane to compensate the spatial walk-off in the $e$ ($o$) arm. The temporal walk-off is due to the different optical paths and velocities for $e$ and $o$ photons. We compensated the total temporal walk-off using a 1.36 (0.45) mm-thick YVO$_4$ cut at $90°$ with respective to the horizontal direction in the transverse plane for the $e$ ($o$) photons.

### S4. Quantum interference between two independent photons

To measure the quantum interference between two single photons from two independent SPDC, we first prepare heralded $e$ photon from each pair in the $\left|+\right\rangle=(\left|H\right\rangle+\left|V\right\rangle)/\sqrt{2}$ state. These two photons simultaneously arrived at a PBS from two orthogonal directions, and only when the two photons emit in different output ports (denoted as 1 and 2, as shown in Fig. 1e), they are detected in the basis of $\left|\pm\right\rangle=(\left|H\right\rangle\pm\left|V\right\rangle)/\sqrt{2}$. For two fully indistinguishable single photons, the output state can be written as a coherent superposition of $\left|H\right\rangle\left|H\right\rangle$ and $\left|V\right\rangle\left|V\right\rangle$, which can be written in the diagonal basis as: $\left|+\right\rangle_1\left|+\right\rangle_2+\left|-\right\rangle_1\left|-\right\rangle_2$. Therefore, the component $|+\rangle_1 |+\rangle_2$ will obtain an interference enhancement, while $|+\rangle_1 |-\rangle_2$ will see a suppression, as evident in Fig.1.

## S5. A comparison with other high-performance SPDC sources

For multi-photon experiment, the SPDC source must simultaneously fulfill the following check list: 1. high brightness (two-photon counts/mW/s), 2. High collection efficiency, 3. High indistinguishability between independently generated photons, and 4. The photons should be excited by ultrafast (typically a few hundred femtoseconds duration) laser pulses to ensure the jitter of the arrival time of the independent photons overlapping on a beam splitter are much smaller than the coherence time of the photons (usually a few nanometers, defined by the narrowband filters). An excellent SPDC calculator for various crystals and parameters was made available online by L. K. Shalm at http://spdcalc.org/, which can be useful for designing optimal SPDC sources for different applications.

**Table S5.1: We make a comparison of the SPDC source used in this work with other high-performance sources.**

| Ref. | Crystal & Thickness | Pump | Wavelength (nm) | Brightness (Hz/mW) | Overall efficiency (collection+detection) | purity |
|---|---|---|---|---|---|---|
| [1] | Type-II BBO, 2 mm | Pulsed, 120 fs | 390: 780+780 | ~350 | ~26.5% | 91% |
| [2] | Type-II BBO, 1 mm | Pulsed, 140 fs | 390: 780+780 | ~2000 | ~29% | 93% |
| [3,4] | Type-II PPKTP, 30 mm | Pulsed, ps | 772: 1544+1544 | ~130 | ~22% | 91% |
| [5] | Type-I BiBO, 0.2 mm | Pulsed, 5 ps | 355: 710+710 | ~1300 | ~75% | ~5% |
| [6] | Type-II PPKTP, 20mm | Pulsed, 160fs | 521: 775+1590 | ~ 334 | ~44% (775 nm) ~7% (1590 nm) | 91% |
| [7] | Type-II PPKTP | Pulsed, 12 ns | 405: 810+810 | - | ~77.4% | - |
| [8] | Type-II PPKTP | Pulsed, ps | 775: 1550+1550 | - | ~75.2% | - |
| This work | Type-II BBO, 2 mm | Pulsed, 150fs | 394: 788+788 | ~2000 | ~42% | 91% |

1   Yao, X.-C. *et al*, Observation of eight-photon entanglement. Nature Photon. 6, 225-228 (2012).

2   Zhang, C. *et al*, Experimental Greenberger-Horne-Zeilinger-type six-photon quantum nonlocality. Phys. Rev. Lett. 115, 260402 (2015).

3   T. Guerreiro *et al*, High efficiency coupling of photon pairs in practice. Opt. Express 21, 27641-27651 (2013).

4   N. Bruno *et al*, Pulsed source of spectrally uncorrelated and indistinguishable photons at telecom wavelengths. Opt. Express 22, 17246-17253 (2014).

5   B. G. Christensen *et al*, Detection-loophole-free test of quantum nonlocality, and applications. Phys. Rev. Lett. 111, 130406 (2013).

6   F. Kaneda *et al*, Heralded single-photon source utilizing highly nondegenerate, spectrally factorable spontaneous parametric downconversion. Opt. Express 24, 10733-10747 (2016).

7   M. Giustina *et al*, Significant-loophole-free test of Bell's theorem with entangled photons. Phys. Rev. Lett. 115, 250401 (2015).

8   L. K. Shalm *et al*, Strong loophole-free test of local realism. Phys. Rev. Lett. 115, 250402 (2015).

## S6. Post-selective coincidence detection

We employ 20 single-photon detectors and a 20-chanel homemade coincidence count module to register a complete set of the 1024 possible combinations of ten-photon coincidence events. As this dual-channel setup can simultaneously measure the two orthogonal polarization states of a single photon by two single-photon detectors, as in the previous eight-photon entanglement experiment [S1], this coincidence detection unit partially eliminate higher-order events. In the ideal case, only one detector in each output mode will be fired. However, if both detectors are fired simultaneously that result in 11-fold coincidence, this indicates that at least one SPDC emits undesired double pairs, which in this way can be detected—but only partially, as limited by the non-unity collection and detection efficiency—and eliminated.

## S7. Experimental parameters and characterization of losses:

(1) The detected count rate of the five pair of entangled photons under 0.57 W pump laser power with 76 MHz repetition rate are: 1.2 MHz, 1.07 MHz, 0.97 MHz, 0.89 MHz, and 0.78 MHz, respectively. This was mainly owing to the attenuation of the pump laser power after passing through each HWP-sandwiched BBO crystal. The attenuation comes from two parts: (a). Total transmission loss at the six crystal surface (two BBOs and one HWP), which is about 5%. (b). the true zero-order HWP, which was optimally designed at 788 nm wavelength, was not a perfect full-wave plate at 394 nm. It rotated the 394 pump laser polarization and caused a loss of 5% of effective power.

(2) Four PBSs. (a) there is only 50% probability that two incoming photons will exit at two different output port of the PBS, at which case the multi-photon entanglement are generated by post-selection. The overall post-selection probability is $(1/2)^4$ in our work using four PBSs. (b) the slight spatial mode mismatch for the two-photon interference at the four PBSs caused an overall loss of about 4%, due to imperfect optical alignment.

(3). Optical transmission. The setup consists of 40 mirrors, 10 lens, 24 wave plates, and 14 PBSs. The mirrors have reflectivity exceeding 99.9% while the lens, wave plates and PBSs are anti-reflection coated with transmission rate over 99.8%. Overall the optical path loss is about 15%.

## S8. Experimentally measured raw data to calculate the fidelity of the ten-photon state

The fidelity of the 10-photon GHZ can be calculated by the expectation values of the average of the population and coherence as $F(\psi^{10}) = (\langle P^{10} \rangle + \langle C^{10} \rangle)/2$, where $P^{10} = (|H\rangle\langle H|)^{\otimes 10} + (|V\rangle\langle V|)^{\otimes 10}$ denotes the population of the $|H\rangle^{\otimes 10}$ and $|V\rangle^{\otimes 10}$ components of the GHZ state, $C^{10} = (1/10)\sum_{k=0}^{k=9}(-1)^k M_{k\pi/10}^{\otimes 10}$ denotes the off-diagonal element of its density matrix and reflects the coherent superposition between the $|H\rangle^{\otimes 10}$ and $|V\rangle^{\otimes 10}$ component of the GHZ state, here $M_{k\pi/10} = \cos(k\pi/10)\sigma_x + \sin(k\pi/10)\sigma_y$. To calculate the expectation value of population $\langle P^{10} \rangle$, we measure each photon of the ten-photon

GHZ state in the basis of $|H\rangle / |V\rangle$ and obtain a complete set of the 1024 possible combinations of ten-photon coincidence events as shown in table S8.1, then we can calculate the probability of $(|H\rangle\langle H|)^{\otimes 10} + (|V\rangle\langle V|)^{\otimes 10}$ and obtain the value of the population $\langle P^{10} \rangle$. To calculate the coherence, we have to measure each photon in the basis of $(|H\rangle \pm e^{ik\pi/10}|V\rangle)/\sqrt{2}$ ($k = 0, 1, 2, 3, \ldots, 9$), which are the eigenstates of the observer $M_{k\pi/10} = \cos(k\pi/10)\sigma_x + \sin(k\pi/10)\sigma_y$ with eigenvalues of $\pm 1$, indicated by the state $|+\rangle / |-\rangle$. Then we can obtain the 1024 possible combinations of ten-photon coincidence events as shown in table S8.2-S8.11, which correspond to a complete set of 1024 eigenstates of $M_{k\pi/10}^{\otimes 10}$ as $|+\rangle^{\otimes 10}$, $|+\rangle^{\otimes 9}|-\rangle$, $\cdots$ $|-\rangle^{\otimes 10}$ and the eigenvalue of $M_{k\pi/10}^{\otimes 10}$ is +1 (with even $|-\rangle$ in the eigenstate ) or -1 (with odd $|-\rangle$ in the eigenstate ). Further, we can get the probability of obtaining the +1 eigenvalue of $M_{k\pi/10}^{\otimes 10}$ denoted by $\mathrm{Pr}_{k+}^{10}$, and -1 eigenvalue of $M_{k\pi/10}^{\otimes 10}$ denoted by $\mathrm{Pr}_{k-}^{10}$. The expectation value of $M_{k\pi/10}^{\otimes 10}$ could be calculated as $\langle M_{k\pi/10}^{\otimes 10}\rangle = \mathrm{Pr}_{k+}^{10} - \mathrm{Pr}_{k-}^{10}$. Next, we can calculate the coherence $\langle C^{10}\rangle = (1/10)\sum_{k=0}^{k=9}(-1)^k\langle M_{k\pi/10}^{\otimes 10}\rangle$.

**TABLE S8.1: The experimentally measured raw data of a complete set of the 1024 possible combinations of ten-photon coincidence events with each photon measured in the basis of $|H\rangle / |V\rangle$ (indicated by the H/V label in the first column and the first line) to obtain $\langle P^{10}\rangle$.** The five H or V label in the first column represent the measurement base choice for the five *o* photons, while the five H or V label in the first line represent the measurement base choice for the five *e* photons.

| | HH HH H | VH HH H | HV HH H | VV HH H | HH VH H | VH VH H | HV VH H | VV VH H | HH HV H | VH HV H | HV HV H | VV HV H | HH VV H | VH VV H | HV VV H | VV VV H |
|---|---|---|---|---|---|---|---|---|---|---|---|---|---|---|---|---|
| HH HH H | 53 | 0 | 1 | 0 | 0 | 0 | 0 | 0 | 0 | 0 | 0 | 0 | 0 | 0 | 0 | 0 |
| VH HH H | 1 | 0 | 0 | 0 | 0 | 0 | 0 | 0 | 0 | 0 | 0 | 0 | 0 | 0 | 0 | 0 |
| HV HH H | 2 | 0 | 0 | 0 | 0 | 0 | 0 | 0 | 0 | 0 | 0 | 0 | 0 | 0 | 0 | 0 |
| VV HH H | 0 | 0 | 1 | 1 | 0 | 0 | 0 | 0 | 0 | 0 | 0 | 0 | 0 | 0 | 0 | 0 |
| HH VH H | 1 | 0 | 0 | 0 | 1 | 0 | 0 | 0 | 0 | 0 | 0 | 0 | 0 | 0 | 0 | 0 |
| VH VH H | 0 | 0 | 0 | 0 | 0 | 0 | 0 | 1 | 0 | 0 | 0 | 0 | 0 | 0 | 0 | 0 |
| HV VH H | 0 | 0 | 0 | 0 | 0 | 0 | 0 | 0 | 0 | 0 | 0 | 0 | 0 | 0 | 0 | 0 |
| VV VH H | 0 | 0 | 1 | 0 | 0 | 0 | 0 | 0 | 0 | 0 | 0 | 0 | 0 | 0 | 0 | 0 |
| HH HV H | 0 | 0 | 0 | 0 | 2 | 0 | 0 | 0 | 0 | 0 | 0 | 0 | 0 | 0 | 0 | 0 |
| VH HV H | 0 | 0 | 0 | 0 | 0 | 0 | 0 | 0 | 0 | 0 | 0 | 0 | 0 | 0 | 0 | 0 |
| HV HV H | 0 | 0 | 0 | 0 | 0 | 0 | 0 | 0 | 0 | 0 | 0 | 0 | 0 | 0 | 0 | 0 |
| VV HV H | 0 | 0 | 0 | 0 | 0 | 0 | 0 | 0 | 0 | 0 | 0 | 0 | 0 | 0 | 0 | 0 |
| HH VV H | 0 | 0 | 0 | 0 | 0 | 0 | 0 | 0 | 0 | 0 | 0 | 0 | 0 | 1 | 0 | 0 |
| VH VV H | 0 | 0 | 0 | 0 | 0 | 2 | 0 | 0 | 0 | 0 | 0 | 0 | 0 | 0 | 1 | 0 |
| HV VV H | 0 | 0 | 0 | 0 | 0 | 0 | 0 | 0 | 0 | 0 | 0 | 0 | 0 | 0 | 0 | 0 |
| VV VV H | 0 | 0 | 0 | 0 | 0 | 0 | 0 | 1 | 0 | 0 | 0 | 0 | 0 | 0 | 0 | 0 |
| HH HH V | 1 | 0 | 0 | 0 | 0 | 0 | 0 | 0 | 0 | 0 | 0 | 0 | 0 | 0 | 0 | 0 |
| VH HH V | 0 | 0 | 0 | 0 | 0 | 0 | 0 | 0 | 0 | 0 | 0 | 0 | 0 | 0 | 0 | 0 |
| HV HH V | 0 | 0 | 0 | 0 | 0 | 0 | 0 | 0 | 0 | 0 | 0 | 0 | 0 | 0 | 0 | 0 |
| VV HH V | 0 | 0 | 0 | 0 | 0 | 0 | 0 | 0 | 0 | 0 | 0 | 0 | 0 | 0 | 0 | 0 |
| HH VH V | 0 | 0 | 0 | 0 | 0 | 0 | 0 | 0 | 0 | 0 | 0 | 0 | 0 | 0 | 0 | 0 |
| VH VH V | 0 | 0 | 0 | 0 | 0 | 0 | 0 | 0 | 0 | 0 | 0 | 0 | 0 | 0 | 0 | 0 |
| HV VH V | 0 | 0 | 0 | 0 | 0 | 0 | 0 | 0 | 0 | 0 | 0 | 0 | 0 | 0 | 0 | 0 |
| VV VH V | 0 | 0 | 0 | 0 | 0 | 0 | 0 | 0 | 0 | 0 | 0 | 0 | 0 | 0 | 0 | 0 |
| HH HV V | 0 | 0 | 0 | 0 | 0 | 0 | 0 | 0 | 1 | 0 | 0 | 0 | 1 | 0 | 0 | 0 |
| VH HV V | 0 | 0 | 0 | 0 | 0 | 0 | 0 | 0 | 0 | 0 | 0 | 0 | 0 | 0 | 0 | 0 |
| HV HV V | 0 | 0 | 0 | 0 | 0 | 0 | 0 | 0 | 0 | 0 | 0 | 0 | 0 | 0 | 0 | 0 |
| VV HV V | 0 | 0 | 0 | 0 | 0 | 0 | 0 | 0 | 0 | 0 | 0 | 0 | 0 | 0 | 0 | 0 |

| HH VV V | 0 | 0 | 0 | 0 | 0 | 0 | 0 | 0 | 0 | 0 | 0 | 0 | 0 | 0 | 0 | 0 |
|---|---|---|---|---|---|---|---|---|---|---|---|---|---|---|---|
| VH VV V | 0 | 0 | 0 | 0 | 0 | 0 | 0 | 0 | 0 | 0 | 0 | 0 | 0 | 1 | 0 | 0 |
| HV VV V | 0 | 0 | 0 | 0 | 0 | 0 | 0 | 0 | 0 | 0 | 0 | 0 | 0 | 0 | 0 | 0 |
| VV VV V | 0 | 0 | 0 | 0 | 0 | 0 | 0 | 0 | 0 | 0 | 0 | 0 | 0 | 0 | 0 | 0 |

|  | HH HH V | VH HH V | HV HH V | VV HH V | HH VH V | VH VH V | HV VH V | VV VH V | HH HV V | VH HV V | HV HV V | VV HV V | HH VV V | VH VV V | HV VV V | VV VV V |
|---|---|---|---|---|---|---|---|---|---|---|---|---|---|---|---|---|
| HH HH H | 0 | 0 | 0 | 0 | 0 | 0 | 0 | 0 | 0 | 0 | 0 | 0 | 0 | 0 | 0 | 0 |
| VH HH H | 0 | 0 | 0 | 0 | 0 | 0 | 0 | 0 | 0 | 0 | 0 | 0 | 0 | 0 | 0 | 0 |
| HV HH H | 1 | 0 | 0 | 0 | 0 | 0 | 0 | 0 | 0 | 0 | 0 | 0 | 0 | 0 | 0 | 0 |
| VV HH H | 0 | 0 | 1 | 0 | 0 | 0 | 0 | 0 | 0 | 0 | 0 | 0 | 0 | 0 | 0 | 0 |
| HH VH H | 0 | 0 | 0 | 0 | 0 | 0 | 0 | 0 | 0 | 0 | 0 | 0 | 0 | 0 | 0 | 0 |
| VH VH H | 0 | 0 | 0 | 0 | 0 | 0 | 0 | 0 | 0 | 0 | 0 | 0 | 0 | 0 | 0 | 0 |
| HV VH H | 0 | 0 | 0 | 0 | 0 | 0 | 0 | 0 | 0 | 0 | 0 | 0 | 0 | 0 | 0 | 0 |
| VV VH H | 0 | 0 | 0 | 0 | 0 | 0 | 0 | 0 | 0 | 0 | 0 | 0 | 0 | 0 | 0 | 0 |
| HH HV H | 0 | 0 | 0 | 0 | 0 | 0 | 0 | 0 | 0 | 0 | 0 | 0 | 0 | 0 | 0 | 0 |
| VH HV H | 0 | 0 | 0 | 0 | 0 | 0 | 0 | 0 | 0 | 0 | 0 | 0 | 0 | 0 | 0 | 0 |
| HV HV H | 0 | 0 | 0 | 0 | 0 | 0 | 0 | 0 | 0 | 0 | 0 | 0 | 0 | 0 | 0 | 0 |
| VV HV H | 0 | 0 | 0 | 0 | 0 | 0 | 0 | 0 | 0 | 0 | 0 | 0 | 0 | 0 | 0 | 0 |
| HH VV H | 0 | 0 | 0 | 0 | 0 | 0 | 0 | 0 | 0 | 0 | 0 | 0 | 0 | 0 | 0 | 0 |
| VH VV H | 0 | 0 | 0 | 0 | 0 | 0 | 0 | 0 | 0 | 0 | 0 | 0 | 0 | 0 | 0 | 0 |
| HV VV H | 0 | 0 | 0 | 0 | 0 | 0 | 0 | 0 | 0 | 0 | 0 | 0 | 0 | 0 | 0 | 0 |
| VV VV H | 0 | 0 | 0 | 0 | 0 | 0 | 0 | 0 | 0 | 0 | 0 | 0 | 0 | 0 | 0 | 0 |
| HH HH V | 1 | 0 | 0 | 0 | 0 | 0 | 0 | 0 | 0 | 0 | 0 | 0 | 0 | 0 | 0 | 0 |
| VH HH V | 0 | 0 | 0 | 0 | 0 | 0 | 0 | 0 | 0 | 0 | 0 | 0 | 0 | 0 | 0 | 0 |
| HV HH V | 1 | 0 | 0 | 0 | 0 | 0 | 0 | 0 | 0 | 0 | 0 | 0 | 0 | 0 | 0 | 0 |
| VV HH V | 0 | 0 | 0 | 0 | 0 | 0 | 0 | 0 | 0 | 0 | 0 | 1 | 0 | 0 | 0 | 0 |
| HH VH V | 0 | 0 | 0 | 0 | 0 | 0 | 0 | 0 | 0 | 0 | 0 | 0 | 0 | 0 | 0 | 0 |
| VH VH V | 0 | 0 | 0 | 0 | 0 | 0 | 0 | 0 | 0 | 0 | 0 | 0 | 0 | 0 | 0 | 0 |
| HV VH V | 0 | 0 | 0 | 0 | 0 | 0 | 0 | 0 | 0 | 0 | 0 | 0 | 0 | 0 | 0 | 0 |
| VV VH V | 0 | 0 | 0 | 0 | 0 | 0 | 0 | 0 | 0 | 0 | 0 | 0 | 0 | 0 | 0 | 0 |
| HH HV V | 0 | 0 | 0 | 0 | 0 | 0 | 0 | 0 | 2 | 0 | 0 | 0 | 2 | 0 | 0 | 0 |
| VH HV V | 0 | 0 | 0 | 0 | 0 | 0 | 0 | 0 | 0 | 0 | 0 | 0 | 0 | 0 | 0 | 0 |
| HV HV V | 0 | 0 | 0 | 0 | 0 | 0 | 0 | 0 | 1 | 0 | 0 | 0 | 3 | 0 | 0 | 0 |
| VV HV V | 0 | 0 | 0 | 0 | 0 | 0 | 0 | 0 | 0 | 1 | 1 | 0 | 0 | 0 | 0 | 0 |
| HH VV V | 0 | 0 | 0 | 0 | 0 | 0 | 0 | 0 | 0 | 0 | 0 | 0 | 1 | 0 | 0 | 0 |
| VH VV V | 0 | 0 | 0 | 0 | 0 | 0 | 0 | 0 | 0 | 0 | 0 | 0 | 0 | 0 | 0 | 0 |
| HV VV V | 0 | 0 | 0 | 0 | 0 | 0 | 0 | 0 | 0 | 0 | 0 | 0 | 0 | 1 | 0 | 0 |
| VV VV V | 0 | 0 | 0 | 1 | 0 | 0 | 0 | 2 | 0 | 0 | 0 | 0 | 0 | 1 | 0 | 49 |

**TABLE S8.2: The experimentally measured raw data of a complete set of the 1024 possible combinations of ten-photon coincidence events with each photon measured in the basis of** $(|H\rangle \pm e^{i0\pi/10}|V\rangle)/\sqrt{2}$ **(indicated by the** $\pm$ **label in the first column and the first line) to obtain** $\langle M_{0\pi/10}^{\otimes 10}\rangle$. **The five + or – label in the first column represent the measurement base choice for the five** $o$ **photons, while the five + or – label in the first line represent the measurement base choice for the five** $e$ **photons.**

| | +++ ++ | - +++ + | +- +++ | -- +++ | ++- ++ | -+- ++ | +-- ++ | --- ++ | +++ -+ | -++- + | +-+- + | --+- + | ++-- + | -+-- + | +--- + | ---- + |
|---|---|---|---|---|---|---|---|---|---|---|---|---|---|---|---|---|
| +++++ | 0 | 0 | 0 | 0 | 0 | 0 | 0 | 0 | 0 | 0 | 0 | 0 | 0 | 0 | 0 | 0 |
| -++++ | 0 | 0 | 0 | 0 | 0 | 0 | 0 | 0 | 0 | 0 | 0 | 0 | 1 | 0 | 0 | 0 |
| +-+++ | 0 | 0 | 0 | 1 | 0 | 0 | 0 | 0 | 0 | 0 | 0 | 0 | 0 | 0 | 0 | 0 |
| --+++ | 0 | 0 | 0 | 0 | 0 | 0 | 0 | 0 | 0 | 0 | 0 | 0 | 0 | 0 | 1 | 1 |
| ++-++ | 0 | 0 | 0 | 0 | 0 | 0 | 0 | 0 | 0 | 0 | 0 | 0 | 1 | 0 | 0 | 0 |
| -+-++ | 1 | 0 | 0 | 0 | 0 | 0 | 0 | 0 | 0 | 0 | 0 | 0 | 0 | 0 | 1 | 0 |
| +--++ | 0 | 0 | 0 | 0 | 0 | 0 | 0 | 0 | 0 | 0 | 0 | 0 | 0 | 0 | 0 | 0 |
| ---++ | 0 | 0 | 0 | 0 | 0 | 0 | 0 | 1 | 0 | 1 | 0 | 0 | 0 | 0 | 0 | 0 |
| +++-+ | 0 | 0 | 0 | 0 | 0 | 0 | 0 | 0 | 0 | 0 | 0 | 0 | 0 | 0 | 0 | 0 |
| -++-+ | 1 | 0 | 0 | 0 | 0 | 0 | 0 | 0 | 0 | 0 | 0 | 0 | 0 | 1 | 0 | 0 |
| +-+-+ | 0 | 1 | 0 | 0 | 0 | 0 | 0 | 0 | 0 | 0 | 0 | 0 | 0 | 0 | 0 | 0 |
| --+-+ | 0 | 0 | 0 | 0 | 0 | 0 | 0 | 0 | 0 | 0 | 0 | 0 | 0 | 0 | 0 | 0 |
| ++--+ | 0 | 0 | 1 | 0 | 0 | 0 | 1 | 0 | 0 | 0 | 0 | 0 | 0 | 0 | 0 | 0 |
| -+--+ | 0 | 0 | 0 | 0 | 0 | 0 | 0 | 0 | 0 | 0 | 0 | 0 | 0 | 0 | 0 | 0 |
| +---+ | 0 | 0 | 1 | 0 | 1 | 0 | 0 | 0 | 0 | 0 | 0 | 0 | 0 | 1 | 0 | 0 |

| | | | | | | | | | | | | | | | |
|---|---|---|---|---|---|---|---|---|---|---|---|---|---|---|---|
| ----+ | 0 | 0 | 0 | 0 | 0 | 0 | 0 | 0 | 0 | 0 | 0 | 0 | 0 | 0 | 0 | 0 |
| +++++ | 0 | 0 | 0 | 1 | 0 | 0 | 0 | 0 | 1 | 1 | 0 | 0 | 0 | 0 | 0 | 0 |
| -+++- | 0 | 0 | 0 | 1 | 0 | 0 | 0 | 0 | 0 | 0 | 0 | 0 | 0 | 0 | 0 | 1 |
| +-++- | 0 | 0 | 0 | 0 | 0 | 0 | 0 | 1 | 0 | 0 | 0 | 0 | 0 | 0 | 0 | 0 |
| --++- | 1 | 0 | 0 | 0 | 0 | 0 | 0 | 0 | 0 | 0 | 0 | 0 | 0 | 0 | 0 | 0 |
| ++++- | 0 | 0 | 0 | 0 | 0 | 0 | 0 | 0 | 0 | 0 | 0 | 1 | 0 | 0 | 1 | 0 |
| -+++- | 0 | 0 | 2 | 0 | 0 | 0 | 0 | 1 | 0 | 1 | 0 | 0 | 1 | 0 | 1 | 0 |
| +--+- | 0 | 1 | 0 | 0 | 0 | 0 | 0 | 0 | 0 | 0 | 0 | 0 | 0 | 0 | 1 | 0 |
| ---+- | 0 | 0 | 0 | 0 | 0 | 1 | 0 | 0 | 0 | 0 | 0 | 0 | 0 | 0 | 0 | 0 |
| +++-- | 0 | 0 | 0 | 0 | 0 | 0 | 0 | 0 | 0 | 0 | 0 | 1 | 0 | 0 | 0 | 0 |
| -++-- | 0 | 1 | 0 | 0 | 0 | 0 | 0 | 0 | 0 | 0 | 0 | 1 | 1 | 0 | 0 | 0 |
| +-+-- | 0 | 0 | 0 | 0 | 0 | 0 | 0 | 0 | 0 | 0 | 1 | 0 | 0 | 0 | 1 | 0 |
| --+-- | 0 | 0 | 0 | 0 | 0 | 0 | 0 | 0 | 0 | 0 | 1 | 0 | 0 | 0 | 0 | 0 |
| ++--- | 0 | 0 | 0 | 0 | 0 | 0 | 0 | 0 | 0 | 0 | 0 | 0 | 0 | 0 | 0 | 0 |
| -+--- | 0 | 0 | 0 | 0 | 0 | 0 | 0 | 0 | 0 | 0 | 0 | 0 | 0 | 0 | 0 | 0 |
| +---- | 0 | 0 | 0 | 0 | 0 | 0 | 0 | 0 | 0 | 0 | 0 | 0 | 0 | 0 | 0 | 0 |
| ----- | 0 | 0 | 0 | 0 | 0 | 0 | 0 | 0 | 0 | 0 | 0 | 1 | 0 | 1 | 0 | 0 |

| | +++ +- | -+++- | +- ++- | -- ++- | ++- +- | -+- +- | +- +- | ---+- | +++ -- | -++- - | +-+- - | --+- - | ++-- - | -+--- | +---- | ----- |
|---|---|---|---|---|---|---|---|---|---|---|---|---|---|---|---|---|
| +++++ | 0 | 0 | 1 | 0 | 1 | 0 | 0 | 0 | 0 | 0 | 0 | 0 | 0 | 0 | 1 | 1 |
| -++++ | 1 | 0 | 0 | 0 | 0 | 0 | 1 | 0 | 0 | 0 | 0 | 0 | 1 | 0 | 0 | 0 |
| +-+++ | 0 | 0 | 0 | 0 | 1 | 1 | 1 | 0 | 0 | 0 | 1 | 0 | 0 | 0 | 0 | 0 |
| --+++ | 0 | 0 | 0 | 0 | 0 | 0 | 0 | 0 | 0 | 0 | 0 | 0 | 0 | 0 | 1 | 1 |
| ++-++ | 0 | 0 | 0 | 0 | 0 | 0 | 0 | 0 | 0 | 0 | 0 | 0 | 0 | 0 | 0 | 0 |
| -+-++ | 0 | 0 | 0 | 0 | 0 | 0 | 0 | 0 | 0 | 0 | 1 | 0 | 0 | 0 | 0 | 0 |
| +--++ | 0 | 0 | 0 | 0 | 0 | 1 | 0 | 0 | 0 | 0 | 0 | 0 | 0 | 0 | 0 | 0 |
| ---++ | 0 | 0 | 0 | 0 | 0 | 0 | 0 | 0 | 0 | 0 | 0 | 0 | 0 | 0 | 0 | 0 |
| +++-+ | 0 | 0 | 0 | 0 | 0 | 0 | 0 | 0 | 0 | 0 | 0 | 0 | 0 | 0 | 0 | 0 |
| -++-+ | 0 | 0 | 0 | 0 | 0 | 0 | 0 | 0 | 0 | 0 | 0 | 0 | 0 | 0 | 0 | 0 |
| +-+-+ | 0 | 1 | 0 | 0 | 0 | 0 | 0 | 0 | 0 | 0 | 0 | 0 | 0 | 1 | 0 | 0 |
| --+-+ | 1 | 0 | 0 | 0 | 0 | 2 | 0 | 0 | 0 | 0 | 0 | 0 | 1 | 0 | 0 | 0 |
| ++--+ | 0 | 0 | 0 | 1 | 0 | 0 | 0 | 0 | 0 | 0 | 0 | 0 | 0 | 0 | 1 | 0 |
| -+--+ | 0 | 0 | 0 | 0 | 1 | 0 | 0 | 0 | 0 | 0 | 0 | 0 | 0 | 0 | 0 | 0 |
| +---+ | 0 | 0 | 0 | 1 | 0 | 0 | 0 | 0 | 0 | 0 | 0 | 0 | 0 | 0 | 0 | 0 |
| ----+ | 0 | 0 | 0 | 0 | 0 | 0 | 0 | 0 | 0 | 0 | 0 | 1 | 0 | 0 | 0 | 0 |
| ++++- | 2 | 0 | 0 | 0 | 0 | 1 | 0 | 0 | 0 | 0 | 0 | 1 | 1 | 0 | 0 | 1 |
| -+++- | 0 | 0 | 0 | 0 | 0 | 0 | 0 | 0 | 1 | 0 | 0 | 0 | 0 | 0 | 0 | 0 |
| +-++- | 0 | 0 | 0 | 0 | 0 | 0 | 0 | 0 | 0 | 0 | 0 | 0 | 0 | 0 | 0 | 0 |
| --++- | 0 | 0 | 0 | 0 | 0 | 0 | 1 | 0 | 0 | 0 | 0 | 0 | 0 | 0 | 0 | 0 |
| ++-+- | 0 | 0 | 0 | 0 | 1 | 0 | 0 | 0 | 0 | 0 | 0 | 0 | 0 | 0 | 0 | 0 |
| -+-+- | 1 | 0 | 0 | 0 | 0 | 1 | 0 | 0 | 0 | 0 | 1 | 0 | 0 | 0 | 0 | 0 |
| +--+- | 0 | 0 | 0 | 0 | 0 | 0 | 0 | 0 | 0 | 0 | 0 | 0 | 0 | 0 | 0 | 1 |
| ---+- | 0 | 0 | 0 | 0 | 0 | 0 | 0 | 0 | 0 | 0 | 0 | 1 | 0 | 0 | 0 | 0 |
| +++-- | 0 | 0 | 1 | 0 | 0 | 0 | 0 | 0 | 0 | 0 | 0 | 0 | 1 | 0 | 0 | 0 |
| -++-- | 1 | 0 | 0 | 0 | 0 | 0 | 0 | 0 | 0 | 0 | 1 | 0 | 0 | 0 | 0 | 0 |
| +-+-- | 1 | 0 | 1 | 0 | 0 | 0 | 1 | 0 | 0 | 0 | 1 | 0 | 0 | 0 | 0 | 0 |
| --+-- | 0 | 0 | 1 | 0 | 0 | 0 | 0 | 0 | 0 | 0 | 0 | 0 | 0 | 0 | 0 | 0 |
| ++--- | 0 | 0 | 0 | 1 | 0 | 0 | 0 | 0 | 0 | 0 | 1 | 0 | 0 | 0 | 0 | 0 |
| -+--- | 0 | 2 | 0 | 0 | 1 | 0 | 0 | 0 | 1 | 0 | 0 | 0 | 0 | 0 | 0 | 0 |
| +---- | 0 | 0 | 0 | 0 | 0 | 0 | 0 | 0 | 1 | 0 | 0 | 0 | 0 | 0 | 0 | 0 |
| ----- | 0 | 0 | 0 | 0 | 0 | 1 | 0 | 0 | 1 | 0 | 1 | 0 | 0 | 0 | 0 | 0 |

**TABLE S8.3: The experimentally measured raw data of a complete set of the 1024 possible combinations of ten-photon coincidence events with each photon measured in the basis of** $(|H\rangle \pm e^{i1\pi/10}|V\rangle)/\sqrt{2}$ **(indicated by the $\pm$ label in the first column and the first line) to obtain** $\left\langle M_{1\pi/10}^{\otimes 10} \right\rangle$**.** The five + or − label in the first column represent the measurement base choice for the five $o$ photons, while the five + or − label in the first line represent the measurement base choice for the five $e$ photons.

| | +++++ | -++++ | +-+++ | --+++ | ++-++ | -+-++ | +--++ | ---++ | +++-+ | -++-+ | +-+-+ | --+-+ | ++--+ | -+--+ | +---+ | ----+ |
|---|---|---|---|---|---|---|---|---|---|---|---|---|---|---|---|---|
| +++++ | 0 | 0 | 0 | 0 | 0 | 0 | 0 | 0 | 0 | 0 | 0 | 0 | 0 | 0 | 0 | 0 |
| -++++ | 0 | 0 | 0 | 1 | 0 | 0 | 0 | 1 | 0 | 0 | 0 | 0 | 1 | 0 | 0 | 0 |
| +-+++ | 0 | 0 | 0 | 0 | 0 | 0 | 0 | 0 | 0 | 0 | 2 | 0 | 0 | 1 | 0 | 0 |
| --+++ | 0 | 0 | 1 | 0 | 1 | 0 | 0 | 0 | 0 | 0 | 0 | 0 | 0 | 0 | 0 | 0 |
| ++-++ | 1 | 1 | 0 | 0 | 0 | 0 | 0 | 1 | 0 | 0 | 0 | 0 | 0 | 0 | 0 | 0 |
| -+-++ | 0 | 0 | 0 | 0 | 0 | 0 | 0 | 1 | 1 | 0 | 0 | 0 | 0 | 0 | 0 | 0 |
| +--++ | 0 | 0 | 0 | 0 | 0 | 0 | 0 | 0 | 0 | 0 | 0 | 1 | 0 | 0 | 0 | 0 |
| ---++ | 0 | 0 | 0 | 0 | 1 | 0 | 0 | 0 | 0 | 0 | 1 | 0 | 0 | 0 | 0 | 0 |
| +++-+ | 0 | 0 | 0 | 0 | 0 | 0 | 0 | 0 | 1 | 0 | 0 | 0 | 0 | 0 | 0 | 1 |
| -++-+ | 0 | 1 | 0 | 0 | 0 | 0 | 0 | 0 | 0 | 0 | 0 | 0 | 1 | 0 | 0 | 0 |
| +-+-+ | 0 | 0 | 1 | 0 | 0 | 0 | 0 | 1 | 0 | 0 | 0 | 0 | 0 | 0 | 1 | 0 |
| --+-+ | 0 | 0 | 0 | 0 | 1 | 0 | 0 | 0 | 0 | 1 | 0 | 0 | 1 | 0 | 0 | 0 |
| ++--+ | 0 | 0 | 0 | 0 | 0 | 0 | 0 | 0 | 0 | 0 | 0 | 0 | 0 | 1 | 0 | 0 |
| -+--+ | 0 | 0 | 0 | 0 | 0 | 0 | 0 | 0 | 0 | 0 | 0 | 0 | 0 | 1 | 0 | 0 |
| +---+ | 0 | 0 | 0 | 0 | 0 | 0 | 0 | 1 | 0 | 2 | 0 | 0 | 0 | 0 | 0 | 0 |
| ----+ | 1 | 0 | 0 | 0 | 1 | 0 | 0 | 0 | 0 | 0 | 0 | 2 | 0 | 1 | 0 | 0 |
| ++++- | 0 | 0 | 0 | 1 | 0 | 0 | 0 | 0 | 0 | 0 | 0 | 0 | 0 | 0 | 0 | 0 |
| -+++- | 0 | 0 | 0 | 0 | 0 | 0 | 0 | 0 | 0 | 0 | 0 | 1 | 0 | 1 | 0 | 0 |
| +-++- | 0 | 0 | 1 | 0 | 1 | 0 | 0 | 1 | 0 | 0 | 0 | 0 | 0 | 0 | 0 | 0 |
| --++- | 0 | 0 | 0 | 0 | 1 | 0 | 0 | 0 | 1 | 0 | 0 | 0 | 0 | 1 | 0 | 0 |
| ++-+- | 0 | 0 | 0 | 0 | 0 | 0 | 0 | 0 | 0 | 0 | 0 | 0 | 0 | 1 | 0 | 0 |
| -+-+- | 0 | 0 | 0 | 1 | 0 | 0 | 0 | 0 | 0 | 0 | 0 | 0 | 0 | 0 | 0 | 0 |
| +--+- | 0 | 0 | 1 | 0 | 0 | 0 | 0 | 0 | 1 | 0 | 0 | 0 | 0 | 0 | 0 | 0 |
| ---+- | 0 | 0 | 0 | 1 | 0 | 0 | 0 | 0 | 1 | 0 | 0 | 0 | 0 | 0 | 0 | 0 |
| +++-- | 0 | 0 | 1 | 0 | 0 | 0 | 0 | 0 | 1 | 0 | 0 | 0 | 0 | 0 | 0 | 0 |
| -++-- | 0 | 0 | 0 | 0 | 0 | 0 | 0 | 0 | 1 | 1 | 2 | 0 | 0 | 0 | 0 | 0 |
| +-+-- | 0 | 0 | 0 | 0 | 0 | 1 | 0 | 0 | 0 | 0 | 0 | 0 | 0 | 0 | 0 | 0 |
| --+-- | 0 | 0 | 0 | 0 | 0 | 0 | 0 | 0 | 0 | 0 | 0 | 0 | 0 | 0 | 0 | 0 |
| ++--- | 0 | 0 | 0 | 1 | 0 | 0 | 0 | 0 | 0 | 1 | 1 | 0 | 0 | 0 | 0 | 0 |
| -+--- | 0 | 0 | 1 | 0 | 0 | 0 | 0 | 0 | 0 | 0 | 1 | 0 | 0 | 0 | 0 | 0 |
| +---- | 0 | 0 | 0 | 0 | 0 | 0 | 0 | 0 | 0 | 0 | 0 | 0 | 0 | 0 | 0 | 0 |
| ----- | 0 | 0 | 0 | 0 | 0 | 0 | 0 | 0 | 0 | 0 | 1 | 0 | 0 | 0 | 2 | 0 |

| | ++++- | -+++- | +-++- | --++- | ++-+- | -+-+- | +--+- | ---+- | +++-- | -++-- | +-+-- | --+-- | ++--- | -+--- | +---- | ----- |
|---|---|---|---|---|---|---|---|---|---|---|---|---|---|---|---|---|
| +++++ | 0 | 0 | 0 | 1 | 0 | 0 | 1 | 0 | 0 | 0 | 0 | 0 | 0 | 0 | 1 | 1 |
| -++++ | 0 | 0 | 0 | 0 | 0 | 0 | 0 | 0 | 0 | 1 | 0 | 0 | 0 | 0 | 0 | 0 |
| +-+++ | 0 | 0 | 1 | 0 | 0 | 1 | 0 | 0 | 0 | 1 | 0 | 0 | 0 | 1 | 0 | 0 |
| --+++ | 0 | 0 | 0 | 0 | 0 | 0 | 0 | 0 | 0 | 1 | 0 | 0 | 0 | 0 | 0 | 0 |
| ++-++ | 0 | 1 | 0 | 1 | 0 | 0 | 0 | 0 | 0 | 0 | 0 | 0 | 0 | 0 | 0 | 0 |
| -+-++ | 0 | 0 | 0 | 0 | 0 | 0 | 0 | 0 | 1 | 0 | 0 | 0 | 0 | 0 | 0 | 0 |
| +--++ | 1 | 1 | 0 | 2 | 0 | 1 | 0 | 0 | 0 | 0 | 0 | 0 | 0 | 0 | 0 | 0 |
| ---++ | 0 | 0 | 0 | 0 | 0 | 0 | 0 | 0 | 0 | 0 | 0 | 1 | 0 | 0 | 0 | 0 |
| +++-+ | 0 | 1 | 0 | 0 | 1 | 0 | 0 | 1 | 1 | 0 | 1 | 0 | 0 | 0 | 1 | 0 |
| -++-+ | 0 | 0 | 0 | 0 | 1 | 0 | 0 | 0 | 0 | 0 | 2 | 0 | 0 | 0 | 0 | 0 |
| +-+-+ | 0 | 0 | 0 | 0 | 0 | 0 | 1 | 0 | 0 | 1 | 0 | 0 | 0 | 0 | 0 | 0 |
| --+-+ | 0 | 0 | 1 | 0 | 0 | 0 | 0 | 1 | 1 | 0 | 0 | 0 | 0 | 0 | 0 | 0 |
| ++--+ | 0 | 0 | 1 | 0 | 0 | 0 | 0 | 1 | 1 | 0 | 0 | 1 | 0 | 1 | 0 | 0 |
| -+--+ | 0 | 0 | 1 | 0 | 0 | 0 | 0 | 1 | 1 | 0 | 0 | 1 | 0 | 0 | 0 | 0 |
| +---+ | 0 | 0 | 0 | 0 | 0 | 0 | 0 | 0 | 0 | 0 | 1 | 0 | 0 | 0 | 0 | 0 |
| ----+ | 0 | 0 | 0 | 0 | 0 | 0 | 0 | 0 | 0 | 0 | 1 | 0 | 0 | 0 | 0 | 0 |
| ++++- | 0 | 1 | 0 | 0 | 0 | 0 | 0 | 1 | 1 | 0 | 0 | 0 | 0 | 0 | 0 | 0 |
| -+++- | 0 | 0 | 0 | 0 | 1 | 0 | 0 | 0 | 0 | 0 | 0 | 0 | 0 | 0 | 0 | 0 |
| +-++- | 0 | 0 | 0 | 0 | 0 | 0 | 0 | 0 | 0 | 0 | 0 | 0 | 0 | 0 | 0 | 0 |
| --++- | 0 | 0 | 0 | 0 | 0 | 0 | 0 | 0 | 0 | 0 | 0 | 0 | 0 | 0 | 0 | 0 |
| ++-+- | 0 | 0 | 0 | 0 | 0 | 0 | 0 | 0 | 0 | 1 | 1 | 0 | 0 | 0 | 1 | 0 |
| -+-+- | 0 | 0 | 0 | 0 | 2 | 0 | 0 | 0 | 0 | 0 | 0 | 0 | 1 | 0 | 0 | 0 |
| +--+- | 0 | 0 | 0 | 0 | 0 | 0 | 1 | 0 | 0 | 0 | 1 | 0 | 0 | 0 | 0 | 0 |
| ---+- | 0 | 0 | 0 | 0 | 0 | 0 | 1 | 0 | 0 | 0 | 0 | 0 | 1 | 0 | 0 | 0 |
| +++-- | 1 | 0 | 0 | 0 | 0 | 0 | 1 | 0 | 0 | 0 | 0 | 0 | 1 | 0 | 0 | 0 |
| -++-- | 1 | 1 | 0 | 0 | 0 | 0 | 0 | 0 | 1 | 0 | 0 | 0 | 0 | 1 | 1 | 0 |
| +-+-- | 0 | 0 | 0 | 0 | 0 | 0 | 0 | 0 | 0 | 0 | 0 | 0 | 0 | 0 | 0 | 0 |
| --+-- | 0 | 0 | 0 | 0 | 0 | 0 | 0 | 0 | 0 | 0 | 0 | 0 | 0 | 0 | 0 | 0 |
| ++--- | 0 | 0 | 0 | 0 | 0 | 1 | 0 | 2 | 0 | 0 | 0 | 0 | 0 | 0 | 1 | 0 |
| -+--- | 0 | 0 | 0 | 0 | 0 | 1 | 0 | 0 | 0 | 0 | 1 | 0 | 0 | 0 | 0 | 0 |
| +---- | 0 | 0 | 0 | 0 | 0 | 0 | 0 | 0 | 0 | 0 | 0 | 0 | 0 | 0 | 0 | 0 |
| ----- | 0 | 0 | 0 | 0 | 0 | 0 | 0 | 0 | 0 | 0 | 0 | 0 | 0 | 0 | 0 | 0 |

**TABLE S8.4: The experimentally measured raw data of a complete set of the 1024 possible combinations of ten-photon coincidence events with each photon measured in the basis of $(|H\rangle \pm e^{i2\pi/10}|V\rangle)/\sqrt{2}$ (indicated by the $\pm$ label in the first column and the first line) to obtain $\left\langle M_{2\pi/10}^{\otimes 10} \right\rangle$.** The five + or – label in the first column represent the measurement base choice for the five $o$ photons, while the five + or – label in the first line represent the measurement base choice for the five $e$ photons.

| | +++++ | -++++ | +-+++ | --+++ | ++-++ | -+-++ | +--++ | ---++ | +++-+ | -++-+ | +-+-+ | --+-+ | ++--+ | -+--+ | +---+ | ----+ |
|---|---|---|---|---|---|---|---|---|---|---|---|---|---|---|---|---|
| +++++ | 0 | 0 | 0 | 0 | 0 | 0 | 0 | 0 | 0 | 0 | 1 | 0 | 0 | 0 | 0 | 0 |
| -++++ | 0 | 0 | 0 | 0 | 0 | 0 | 0 | 0 | 0 | 0 | 0 | 0 | 0 | 0 | 0 | 1 |
| +-+++ | 0 | 0 | 1 | 0 | 0 | 0 | 0 | 0 | 0 | 0 | 0 | 0 | 0 | 1 | 1 | 0 |
| --+++ | 1 | 0 | 0 | 0 | 0 | 0 | 0 | 0 | 0 | 0 | 0 | 0 | 1 | 0 | 0 | 0 |
| ++-++ | 0 | 0 | 1 | 0 | 0 | 0 | 0 | 0 | 0 | 0 | 0 | 0 | 0 | 1 | 0 | 0 |
| -+-++ | 0 | 0 | 1 | 0 | 0 | 0 | 1 | 0 | 0 | 1 | 0 | 1 | 0 | 0 | 0 | 0 |
| +--++ | 0 | 0 | 0 | 0 | 0 | 1 | 0 | 0 | 0 | 0 | 0 | 0 | 0 | 0 | 0 | 0 |
| ---++ | 0 | 0 | 1 | 0 | 0 | 0 | 0 | 0 | 0 | 0 | 0 | 0 | 0 | 0 | 0 | 0 |
| +++-+ | 0 | 0 | 0 | 0 | 0 | 0 | 0 | 0 | 0 | 0 | 0 | 0 | 0 | 1 | 0 | 0 |
| -++-+ | 0 | 0 | 0 | 0 | 0 | 0 | 0 | 0 | 0 | 0 | 0 | 0 | 0 | 0 | 0 | 0 |
| +-+-+ | 0 | 0 | 0 | 0 | 1 | 0 | 1 | 0 | 0 | 0 | 1 | 0 | 0 | 0 | 0 | 0 |
| --+-+ | 0 | 1 | 0 | 0 | 0 | 0 | 0 | 0 | 0 | 0 | 0 | 0 | 1 | 0 | 0 | 0 |
| ++--+ | 0 | 0 | 0 | 1 | 0 | 0 | 0 | 0 | 0 | 0 | 0 | 0 | 0 | 0 | 0 | 0 |
| -+--+ | 0 | 0 | 0 | 1 | 1 | 0 | 0 | 0 | 0 | 1 | 0 | 0 | 0 | 0 | 0 | 0 |
| +---+ | 0 | 0 | 0 | 0 | 0 | 0 | 0 | 0 | 0 | 0 | 0 | 0 | 0 | 0 | 0 | 0 |
| ----+ | 0 | 0 | 0 | 0 | 1 | 0 | 0 | 0 | 0 | 0 | 0 | 0 | 0 | 0 | 0 | 0 |
| ++++- | 0 | 0 | 0 | 0 | 0 | 2 | 0 | 0 | 2 | 0 | 0 | 0 | 0 | 0 | 0 | 0 |
| -+++- | 0 | 0 | 0 | 0 | 0 | 0 | 0 | 0 | 0 | 0 | 1 | 0 | 0 | 0 | 0 | 0 |
| +-++- | 0 | 0 | 0 | 0 | 0 | 1 | 1 | 0 | 0 | 0 | 0 | 0 | 0 | 0 | 0 | 0 |
| --++- | 0 | 0 | 1 | 0 | 0 | 0 | 0 | 0 | 0 | 0 | 0 | 0 | 0 | 1 | 0 | 0 |
| ++-+- | 0 | 0 | 0 | 0 | 0 | 1 | 0 | 0 | 0 | 0 | 0 | 0 | 0 | 1 | 0 | 0 |
| -+-+- | 0 | 0 | 0 | 0 | 0 | 0 | 0 | 1 | 0 | 0 | 1 | 0 | 0 | 0 | 0 | 0 |
| +--+- | 0 | 0 | 1 | 0 | 0 | 1 | 0 | 0 | 0 | 1 | 0 | 0 | 0 | 0 | 0 | 0 |
| ---+- | 0 | 0 | 0 | 0 | 0 | 0 | 0 | 0 | 0 | 1 | 1 | 0 | 0 | 0 | 0 | 0 |
| +++-- | 0 | 0 | 0 | 0 | 0 | 0 | 0 | 0 | 1 | 0 | 0 | 0 | 0 | 0 | 0 | 0 |
| -++-- | 0 | 0 | 0 | 0 | 0 | 0 | 0 | 0 | 0 | 0 | 1 | 0 | 0 | 0 | 0 | 0 |
| +-+-- | 0 | 0 | 0 | 1 | 0 | 0 | 0 | 0 | 1 | 0 | 0 | 2 | 1 | 0 | 0 | 0 |
| --+-- | 0 | 0 | 0 | 0 | 0 | 0 | 0 | 1 | 0 | 0 | 0 | 0 | 0 | 0 | 0 | 0 |
| ++--- | 0 | 0 | 1 | 0 | 0 | 0 | 0 | 0 | 1 | 0 | 0 | 0 | 0 | 0 | 0 | 0 |
| -+--- | 0 | 0 | 0 | 0 | 0 | 0 | 0 | 1 | 0 | 0 | 0 | 0 | 0 | 0 | 1 | 0 |
| +---- | 0 | 0 | 0 | 0 | 0 | 0 | 1 | 0 | 0 | 0 | 0 | 0 | 1 | 0 | 0 | 0 |
| ----- | 0 | 0 | 0 | 0 | 0 | 0 | 0 | 1 | 0 | 0 | 0 | 0 | 0 | 0 | 0 | 0 |

| | ++++- | -+++- | +-++- | --++- | ++-+- | -+-+- | +--+- | ---+- | +++-- | -++-- | +-+-- | --+-- | ++--- | -+--- | +---- | ----- |
|---|---|---|---|---|---|---|---|---|---|---|---|---|---|---|---|---|
| +++++ | 0 | 1 | 0 | 0 | 0 | 0 | 0 | 0 | 0 | 0 | 0 | 0 | 0 | 0 | 0 | 0 |
| -++++ | 0 | 0 | 0 | 0 | 0 | 1 | 0 | 0 | 0 | 0 | 1 | 0 | 0 | 0 | 0 | 0 |
| +-+++ | 0 | 0 | 0 | 0 | 0 | 0 | 1 | 1 | 0 | 0 | 0 | 0 | 1 | 0 | 0 | 1 |
| --+++ | 0 | 0 | 1 | 0 | 1 | 0 | 0 | 0 | 0 | 0 | 0 | 0 | 0 | 0 | 0 | 0 |
| ++-++ | 0 | 0 | 0 | 1 | 0 | 0 | 0 | 0 | 0 | 0 | 0 | 0 | 0 | 0 | 0 | 0 |
| -+-++ | 0 | 0 | 0 | 0 | 1 | 1 | 0 | 0 | 1 | 0 | 0 | 0 | 0 | 1 | 0 | 1 |
| +--++ | 0 | 0 | 0 | 0 | 0 | 1 | 0 | 0 | 0 | 0 | 0 | 0 | 0 | 0 | 1 | 0 |
| ---++ | 0 | 0 | 0 | 0 | 0 | 0 | 1 | 0 | 0 | 1 | 0 | 0 | 0 | 0 | 0 | 1 |
| +++-+ | 1 | 0 | 0 | 0 | 0 | 0 | 0 | 0 | 0 | 0 | 0 | 0 | 0 | 0 | 0 | 0 |
| -++-+ | 0 | 0 | 0 | 0 | 0 | 0 | 0 | 0 | 0 | 0 | 1 | 0 | 0 | 0 | 0 | 0 |
| +-+-+ | 0 | 0 | 0 | 0 | 0 | 0 | 0 | 0 | 0 | 0 | 0 | 0 | 0 | 0 | 0 | 0 |
| --+-+ | 0 | 0 | 0 | 0 | 0 | 0 | 0 | 0 | 0 | 0 | 0 | 0 | 0 | 0 | 0 | 0 |
| ++--+ | 0 | 0 | 0 | 0 | 0 | 0 | 0 | 0 | 0 | 0 | 0 | 0 | 0 | 0 | 0 | 0 |

| | | | | | | | | | | | | | | | |
|---|---|---|---|---|---|---|---|---|---|---|---|---|---|---|---|
| -+--+ | 0 | 0 | 1 | 1 | 0 | 0 | 1 | 0 | 1 | 0 | 1 | 0 | 0 | 0 | 0 | 0 |
| +--+ | 0 | 0 | 0 | 0 | 0 | 1 | 0 | 0 | 0 | 0 | 0 | 0 | 0 | 0 | 0 | 0 |
| ----+ | 0 | 0 | 0 | 0 | 0 | 0 | 0 | 0 | 0 | 0 | 0 | 0 | 0 | 0 | 0 | 1 |
| ++++- | 0 | 0 | 0 | 0 | 0 | 0 | 0 | 0 | 0 | 0 | 0 | 0 | 0 | 0 | 0 | 0 |
| -+++- | 0 | 0 | 0 | 0 | 0 | 0 | 0 | 0 | 2 | 0 | 0 | 0 | 0 | 0 | 0 | 0 |
| +-++- | 0 | 0 | 1 | 0 | 0 | 0 | 0 | 0 | 0 | 0 | 0 | 0 | 0 | 0 | 0 | 0 |
| --++- | 0 | 0 | 0 | 0 | 0 | 0 | 1 | 0 | 0 | 0 | 0 | 0 | 0 | 0 | 0 | 0 |
| ++-+- | 1 | 0 | 0 | 0 | 0 | 0 | 0 | 0 | 0 | 0 | 0 | 0 | 0 | 0 | 1 | 0 |
| -+-+- | 0 | 0 | 0 | 0 | 0 | 0 | 0 | 0 | 0 | 0 | 0 | 0 | 0 | 0 | 0 | 0 |
| +--+- | 0 | 0 | 0 | 1 | 0 | 0 | 0 | 0 | 0 | 0 | 0 | 0 | 1 | 0 | 0 | 2 |
| ---+- | 0 | 0 | 1 | 0 | 1 | 0 | 0 | 0 | 0 | 0 | 0 | 0 | 0 | 0 | 0 | 0 |
| +++-- | 0 | 0 | 0 | 0 | 0 | 0 | 0 | 0 | 1 | 0 | 0 | 0 | 1 | 0 | 0 | 0 |
| -++-- | 0 | 0 | 0 | 0 | 0 | 0 | 1 | 0 | 0 | 0 | 0 | 0 | 0 | 0 | 0 | 0 |
| +-+-- | 1 | 0 | 0 | 1 | 0 | 0 | 0 | 0 | 0 | 0 | 0 | 0 | 0 | 0 | 0 | 0 |
| --+-- | 0 | 0 | 0 | 0 | 0 | 0 | 0 | 0 | 0 | 0 | 0 | 0 | 0 | 1 | 0 | 0 |
| ++--- | 0 | 0 | 0 | 0 | 0 | 0 | 0 | 0 | 1 | 0 | 0 | 0 | 1 | 0 | 0 | 0 |
| -+--- | 0 | 0 | 0 | 1 | 0 | 0 | 0 | 0 | 0 | 0 | 0 | 0 | 0 | 0 | 0 | 0 |
| +---- | 0 | 1 | 0 | 0 | 1 | 0 | 0 | 1 | 1 | 0 | 0 | 0 | 0 | 1 | 0 | 0 |
| ----- | 0 | 0 | 0 | 1 | 0 | 0 | 1 | 0 | 0 | 0 | 0 | 1 | 0 | 0 | 0 | 0 |

**TABLE S8.5: The experimentally measured raw data of a complete set of the 1024 possible combinations of ten-photon coincidence events with each photon measured in the basis of $(|H\rangle \pm e^{i3\pi/10}|V\rangle)/\sqrt{2}$ (indicated by the $\pm$ label in the first column and the first line) to obtain $\left\langle M_{3\pi/10}^{\otimes 10}\right\rangle$. The five + or – label in the first column represent the measurement base choice for the five $o$ photons, while the five + or – label in the first line represent the measurement base choice for the five $e$ photons.**

| | +++ ++ | - +++ + | +- +++ | -- +++ | ++- ++ | -+- ++ | +-- ++ | --- ++ | +++ -+ | -++- + | +-+- + | --+- + | ++-- + | -+-- + | +--- + | ----+ |
|---|---|---|---|---|---|---|---|---|---|---|---|---|---|---|---|---|
| +++++ | 0 | 1 | 0 | 0 | 0 | 0 | 0 | 1 | 0 | 1 | 0 | 0 | 0 | 0 | 1 | 0 |
| -++++ | 0 | 0 | 0 | 0 | 0 | 0 | 1 | 0 | 0 | 0 | 0 | 0 | 0 | 0 | 0 | 0 |
| +-+++ | 0 | 0 | 0 | 0 | 0 | 0 | 0 | 1 | 0 | 0 | 0 | 0 | 0 | 0 | 0 | 1 |
| --+++ | 0 | 0 | 0 | 0 | 0 | 0 | 0 | 0 | 0 | 0 | 0 | 0 | 0 | 0 | 0 | 0 |
| ++-++ | 1 | 0 | 0 | 0 | 0 | 0 | 0 | 0 | 0 | 0 | 0 | 0 | 0 | 0 | 0 | 0 |
| -+-++ | 0 | 0 | 0 | 0 | 0 | 0 | 0 | 0 | 0 | 0 | 0 | 1 | 0 | 1 | 0 | 0 |
| +--++ | 0 | 0 | 0 | 0 | 0 | 0 | 0 | 1 | 1 | 0 | 0 | 0 | 1 | 0 | 0 | 0 |
| ---++ | 0 | 1 | 0 | 0 | 0 | 1 | 0 | 0 | 0 | 0 | 0 | 0 | 0 | 0 | 0 | 0 |
| +++-+ | 1 | 0 | 0 | 0 | 0 | 1 | 0 | 0 | 1 | 2 | 0 | 0 | 0 | 0 | 0 | 0 |
| -++-+ | 0 | 0 | 0 | 0 | 1 | 0 | 0 | 0 | 0 | 0 | 0 | 0 | 0 | 0 | 1 | 0 |
| +-+-+ | 0 | 0 | 0 | 0 | 0 | 0 | 0 | 0 | 0 | 0 | 0 | 0 | 0 | 0 | 0 | 0 |
| --+-+ | 0 | 0 | 0 | 0 | 0 | 0 | 0 | 0 | 1 | 0 | 0 | 0 | 0 | 0 | 1 | 0 |
| ++--+ | 0 | 0 | 1 | 0 | 0 | 0 | 0 | 0 | 0 | 0 | 0 | 0 | 0 | 0 | 1 | 0 |
| -+--+ | 1 | 0 | 0 | 0 | 0 | 0 | 0 | 0 | 1 | 0 | 0 | 0 | 0 | 0 | 0 | 0 |
| +---+ | 1 | 0 | 0 | 0 | 0 | 0 | 0 | 0 | 0 | 0 | 0 | 0 | 1 | 0 | 0 | 0 |
| ----+ | 0 | 0 | 0 | 0 | 0 | 0 | 0 | 0 | 0 | 0 | 1 | 0 | 0 | 0 | 1 | 0 |
| ++++- | 0 | 0 | 0 | 0 | 0 | 0 | 0 | 0 | 0 | 0 | 1 | 0 | 0 | 0 | 0 | 0 |
| -+++- | 0 | 0 | 0 | 0 | 0 | 0 | 1 | 0 | 0 | 0 | 0 | 0 | 0 | 0 | 0 | 0 |
| +-++- | 0 | 0 | 0 | 0 | 0 | 0 | 0 | 1 | 0 | 0 | 0 | 0 | 0 | 0 | 0 | 0 |
| --++- | 0 | 0 | 0 | 0 | 0 | 0 | 0 | 0 | 0 | 1 | 0 | 0 | 0 | 0 | 0 | 0 |
| ++-+- | 2 | 0 | 0 | 0 | 0 | 0 | 0 | 0 | 0 | 0 | 0 | 0 | 0 | 0 | 0 | 0 |
| -+-+- | 1 | 0 | 0 | 0 | 0 | 0 | 1 | 0 | 0 | 0 | 0 | 0 | 2 | 0 | 0 | 0 |
| +--+- | 0 | 0 | 0 | 0 | 0 | 0 | 0 | 0 | 0 | 0 | 1 | 0 | 0 | 0 | 0 | 0 |
| ---+- | 0 | 0 | 1 | 0 | 0 | 0 | 0 | 0 | 0 | 0 | 0 | 0 | 0 | 0 | 0 | 0 |
| +++-- | 0 | 0 | 0 | 0 | 0 | 0 | 0 | 1 | 0 | 0 | 1 | 0 | 0 | 0 | 0 | 0 |
| -++-- | 0 | 0 | 0 | 0 | 0 | 0 | 0 | 0 | 0 | 0 | 0 | 0 | 0 | 0 | 0 | 0 |
| +-+-- | 0 | 0 | 0 | 0 | 0 | 0 | 0 | 0 | 0 | 0 | 0 | 0 | 0 | 0 | 0 | 0 |
| --+-- | 0 | 1 | 1 | 0 | 0 | 0 | 0 | 0 | 3 | 0 | 0 | 0 | 0 | 0 | 0 | 0 |
| ++--- | 0 | 0 | 0 | 0 | 1 | 0 | 0 | 0 | 0 | 0 | 0 | 0 | 1 | 0 | 0 | 0 |
| -+--- | 1 | 0 | 0 | 0 | 0 | 0 | 0 | 0 | 0 | 0 | 0 | 0 | 0 | 1 | 0 | 0 |
| +---- | 0 | 0 | 0 | 0 | 0 | 0 | 0 | 0 | 0 | 0 | 0 | 0 | 0 | 0 | 0 | 0 |
| ----- | 1 | 0 | 0 | 0 | 0 | 0 | 0 | 0 | 0 | 0 | 2 | 0 | 0 | 0 | 0 | 0 |

| | +++ +- | - +++ - | +- ++- | -- ++- | ++- +- | -+- +- | +-- +- | ---+- | +++ -- | -++- - | +-+- - | --+- - | ++-- - | -+-- - | +---- | +--- | ----- |
|---|---|---|---|---|---|---|---|---|---|---|---|---|---|---|---|---|---|
| +++++ | 0 | 0 | 0 | 0 | 0 | 0 | 0 | 0 | 0 | 1 | 0 | 1 | 0 | 0 | 0 | 0 | 0 |
| -++++ | 0 | 0 | 0 | 1 | 0 | 0 | 0 | 0 | 1 | 0 | 0 | 0 | 0 | 0 | 0 | 0 | 0 |
| +-+++ | 0 | 0 | 0 | 0 | 0 | 0 | 0 | 0 | 0 | 0 | 0 | 0 | 0 | 0 | 0 | 0 | 0 |
| --+++ | 0 | 0 | 0 | 0 | 0 | 0 | 1 | 0 | 0 | 0 | 0 | 0 | 0 | 0 | 0 | 0 | 0 |
| ++-++ | 0 | 0 | 0 | 0 | 0 | 0 | 0 | 0 | 0 | 0 | 0 | 0 | 0 | 0 | 1 | 0 | 0 |
| -+-++ | 1 | 1 | 0 | 0 | 0 | 0 | 0 | 0 | 0 | 0 | 0 | 0 | 1 | 0 | 0 | 0 | 0 |
| +--++ | 0 | 0 | 0 | 1 | 0 | 0 | 0 | 0 | 0 | 0 | 0 | 0 | 1 | 0 | 0 | 0 | 0 |
| ---++ | 0 | 2 | 1 | 0 | 0 | 0 | 0 | 0 | 1 | 0 | 0 | 0 | 0 | 0 | 0 | 0 | 0 |
| +++-+ | 0 | 0 | 0 | 1 | 0 | 0 | 0 | 0 | 0 | 0 | 0 | 0 | 0 | 0 | 0 | 0 | 0 |
| -++-+ | 1 | 0 | 0 | 0 | 0 | 0 | 0 | 0 | 0 | 1 | 0 | 0 | 0 | 0 | 0 | 0 | 0 |
| +-+-+ | 0 | 0 | 0 | 0 | 0 | 0 | 0 | 0 | 0 | 0 | 0 | 0 | 0 | 0 | 0 | 0 | 1 |
| --+-+ | 0 | 0 | 0 | 0 | 0 | 0 | 0 | 1 | 0 | 0 | 0 | 0 | 0 | 1 | 0 | 0 | 1 |
| ++--+ | 0 | 0 | 0 | 0 | 0 | 0 | 0 | 0 | 0 | 0 | 0 | 0 | 0 | 0 | 0 | 0 | 0 |
| -+--+ | 0 | 0 | 1 | 0 | 0 | 1 | 0 | 0 | 0 | 0 | 0 | 0 | 0 | 0 | 0 | 0 | 0 |
| +---+ | 0 | 0 | 0 | 0 | 0 | 0 | 0 | 1 | 0 | 0 | 1 | 0 | 0 | 0 | 0 | 0 | 0 |
| ----+ | 0 | 0 | 0 | 0 | 0 | 1 | 0 | 0 | 0 | 1 | 0 | 0 | 0 | 0 | 0 | 0 | 0 |
| ++++- | 0 | 0 | 2 | 0 | 0 | 0 | 0 | 0 | 1 | 0 | 0 | 0 | 0 | 0 | 0 | 0 | 0 |
| -+++- | 0 | 0 | 0 | 0 | 0 | 1 | 0 | 0 | 0 | 0 | 0 | 0 | 0 | 0 | 0 | 0 | 0 |
| +-++- | 0 | 0 | 0 | 0 | 0 | 0 | 0 | 0 | 0 | 0 | 0 | 0 | 0 | 0 | 0 | 0 | 0 |
| --++- | 0 | 0 | 0 | 1 | 0 | 0 | 0 | 0 | 0 | 0 | 0 | 0 | 1 | 0 | 1 | 0 | 0 |
| ++-+- | 0 | 0 | 0 | 0 | 0 | 0 | 0 | 0 | 0 | 0 | 0 | 0 | 0 | 0 | 0 | 0 | 2 |
| -+-+- | 1 | 1 | 0 | 0 | 0 | 0 | 0 | 0 | 0 | 0 | 0 | 0 | 0 | 0 | 0 | 0 | 0 |
| +--+- | 0 | 0 | 0 | 0 | 0 | 0 | 0 | 0 | 0 | 0 | 0 | 0 | 0 | 0 | 0 | 0 | 0 |
| ---+- | 0 | 0 | 0 | 0 | 0 | 0 | 0 | 0 | 0 | 0 | 0 | 0 | 0 | 0 | 1 | 0 | 0 |
| +++-- | 0 | 0 | 0 | 0 | 0 | 0 | 0 | 0 | 0 | 1 | 0 | 0 | 0 | 0 | 0 | 0 | 0 |
| -++-- | 0 | 0 | 0 | 0 | 0 | 0 | 0 | 1 | 0 | 0 | 0 | 0 | 0 | 0 | 0 | 0 | 0 |
| +-+-- | 0 | 0 | 0 | 0 | 0 | 0 | 0 | 0 | 0 | 1 | 0 | 0 | 0 | 0 | 0 | 0 | 0 |
| --+-- | 0 | 0 | 0 | 0 | 0 | 0 | 0 | 0 | 0 | 0 | 0 | 0 | 0 | 0 | 0 | 0 | 0 |
| ++--- | 0 | 0 | 0 | 0 | 0 | 0 | 0 | 0 | 0 | 1 | 0 | 0 | 0 | 0 | 0 | 0 | 0 |
| -+--- | 0 | 0 | 0 | 1 | 0 | 1 | 0 | 0 | 0 | 0 | 0 | 0 | 0 | 0 | 0 | 0 | 0 |
| +---- | 0 | 1 | 0 | 0 | 0 | 0 | 0 | 0 | 0 | 1 | 0 | 0 | 0 | 1 | 0 | 0 | 0 |
| ----- | 0 | 0 | 0 | 0 | 0 | 0 | 0 | 0 | 0 | 0 | 0 | 0 | 0 | 0 | 1 | 0 |

**TABLE S8.6: The experimentally measured raw data of a complete set of the 1024 possible combinations of ten-photon coincidence events with each photon measured in the basis of** $(|H\rangle \pm e^{i4\pi/10}|V\rangle)/\sqrt{2}$ **(indicated by the $\pm$ label in the first column and the first line) to obtain** $\left\langle M_{4\pi/10}^{\otimes 10} \right\rangle$. The five + or – label in the first column represent the measurement base choice for the five *o* photons, while the five + or – label in the first line represent the measurement base choice for the five *e* photons.

| | +++ ++ | - +++ + | +- +++ | -- +++ | ++- ++ | -+- ++ | +-- ++ | --- ++ | +++ -+ | -+- -+ | +-+- + | --+- + | ++-- + | -+-- + | +--- + | ----+ |
|---|---|---|---|---|---|---|---|---|---|---|---|---|---|---|---|---|
| +++++ | 0 | 0 | 0 | 0 | 0 | 0 | 0 | 0 | 0 | 0 | 1 | 0 | 0 | 0 | 0 | 0 |
| -++++ | 0 | 0 | 0 | 0 | 0 | 0 | 0 | 0 | 0 | 0 | 0 | 1 | 0 | 0 | 0 | 0 |
| +-+++ | 0 | 0 | 0 | 0 | 0 | 0 | 0 | 0 | 0 | 0 | 0 | 0 | 0 | 0 | 0 | 0 |
| --+++ | 0 | 0 | 0 | 0 | 0 | 1 | 0 | 0 | 0 | 0 | 0 | 0 | 1 | 0 | 1 |
| ++-++ | 0 | 0 | 0 | 0 | 0 | 0 | 0 | 0 | 0 | 0 | 0 | 0 | 0 | 1 | 0 | 0 |
| -+-++ | 0 | 0 | 0 | 0 | 0 | 0 | 0 | 0 | 0 | 0 | 0 | 0 | 0 | 0 | 0 | 0 |
| +--++ | 0 | 0 | 0 | 0 | 0 | 0 | 1 | 0 | 0 | 0 | 1 | 0 | 0 | 0 | 0 | 0 |
| ---++ | 0 | 0 | 0 | 0 | 0 | 0 | 0 | 0 | 0 | 0 | 1 | 1 | 0 | 0 | 0 | 0 |
| +++-+ | 0 | 0 | 0 | 0 | 0 | 0 | 0 | 0 | 0 | 0 | 0 | 0 | 0 | 2 | 0 | 0 |
| -++-+ | 0 | 0 | 0 | 0 | 0 | 0 | 0 | 0 | 0 | 0 | 0 | 0 | 0 | 0 | 0 | 0 |
| +-+-+ | 1 | 0 | 0 | 0 | 0 | 0 | 0 | 0 | 0 | 0 | 2 | 0 | 0 | 0 | 0 | 0 |
| --+-+ | 0 | 0 | 0 | 0 | 0 | 0 | 0 | 0 | 0 | 1 | 0 | 0 | 0 | 0 | 0 | 0 |
| ++--+ | 1 | 0 | 0 | 0 | 0 | 1 | 0 | 0 | 0 | 0 | 0 | 0 | 0 | 0 | 0 | 1 |
| -+--+ | 0 | 0 | 0 | 0 | 2 | 0 | 0 | 0 | 0 | 0 | 1 | 0 | 0 | 0 | 0 | 0 |
| +---+ | 0 | 0 | 1 | 0 | 0 | 0 | 0 | 0 | 0 | 0 | 0 | 0 | 0 | 0 | 0 | 0 |
| ----+ | 0 | 0 | 0 | 0 | 0 | 0 | 0 | 0 | 0 | 0 | 0 | 0 | 0 | 0 | 0 | 0 |
| +++++ | 0 | 0 | 0 | 0 | 0 | 0 | 0 | 1 | 0 | 0 | 0 | 0 | 0 | 0 | 0 | 0 |
| -+++- | 0 | 0 | 0 | 0 | 0 | 0 | 0 | 0 | 0 | 0 | 0 | 0 | 0 | 0 | 0 | 0 |
| +-++- | 0 | 0 | 0 | 0 | 1 | 0 | 0 | 0 | 0 | 0 | 0 | 0 | 0 | 0 | 0 | 0 |
| --++- | 1 | 1 | 0 | 0 | 0 | 0 | 0 | 1 | 0 | 0 | 0 | 0 | 0 | 0 | 0 | 0 |
| ++-+- | 0 | 0 | 0 | 0 | 0 | 1 | 0 | 0 | 0 | 0 | 0 | 0 | 0 | 0 | 1 | 1 |
| -+-+- | 0 | 0 | 0 | 0 | 0 | 0 | 0 | 0 | 2 | 0 | 0 | 0 | 0 | 0 | 0 | 0 |
| +--+- | 0 | 0 | 0 | 0 | 0 | 0 | 0 | 0 | 0 | 0 | 0 | 0 | 0 | 0 | 0 | 0 |
| ---+- | 0 | 0 | 0 | 0 | 0 | 0 | 0 | 0 | 0 | 0 | 0 | 0 | 1 | 0 | 0 | 0 |
| +++-- | 1 | 0 | 0 | 0 | 0 | 0 | 0 | 0 | 0 | 0 | 0 | 0 | 0 | 0 | 0 | 0 |
| -++-- | 0 | 0 | 0 | 0 | 0 | 0 | 0 | 0 | 0 | 0 | 0 | 0 | 0 | 0 | 0 | 0 |
| +-+-- | 0 | 0 | 0 | 1 | 0 | 0 | 0 | 0 | 0 | 1 | 0 | 0 | 0 | 0 | 0 | 0 |
| --+-- | 0 | 0 | 0 | 0 | 0 | 0 | 0 | 0 | 0 | 1 | 0 | 0 | 0 | 0 | 0 | 1 |
| ++--- | 0 | 0 | 0 | 0 | 0 | 0 | 0 | 0 | 0 | 0 | 0 | 0 | 0 | 2 | 0 | 0 |
| -+--- | 1 | 0 | 0 | 1 | 0 | 0 | 0 | 0 | 0 | 0 | 1 | 0 | 0 | 0 | 0 | 1 |
| +---- | 0 | 0 | 0 | 0 | 0 | 0 | 0 | 1 | 0 | 0 | 0 | 1 | 0 | 0 | 0 | 0 |
| ----- | 0 | 1 | 0 | 0 | 0 | 0 | 0 | 0 | 0 | 0 | 0 | 0 | 0 | 0 | 0 | 0 |

| | +++ +- | - +++ - | +- ++- | -- ++- | ++- +- | -+- +- | +- +- | ---+- | +++ -- | -++- - | +-+- - | --+- | ++-- - | -+--- | +---- | ----- |
|---|---|---|---|---|---|---|---|---|---|---|---|---|---|---|---|---|
| +++++ | 0 | 0 | 0 | 0 | 1 | 0 | 0 | 0 | 1 | 0 | 0 | 0 | 1 | 0 | 0 | 0 |
| -++++ | 0 | 0 | 0 | 0 | 0 | 0 | 0 | 0 | 0 | 0 | 0 | 0 | 0 | 0 | 0 | 0 |
| +-+++ | 1 | 0 | 0 | 0 | 0 | 0 | 0 | 0 | 1 | 0 | 0 | 0 | 0 | 0 | 0 | 0 |
| --+++ | 0 | 1 | 0 | 0 | 1 | 0 | 0 | 0 | 0 | 0 | 0 | 0 | 0 | 1 | 0 | 0 |
| ++-++ | 0 | 0 | 0 | 0 | 0 | 0 | 0 | 0 | 0 | 0 | 0 | 0 | 0 | 0 | 0 | 0 |
| -+-++ | 0 | 0 | 0 | 0 | 0 | 0 | 0 | 0 | 0 | 0 | 0 | 0 | 0 | 0 | 1 | 0 |
| +--++ | 0 | 0 | 0 | 0 | 0 | 0 | 0 | 0 | 0 | 0 | 0 | 0 | 0 | 0 | 0 | 0 |
| ---++ | 0 | 0 | 0 | 0 | 0 | 0 | 0 | 0 | 0 | 0 | 0 | 1 | 0 | 0 | 0 | 0 |
| +++-+ | 0 | 0 | 0 | 0 | 1 | 0 | 0 | 0 | 0 | 0 | 0 | 0 | 1 | 0 | 0 | 0 |
| -++-+ | 0 | 0 | 0 | 0 | 0 | 0 | 0 | 0 | 0 | 0 | 0 | 1 | 0 | 0 | 0 | 0 |
| +-+-+ | 0 | 0 | 0 | 0 | 1 | 0 | 0 | 0 | 0 | 0 | 0 | 0 | 0 | 0 | 0 | 0 |
| --+-+ | 0 | 0 | 0 | 0 | 1 | 0 | 0 | 0 | 0 | 0 | 0 | 0 | 0 | 0 | 0 | 0 |
| ++--+ | 0 | 0 | 1 | 1 | 0 | 0 | 0 | 0 | 0 | 0 | 0 | 0 | 0 | 0 | 0 | 0 |
| -+--+ | 0 | 0 | 0 | 0 | 0 | 0 | 0 | 0 | 0 | 1 | 0 | 0 | 0 | 0 | 0 | 0 |
| +---+ | 0 | 0 | 0 | 0 | 1 | 0 | 0 | 0 | 0 | 0 | 0 | 0 | 1 | 0 | 0 | 0 |
| ----+ | 0 | 0 | 0 | 0 | 0 | 0 | 0 | 0 | 0 | 0 | 0 | 0 | 0 | 0 | 0 | 0 |
| ++++- | 0 | 0 | 0 | 0 | 0 | 0 | 0 | 0 | 0 | 0 | 0 | 0 | 0 | 0 | 0 | 1 |
| -+++- | 0 | 0 | 0 | 1 | 0 | 0 | 0 | 0 | 1 | 0 | 1 | 1 | 0 | 0 | 0 | 0 |
| +-++- | 0 | 0 | 0 | 0 | 0 | 0 | 0 | 0 | 0 | 0 | 0 | 0 | 0 | 0 | 1 | 0 |
| --++- | 0 | 0 | 0 | 0 | 0 | 0 | 0 | 0 | 0 | 0 | 0 | 0 | 0 | 0 | 0 | 0 |
| ++-+- | 0 | 0 | 0 | 0 | 0 | 0 | 0 | 0 | 0 | 0 | 0 | 0 | 0 | 0 | 1 | 0 |
| -+-+- | 0 | 0 | 0 | 0 | 0 | 1 | 0 | 0 | 0 | 0 | 0 | 0 | 0 | 0 | 0 | 0 |
| +--+- | 0 | 0 | 0 | 0 | 0 | 0 | 0 | 0 | 0 | 0 | 0 | 0 | 0 | 0 | 0 | 0 |
| ---+- | 0 | 0 | 1 | 1 | 0 | 0 | 0 | 0 | 1 | 0 | 0 | 0 | 0 | 0 | 0 | 0 |
| +++-- | 0 | 0 | 0 | 0 | 0 | 0 | 0 | 0 | 0 | 0 | 0 | 0 | 0 | 1 | 0 | 0 |
| -++-- | 1 | 0 | 0 | 1 | 1 | 0 | 0 | 0 | 0 | 1 | 0 | 0 | 0 | 0 | 0 | 0 |
| +-+-- | 0 | 0 | 0 | 0 | 0 | 0 | 0 | 0 | 0 | 0 | 0 | 0 | 0 | 0 | 0 | 0 |
| --+-- | 0 | 0 | 0 | 0 | 0 | 0 | 0 | 1 | 0 | 0 | 0 | 0 | 0 | 0 | 0 | 0 |
| ++--- | 1 | 0 | 0 | 0 | 0 | 0 | 0 | 0 | 0 | 0 | 0 | 0 | 0 | 0 | 0 | 0 |
| -+--- | 0 | 0 | 0 | 0 | 0 | 0 | 0 | 1 | 0 | 0 | 0 | 1 | 0 | 0 | 0 | 0 |
| +---- | 1 | 0 | 0 | 0 | 0 | 0 | 0 | 0 | 0 | 0 | 0 | 0 | 1 | 0 | 0 | 0 |
| ----- | 0 | 0 | 0 | 1 | 1 | 0 | 0 | 0 | 0 | 0 | 0 | 0 | 0 | 0 | 0 | 0 |

**TABLE S8.7: The experimentally measured raw data of a complete set of the 1024 possible combinations of ten-photon coincidence events with each photon measured in the basis of** $(|H\rangle \pm e^{i5\pi/10}|V\rangle)/\sqrt{2}$ **(indicated by the** $\pm$ **label in the first column and the first line) to obtain** $\left\langle M_{5\pi/10}^{\otimes 10} \right\rangle$. **The five + or – label in the first column represent the measurement base choice for the five** *o* **photons, while the five + or – label in the first line represent the measurement base choice for the five** *e* **photons.**

| | +++ ++ | - +++ + | +- +++ | -- +++ | ++- ++ | -+- ++ | +-- ++ | --- ++ | +++ -+ | -++- + | +-+- + | --+- + | ++-- + | -+-- + | +--- + | ----+ |
|---|---|---|---|---|---|---|---|---|---|---|---|---|---|---|---|---|
| +++++ | 0 | 1 | 0 | 0 | 0 | 0 | 0 | 0 | 0 | 0 | 0 | 0 | 0 | 0 | 1 | 1 |
| -++++ | 0 | 0 | 0 | 1 | 0 | 1 | 0 | 0 | 0 | 0 | 0 | 0 | 0 | 0 | 0 | 0 |
| +-+++ | 0 | 0 | 1 | 0 | 0 | 0 | 0 | 0 | 0 | 0 | 0 | 0 | 0 | 0 | 0 | 0 |
| --+++ | 1 | 0 | 0 | 0 | 0 | 0 | 0 | 0 | 1 | 0 | 0 | 0 | 0 | 0 | 0 | 1 |
| ++-++ | 0 | 0 | 0 | 0 | 1 | 0 | 0 | 0 | 0 | 0 | 1 | 0 | 0 | 0 | 0 | 0 |
| -+-++ | 0 | 0 | 0 | 0 | 0 | 0 | 0 | 0 | 0 | 0 | 0 | 0 | 0 | 1 | 1 | 0 |
| +--++ | 0 | 0 | 1 | 0 | 0 | 0 | 0 | 1 | 0 | 0 | 0 | 0 | 0 | 0 | 0 | 0 |
| ---++ | 0 | 0 | 0 | 0 | 0 | 0 | 0 | 0 | 0 | 0 | 0 | 0 | 0 | 0 | 0 | 0 |
| +++-+ | 1 | 0 | 0 | 0 | 0 | 0 | 0 | 0 | 0 | 0 | 0 | 0 | 0 | 0 | 0 | 0 |
| -++-+ | 0 | 2 | 0 | 0 | 0 | 0 | 0 | 1 | 0 | 0 | 0 | 0 | 0 | 0 | 0 | 0 |
| +-+-+ | 0 | 1 | 0 | 0 | 1 | 0 | 0 | 1 | 0 | 0 | 0 | 0 | 0 | 0 | 0 | 0 |
| --+-+ | 0 | 0 | 0 | 0 | 0 | 0 | 1 | 0 | 0 | 0 | 0 | 0 | 0 | 0 | 0 | 0 |
| ++--+ | 0 | 0 | 0 | 0 | 0 | 0 | 0 | 0 | 0 | 0 | 0 | 0 | 1 | 0 | 0 | 0 |
| -+--+ | 0 | 0 | 1 | 0 | 1 | 0 | 0 | 0 | 0 | 0 | 0 | 0 | 0 | 0 | 0 | 0 |
| +---+ | 1 | 0 | 0 | 0 | 0 | 0 | 0 | 0 | 0 | 0 | 0 | 0 | 0 | 0 | 0 | 0 |
| ----+ | 0 | 0 | 0 | 0 | 0 | 0 | 0 | 0 | 0 | 0 | 0 | 0 | 0 | 0 | 0 | 0 |
| ++++- | 0 | 0 | 0 | 1 | 0 | 0 | 0 | 0 | 0 | 0 | 0 | 0 | 0 | 0 | 0 | 1 |
| -+++- | 0 | 0 | 0 | 0 | 1 | 0 | 0 | 0 | 1 | 0 | 0 | 0 | 0 | 0 | 0 | 0 |
| +-++- | 0 | 0 | 0 | 0 | 0 | 0 | 0 | 0 | 0 | 0 | 0 | 0 | 0 | 0 | 0 | 0 |
| --++- | 0 | 0 | 1 | 0 | 0 | 0 | 0 | 0 | 0 | 0 | 0 | 0 | 0 | 0 | 0 | 1 |
| ++-+- | 0 | 0 | 0 | 0 | 0 | 0 | 0 | 0 | 0 | 0 | 1 | 0 | 0 | 0 | 0 | 0 |
| -+-+- | 0 | 0 | 0 | 0 | 0 | 0 | 1 | 1 | 0 | 0 | 0 | 0 | 0 | 0 | 0 | 0 |
| +--+- | 0 | 0 | 0 | 0 | 0 | 0 | 0 | 0 | 1 | 0 | 0 | 0 | 0 | 0 | 0 | 0 |
| ---+- | 0 | 0 | 0 | 0 | 0 | 0 | 0 | 0 | 0 | 0 | 0 | 0 | 0 | 0 | 0 | 0 |
| +++-- | 0 | 1 | 0 | 0 | 0 | 0 | 0 | 0 | 0 | 0 | 0 | 0 | 0 | 0 | 0 | 0 |
| -++-- | 0 | 1 | 0 | 0 | 0 | 1 | 0 | 0 | 0 | 0 | 0 | 0 | 0 | 0 | 0 | 0 |
| +-+-- | 0 | 0 | 0 | 0 | 0 | 0 | 0 | 0 | 1 | 0 | 0 | 0 | 0 | 0 | 0 | 0 |
| --+-- | 0 | 0 | 0 | 0 | 0 | 0 | 0 | 0 | 1 | 0 | 0 | 1 | 0 | 0 | 1 | 0 |
| ++--- | 0 | 0 | 0 | 0 | 0 | 1 | 0 | 0 | 0 | 0 | 0 | 0 | 0 | 0 | 0 | 0 |
| -+--- | 0 | 0 | 0 | 0 | 0 | 0 | 0 | 0 | 0 | 0 | 0 | 1 | 0 | 1 | 1 | 0 |
| +---- | 0 | 0 | 0 | 0 | 0 | 0 | 0 | 0 | 1 | 0 | 0 | 0 | 0 | 0 | 0 | 0 |
| ----- | 0 | 0 | 0 | 0 | 0 | 0 | 2 | 0 | 0 | 0 | 0 | 0 | 0 | 0 | 0 | 0 |

| | +++ +- | - +++ - | +- ++- | -- ++- | ++- +- | -+- +- | +-- +- | ---+- | +++ -- | -++ -- | +-+- - | --+-- | ++-- - | -+--- | +---- | ----- |
|---|---|---|---|---|---|---|---|---|---|---|---|---|---|---|---|---|
| +++++ | 0 | 0 | 0 | 0 | 0 | 0 | 0 | 0 | 0 | 0 | 0 | 0 | 0 | 0 | 0 | 0 |
| -++++ | 0 | 0 | 0 | 0 | 1 | 0 | 0 | 0 | 0 | 0 | 0 | 0 | 0 | 1 | 0 | 0 |
| +-+++ | 0 | 0 | 0 | 0 | 1 | 0 | 0 | 0 | 0 | 0 | 1 | 0 | 0 | 1 | 0 | 0 |
| --+++ | 0 | 1 | 0 | 0 | 0 | 1 | 0 | 0 | 0 | 0 | 1 | 0 | 0 | 0 | 0 | 0 |
| ++-++ | 0 | 0 | 0 | 1 | 1 | 0 | 0 | 0 | 1 | 0 | 0 | 0 | 0 | 1 | 0 | 0 |
| -+-++ | 0 | 0 | 0 | 0 | 0 | 1 | 0 | 0 | 0 | 0 | 0 | 0 | 0 | 0 | 0 | 0 |
| +--++ | 0 | 0 | 0 | 1 | 0 | 0 | 0 | 0 | 0 | 0 | 0 | 0 | 0 | 0 | 0 | 0 |
| ---++ | 0 | 0 | 0 | 1 | 0 | 0 | 0 | 1 | 0 | 0 | 1 | 0 | 0 | 0 | 0 | 0 |
| +++-+ | 0 | 0 | 0 | 0 | 1 | 0 | 0 | 1 | 0 | 0 | 0 | 0 | 0 | 0 | 0 | 0 |
| -++-+ | 0 | 0 | 0 | 1 | 0 | 0 | 0 | 0 | 0 | 0 | 1 | 0 | 0 | 0 | 0 | 0 |
| +-+-+ | 0 | 0 | 0 | 0 | 0 | 0 | 0 | 0 | 0 | 0 | 0 | 0 | 0 | 0 | 1 | 0 |
| --+-+ | 0 | 0 | 0 | 0 | 0 | 0 | 0 | 0 | 0 | 0 | 0 | 0 | 0 | 0 | 0 | 0 |
| ++--+ | 0 | 0 | 0 | 0 | 0 | 0 | 0 | 0 | 0 | 0 | 0 | 0 | 0 | 0 | 0 | 0 |
| -+--+ | 0 | 0 | 0 | 1 | 0 | 0 | 0 | 0 | 0 | 0 | 0 | 0 | 0 | 0 | 0 | 0 |
| +---+ | 0 | 0 | 0 | 0 | 0 | 0 | 0 | 0 | 0 | 0 | 0 | 0 | 0 | 0 | 0 | 0 |
| ----+ | 0 | 0 | 0 | 0 | 0 | 0 | 0 | 0 | 0 | 0 | 0 | 0 | 0 | 0 | 0 | 0 |
| ++++- | 0 | 0 | 1 | 0 | 0 | 0 | 1 | 0 | 0 | 0 | 0 | 0 | 0 | 0 | 0 | 0 |
| -+++- | 0 | 0 | 0 | 1 | 0 | 0 | 0 | 0 | 1 | 0 | 0 | 0 | 0 | 0 | 0 | 0 |
| +-++- | 0 | 0 | 1 | 0 | 1 | 1 | 0 | 0 | 0 | 0 | 0 | 0 | 0 | 0 | 1 | 1 |
| --++- | 0 | 0 | 0 | 1 | 0 | 0 | 0 | 2 | 0 | 0 | 1 | 1 | 0 | 0 | 0 | 0 |
| ++-+- | 0 | 0 | 1 | 0 | 1 | 1 | 0 | 0 | 0 | 0 | 0 | 0 | 0 | 0 | 1 | 1 |
| -+-+- | 0 | 0 | 0 | 1 | 0 | 0 | 0 | 2 | 0 | 1 | 1 | 1 | 0 | 0 | 0 | 0 |
| +--+- | 0 | 0 | 0 | 0 | 0 | 0 | 0 | 0 | 1 | 1 | 0 | 0 | 0 | 0 | 0 | 0 |
| ---+- | 0 | 0 | 0 | 1 | 0 | 1 | 0 | 0 | 0 | 0 | 0 | 0 | 0 | 0 | 0 | 0 |
| +++-- | 0 | 0 | 0 | 0 | 0 | 1 | 0 | 0 | 0 | 0 | 2 | 0 | 1 | 0 | 0 | 0 |
| -++-- | 0 | 0 | 0 | 0 | 0 | 0 | 0 | 0 | 0 | 0 | 0 | 0 | 0 | 1 | 0 | 0 |
| +-+-- | 0 | 1 | 0 | 1 | 0 | 0 | 0 | 0 | 0 | 0 | 0 | 1 | 0 | 0 | 0 | 0 |
| --+-- | 0 | 0 | 0 | 0 | 0 | 0 | 0 | 0 | 0 | 0 | 0 | 0 | 0 | 0 | 0 | 0 |
| ++--- | 1 | 0 | 0 | 0 | 0 | 0 | 0 | 0 | 1 | 0 | 0 | 0 | 0 | 0 | 1 | 0 |
| -+--- | 0 | 0 | 0 | 0 | 0 | 0 | 0 | 0 | 0 | 0 | 0 | 0 | 0 | 0 | 0 | 0 |

| | | | | | | | | | | | | | | | |
|---|---|---|---|---|---|---|---|---|---|---|---|---|---|---|---|
| +---- | 0 | 0 | 0 | 0 | 0 | 0 | 2 | 0 | 0 | 0 | 0 | 0 | 0 | 0 | 0 | 0 |
| ----- | 0 | 0 | 0 | 1 | 0 | 0 | 0 | 0 | 0 | 0 | 0 | 1 | 0 | 0 | 0 | 0 |

**TABLE S8.8: The experimentally measured raw data of a complete set of the 1024 possible combinations of ten-photon coincidence events with each photon measured in the basis of** $(|H\rangle \pm e^{i6\pi/10}|V\rangle)/\sqrt{2}$ **(indicated by the $\pm$ label in the first column and the first line) to obtain** $\left\langle M_{6\pi/10}^{\otimes 10} \right\rangle$**.** The five + or – label in the first column represent the measurement base choice for the five *o* photons, while the five + or – label in the first line represent the measurement base choice for the five *e* photons.

| | +++++ | -++++ | +-+++ | --+++ | ++-++ | -+-++ | +--++ | ---++ | +++-+ | -++-+ | +-+-+ | --+-+ | ++--+ | -+--+ | +---+ | ----+ |
|---|---|---|---|---|---|---|---|---|---|---|---|---|---|---|---|---|
| +++++ | 0 | 0 | 0 | 0 | 0 | 0 | 0 | 0 | 0 | 0 | 0 | 0 | 0 | 0 | 0 | 0 |
| -++++ | 0 | 0 | 0 | 1 | 1 | 0 | 0 | 0 | 0 | 0 | 0 | 0 | 0 | 0 | 0 | 0 |
| +-+++ | 0 | 0 | 0 | 0 | 0 | 0 | 0 | 0 | 0 | 0 | 0 | 0 | 0 | 2 | 0 | 0 |
| --+++ | 0 | 0 | 0 | 0 | 0 | 0 | 0 | 0 | 0 | 0 | 0 | 0 | 0 | 0 | 0 | 0 |
| ++-++ | 0 | 0 | 0 | 0 | 0 | 0 | 0 | 0 | 0 | 0 | 0 | 0 | 0 | 0 | 0 | 0 |
| -+-++ | 0 | 0 | 0 | 0 | 0 | 1 | 0 | 0 | 0 | 0 | 2 | 0 | 0 | 0 | 0 | 0 |
| +--++ | 0 | 0 | 0 | 0 | 0 | 1 | 1 | 0 | 1 | 0 | 1 | 0 | 0 | 0 | 0 | 0 |
| ---++ | 0 | 0 | 0 | 0 | 0 | 0 | 1 | 0 | 0 | 0 | 0 | 0 | 0 | 0 | 0 | 0 |
| +++-+ | 0 | 0 | 0 | 0 | 0 | 0 | 0 | 0 | 0 | 1 | 0 | 0 | 0 | 0 | 0 | 1 |
| -++-+ | 0 | 0 | 0 | 0 | 0 | 0 | 0 | 0 | 0 | 0 | 0 | 1 | 0 | 0 | 0 | 0 |
| +-+-+ | 0 | 0 | 0 | 0 | 1 | 0 | 2 | 0 | 0 | 0 | 0 | 0 | 0 | 0 | 0 | 0 |
| --+-+ | 0 | 0 | 0 | 0 | 0 | 0 | 0 | 0 | 0 | 0 | 0 | 0 | 0 | 0 | 0 | 1 |
| ++--+ | 0 | 0 | 0 | 0 | 1 | 0 | 0 | 0 | 0 | 0 | 0 | 0 | 0 | 0 | 0 | 0 |
| -+--+ | 0 | 0 | 1 | 0 | 0 | 0 | 0 | 0 | 0 | 0 | 0 | 0 | 0 | 0 | 0 | 0 |
| +---+ | 0 | 0 | 0 | 0 | 0 | 0 | 0 | 1 | 1 | 0 | 0 | 0 | 0 | 1 | 1 | 0 |
| ----+ | 1 | 0 | 0 | 0 | 0 | 0 | 0 | 0 | 0 | 0 | 1 | 0 | 0 | 0 | 0 | 1 |
| ++++- | 0 | 0 | 0 | 0 | 0 | 0 | 0 | 0 | 0 | 0 | 0 | 0 | 0 | 1 | 0 | 0 |
| -+++- | 0 | 0 | 0 | 1 | 0 | 0 | 0 | 0 | 1 | 0 | 0 | 0 | 0 | 0 | 0 | 0 |
| +-++- | 0 | 0 | 0 | 0 | 0 | 0 | 1 | 0 | 0 | 1 | 0 | 1 | 0 | 0 | 0 | 0 |
| --++- | 0 | 0 | 0 | 0 | 0 | 0 | 0 | 0 | 0 | 0 | 0 | 0 | 0 | 0 | 0 | 0 |
| ++-+- | 0 | 0 | 0 | 0 | 0 | 0 | 0 | 0 | 0 | 0 | 0 | 0 | 0 | 0 | 0 | 0 |
| -+-+- | 1 | 0 | 0 | 0 | 0 | 0 | 0 | 0 | 0 | 2 | 0 | 0 | 0 | 0 | 1 | 0 |
| +--+- | 0 | 0 | 0 | 0 | 0 | 0 | 0 | 0 | 0 | 0 | 0 | 0 | 0 | 0 | 0 | 0 |
| ---+- | 0 | 0 | 0 | 0 | 0 | 0 | 0 | 0 | 0 | 1 | 0 | 0 | 0 | 0 | 0 | 0 |
| +++-- | 1 | 1 | 0 | 0 | 0 | 0 | 0 | 0 | 1 | 0 | 0 | 0 | 0 | 0 | 0 | 0 |
| -++-- | 0 | 0 | 0 | 0 | 0 | 0 | 0 | 0 | 0 | 0 | 0 | 0 | 0 | 0 | 0 | 0 |
| +-+-- | 0 | 0 | 0 | 0 | 0 | 0 | 0 | 0 | 1 | 0 | 0 | 0 | 0 | 1 | 1 | 0 |
| --+-- | 0 | 0 | 1 | 0 | 0 | 0 | 0 | 0 | 0 | 0 | 0 | 0 | 0 | 0 | 0 | 0 |
| ++--- | 0 | 0 | 0 | 0 | 0 | 0 | 0 | 0 | 0 | 0 | 0 | 0 | 0 | 0 | 0 | 0 |
| -+--- | 0 | 0 | 0 | 0 | 0 | 0 | 0 | 0 | 0 | 0 | 0 | 0 | 0 | 0 | 0 | 0 |
| +---- | 0 | 0 | 0 | 0 | 0 | 0 | 0 | 0 | 0 | 0 | 0 | 0 | 0 | 0 | 0 | 0 |
| ----- | 0 | 0 | 0 | 0 | 0 | 0 | 0 | 0 | 0 | 0 | 0 | 0 | 0 | 0 | 1 | 0 |

| | ++++- | -+++- | +-++- | --++- | ++-+- | -+-+- | +--+- | ---+- | +++-- | -++-- | +-+-- | --+-- | ++--- | -+--- | +---- | ----- |
|---|---|---|---|---|---|---|---|---|---|---|---|---|---|---|---|---|
| +++++ | 0 | 0 | 2 | 0 | 0 | 0 | 0 | 0 | 0 | 0 | 0 | 0 | 0 | 0 | 0 | 0 |
| -++++ | 0 | 0 | 0 | 0 | 0 | 0 | 0 | 0 | 0 | 0 | 1 | 0 | 0 | 0 | 0 | 0 |
| +-+++ | 1 | 0 | 1 | 0 | 0 | 0 | 0 | 0 | 0 | 0 | 2 | 0 | 0 | 0 | 0 | 0 |
| --+++ | 0 | 0 | 0 | 0 | 0 | 0 | 0 | 0 | 1 | 0 | 0 | 1 | 0 | 0 | 0 | 0 |
| ++-++ | 1 | 0 | 0 | 0 | 0 | 0 | 0 | 1 | 0 | 1 | 1 | 0 | 0 | 0 | 0 | 0 |
| -+-++ | 0 | 0 | 0 | 0 | 0 | 0 | 0 | 1 | 1 | 0 | 0 | 0 | 0 | 0 | 0 | 0 |
| +--++ | 0 | 1 | 0 | 0 | 0 | 0 | 0 | 0 | 0 | 0 | 0 | 0 | 0 | 0 | 0 | 0 |
| ---++ | 0 | 1 | 0 | 0 | 0 | 0 | 0 | 0 | 0 | 0 | 0 | 1 | 0 | 0 | 0 | 0 |
| +++-+ | 0 | 0 | 0 | 0 | 0 | 0 | 0 | 0 | 0 | 0 | 0 | 0 | 0 | 1 | 1 | 0 |
| -++-+ | 0 | 0 | 0 | 0 | 0 | 0 | 0 | 1 | 1 | 0 | 0 | 0 | 0 | 0 | 0 | 0 |
| +-+-+ | 0 | 0 | 0 | 0 | 0 | 0 | 0 | 0 | 0 | 0 | 0 | 0 | 0 | 0 | 0 | 0 |
| --+-+ | 0 | 0 | 0 | 0 | 0 | 0 | 1 | 0 | 0 | 0 | 0 | 0 | 0 | 0 | 0 | 0 |
| ++--+ | 0 | 0 | 0 | 0 | 0 | 0 | 0 | 0 | 0 | 0 | 1 | 0 | 0 | 0 | 0 | 0 |

| | | | | | | | | | | | | | | | |
|---|---|---|---|---|---|---|---|---|---|---|---|---|---|---|---|
| -+--+ | 0 | 0 | 0 | 0 | 0 | 0 | 0 | 0 | 0 | 0 | 0 | 0 | 0 | 0 | 0 |
| +--+ | 0 | 0 | 0 | 0 | 0 | 0 | 0 | 0 | 0 | 1 | 0 | 0 | 0 | 0 | 0 |
| ----+ | 0 | 0 | 0 | 0 | 0 | 0 | 0 | 0 | 0 | 0 | 0 | 0 | 0 | 0 | 0 |
| ++++- | 0 | 1 | 0 | 0 | 0 | 0 | 0 | 0 | 1 | 0 | 0 | 0 | 0 | 0 | 0 |
| -+++- | 0 | 0 | 0 | 0 | 0 | 0 | 0 | 1 | 0 | 0 | 0 | 0 | 0 | 1 | 0 |
| +-++- | 0 | 0 | 0 | 0 | 0 | 1 | 0 | 0 | 0 | 0 | 0 | 0 | 0 | 0 | 2 | 1 |
| --++- | 0 | 0 | 0 | 0 | 0 | 0 | 0 | 0 | 0 | 0 | 0 | 0 | 0 | 0 | 0 |
| ++-+- | 0 | 0 | 0 | 0 | 0 | 0 | 0 | 0 | 0 | 1 | 1 | 0 | 0 | 1 | 0 |
| -+-+- | 1 | 0 | 0 | 0 | 0 | 0 | 0 | 0 | 0 | 1 | 0 | 0 | 0 | 0 | 0 |
| +--+- | 0 | 0 | 0 | 0 | 0 | 0 | 0 | 0 | 0 | 0 | 0 | 0 | 0 | 0 | 0 |
| ---+- | 0 | 1 | 0 | 0 | 0 | 0 | 0 | 0 | 0 | 1 | 0 | 0 | 0 | 0 | 0 |
| +++-- | 0 | 0 | 1 | 0 | 0 | 0 | 0 | 0 | 0 | 0 | 0 | 0 | 1 | 0 | 0 |
| -++-- | 0 | 0 | 0 | 0 | 0 | 0 | 0 | 0 | 1 | 0 | 0 | 0 | 0 | 0 | 0 |
| +-+-- | 0 | 0 | 0 | 0 | 0 | 0 | 0 | 0 | 0 | 0 | 0 | 0 | 0 | 0 | 0 |
| --+-- | 0 | 0 | 0 | 0 | 1 | 0 | 0 | 0 | 0 | 0 | 0 | 0 | 0 | 0 | 0 |
| ++--- | 0 | 0 | 0 | 0 | 0 | 1 | 0 | 0 | 0 | 0 | 0 | 0 | 0 | 0 | 0 |
| -+--- | 0 | 1 | 0 | 0 | 0 | 1 | 0 | 0 | 2 | 1 | 0 | 0 | 0 | 0 | 0 |
| +---- | 0 | 0 | 0 | 0 | 0 | 1 | 0 | 0 | 0 | 0 | 0 | 0 | 0 | 0 | 0 |
| ----- | 0 | 0 | 0 | 0 | 0 | 1 | 0 | 0 | 0 | 0 | 0 | 0 | 0 | 0 | 0 |

**TABLE S8.9: The experimentally measured raw data of a complete set of the 1024 possible combinations of ten-photon coincidence events with each photon measured in the basis of $(|H\rangle \pm e^{i7\pi/10}|V\rangle)/\sqrt{2}$ (indicated by the $\pm$ label in the first column and the first line) to obtain $\langle M_{7\pi/10}^{\otimes 10}\rangle$.** The five + or – label in the first column represent the measurement base choice for the five *o* photons, while the five + or – label in the first line represent the measurement base choice for the five *e* photons.

| | +++++ | -++++ | +-+++ | --+++ | ++-++ | -+-++ | +--++ | ---++ | +++-+ | -++-+ | +-+-+ | --+-+ | ++--+ | -+--+ | +---+ | ----+ |
|---|---|---|---|---|---|---|---|---|---|---|---|---|---|---|---|---|
| +++++ | 0 | 0 | 0 | 0 | 0 | 0 | 0 | 0 | 1 | 0 | 0 | 0 | 0 | 0 | 0 | 0 |
| -++++ | 0 | 0 | 1 | 0 | 0 | 1 | 0 | 0 | 0 | 0 | 0 | 0 | 0 | 0 | 0 | 0 |
| +-+++ | 0 | 0 | 0 | 1 | 0 | 0 | 0 | 0 | 0 | 0 | 1 | 1 | 0 | 0 | 0 | 1 |
| --+++ | 1 | 1 | 0 | 0 | 1 | 0 | 0 | 0 | 0 | 0 | 0 | 0 | 0 | 0 | 0 | 0 |
| ++-++ | 0 | 0 | 1 | 0 | 0 | 0 | 1 | 0 | 0 | 0 | 0 | 0 | 0 | 0 | 0 | 0 |
| -+-++ | 0 | 0 | 1 | 0 | 0 | 0 | 0 | 0 | 0 | 0 | 0 | 0 | 0 | 0 | 0 | 0 |
| +--++ | 0 | 0 | 0 | 1 | 1 | 0 | 0 | 0 | 0 | 1 | 0 | 1 | 0 | 0 | 0 | 0 |
| ---++ | 0 | 0 | 0 | 0 | 0 | 0 | 1 | 0 | 0 | 0 | 0 | 1 | 0 | 0 | 0 | 0 |
| +++-+ | 1 | 0 | 0 | 0 | 0 | 0 | 0 | 0 | 0 | 0 | 1 | 1 | 0 | 0 | 0 | 0 |
| -++-+ | 0 | 1 | 0 | 0 | 0 | 0 | 0 | 0 | 0 | 0 | 0 | 0 | 0 | 0 | 0 | 0 |
| +-+-+ | 0 | 0 | 0 | 0 | 0 | 0 | 0 | 0 | 0 | 0 | 0 | 0 | 0 | 0 | 0 | 0 |
| --+-+ | 0 | 0 | 0 | 0 | 0 | 0 | 0 | 0 | 0 | 2 | 0 | 0 | 0 | 0 | 0 | 0 |
| ++--+ | 0 | 0 | 0 | 0 | 0 | 0 | 0 | 0 | 0 | 0 | 0 | 0 | 0 | 0 | 1 | 0 |
| -+--+ | 0 | 0 | 0 | 0 | 0 | 0 | 0 | 0 | 1 | 0 | 1 | 0 | 1 | 0 | 0 | 0 |
| +---+ | 0 | 0 | 0 | 0 | 0 | 0 | 0 | 0 | 0 | 0 | 0 | 0 | 0 | 0 | 0 | 0 |
| ----+ | 0 | 0 | 0 | 0 | 0 | 0 | 0 | 0 | 1 | 0 | 0 | 0 | 0 | 1 | 0 | 0 |
| ++++- | 0 | 0 | 0 | 0 | 0 | 0 | 0 | 0 | 0 | 0 | 0 | 0 | 0 | 0 | 0 | 0 |
| -+++- | 0 | 0 | 0 | 0 | 0 | 0 | 0 | 0 | 1 | 0 | 0 | 0 | 0 | 0 | 0 | 0 |
| +-++- | 0 | 0 | 3 | 0 | 0 | 0 | 0 | 0 | 1 | 1 | 0 | 0 | 0 | 0 | 0 | 0 |
| --++- | 0 | 0 | 0 | 0 | 0 | 0 | 0 | 0 | 0 | 1 | 1 | 0 | 0 | 0 | 0 | 1 |
| ++-+- | 0 | 0 | 0 | 0 | 0 | 0 | 0 | 0 | 0 | 0 | 0 | 0 | 0 | 0 | 0 | 0 |
| -+-+- | 1 | 1 | 0 | 0 | 0 | 0 | 0 | 1 | 0 | 0 | 0 | 1 | 0 | 0 | 0 | 0 |
| +--+- | 1 | 0 | 0 | 0 | 0 | 0 | 0 | 0 | 1 | 0 | 0 | 0 | 0 | 0 | 0 | 0 |
| ---+- | 0 | 0 | 1 | 0 | 0 | 0 | 0 | 0 | 0 | 0 | 0 | 0 | 1 | 0 | 0 | 0 |
| +++-- | 1 | 0 | 1 | 0 | 0 | 0 | 0 | 1 | 0 | 1 | 0 | 1 | 0 | 1 | 0 | 0 |
| -++-- | 0 | 0 | 0 | 0 | 0 | 0 | 0 | 0 | 0 | 2 | 1 | 0 | 1 | 0 | 0 | 0 |
| +-+-- | 0 | 0 | 0 | 1 | 0 | 0 | 0 | 1 | 1 | 0 | 1 | 0 | 0 | 0 | 0 | 1 |
| --+-- | 0 | 0 | 0 | 0 | 0 | 0 | 0 | 0 | 0 | 0 | 0 | 0 | 0 | 0 | 0 | 0 |
| ++--- | 0 | 0 | 0 | 0 | 0 | 1 | 0 | 0 | 0 | 0 | 0 | 0 | 0 | 0 | 0 | 0 |
| -+--- | 0 | 0 | 0 | 0 | 0 | 0 | 0 | 0 | 0 | 0 | 0 | 0 | 0 | 0 | 0 | 0 |
| +---- | 0 | 0 | 0 | 0 | 0 | 0 | 0 | 0 | 0 | 0 | 0 | 0 | 1 | 0 | 0 | 0 |
| ----- | 0 | 0 | 0 | 1 | 0 | 0 | 1 | 0 | 0 | 0 | 0 | 0 | 0 | 0 | 0 | 0 |

| | +++ +- | - +++ - | +- ++- | -- ++- | ++- +- | -+- +- | +-- +- | ---+- | +++ -- | -++- -- | +-+- - | --+-- | ++-- - | -+--- | +---- | ----- |
|---|---|---|---|---|---|---|---|---|---|---|---|---|---|---|---|---|
| +++++ | 0 | 0 | 0 | 0 | 0 | 0 | 0 | 0 | 0 | 0 | 0 | 0 | 0 | 0 | 0 | 0 |
| -++++ | 0 | 0 | 0 | 0 | 0 | 0 | 0 | 1 | 0 | 0 | 0 | 0 | 0 | 0 | 1 | 0 |
| +-+++ | 0 | 0 | 1 | 0 | 0 | 0 | 0 | 0 | 1 | 0 | 0 | 0 | 0 | 0 | 0 | 0 |
| --+++ | 0 | 0 | 0 | 0 | 0 | 0 | 0 | 0 | 0 | 0 | 0 | 0 | 0 | 0 | 0 | 0 |
| ++-++ | 0 | 0 | 0 | 0 | 0 | 0 | 0 | 0 | 0 | 0 | 0 | 0 | 0 | 0 | 0 | 0 |
| -+-++ | 1 | 0 | 0 | 0 | 1 | 0 | 0 | 0 | 0 | 0 | 0 | 0 | 0 | 0 | 0 | 0 |
| +--++ | 0 | 0 | 0 | 0 | 0 | 0 | 0 | 0 | 0 | 0 | 0 | 1 | 0 | 1 | 0 | 0 |
| ---++ | 0 | 2 | 0 | 0 | 0 | 0 | 0 | 0 | 0 | 0 | 0 | 0 | 0 | 0 | 1 | 0 |
| +++-+ | 1 | 0 | 0 | 0 | 0 | 0 | 0 | 0 | 0 | 0 | 0 | 0 | 0 | 0 | 1 | 0 |
| -++-+ | 0 | 0 | 1 | 0 | 0 | 0 | 0 | 0 | 0 | 0 | 1 | 0 | 0 | 0 | 0 | 0 |
| +-+-+ | 0 | 0 | 0 | 2 | 0 | 0 | 0 | 0 | 0 | 0 | 0 | 0 | 1 | 1 | 0 | 0 |
| --+-+ | 0 | 0 | 0 | 0 | 0 | 0 | 0 | 0 | 0 | 0 | 0 | 0 | 1 | 1 | 0 | 0 |
| ++--+ | 0 | 0 | 0 | 0 | 0 | 0 | 0 | 0 | 0 | 0 | 0 | 0 | 0 | 0 | 0 | 0 |
| -+--+ | 0 | 0 | 0 | 0 | 0 | 0 | 0 | 0 | 0 | 0 | 0 | 0 | 0 | 2 | 0 | 1 |
| +---+ | 0 | 0 | 0 | 0 | 0 | 0 | 0 | 0 | 0 | 0 | 0 | 0 | 0 | 0 | 1 | 1 |
| ----+ | 0 | 0 | 0 | 0 | 0 | 0 | 0 | 0 | 0 | 0 | 0 | 0 | 0 | 0 | 0 | 1 |
| ++++- | 0 | 0 | 0 | 0 | 0 | 0 | 0 | 1 | 0 | 0 | 0 | 0 | 0 | 1 | 0 | 0 |
| -+++- | 0 | 0 | 0 | 0 | 0 | 1 | 0 | 0 | 0 | 0 | 0 | 0 | 0 | 0 | 0 | 0 |
| +-++- | 0 | 0 | 1 | 0 | 0 | 0 | 0 | 0 | 1 | 1 | 0 | 0 | 0 | 1 | 0 | 0 |
| --++- | 0 | 0 | 0 | 1 | 0 | 0 | 0 | 0 | 1 | 0 | 0 | 0 | 0 | 0 | 1 | 0 |
| ++-+- | 0 | 0 | 1 | 0 | 0 | 1 | 0 | 0 | 0 | 0 | 1 | 0 | 0 | 0 | 1 | 0 |
| -+-+- | 0 | 0 | 0 | 0 | 0 | 0 | 0 | 0 | 0 | 0 | 0 | 0 | 0 | 0 | 0 | 0 |
| +--+- | 0 | 0 | 0 | 0 | 0 | 0 | 0 | 0 | 1 | 0 | 0 | 0 | 0 | 0 | 0 | 0 |
| ---+- | 0 | 0 | 0 | 0 | 0 | 0 | 0 | 0 | 0 | 0 | 0 | 0 | 0 | 0 | 0 | 1 |
| +++-- | 1 | 0 | 1 | 0 | 0 | 1 | 0 | 0 | 0 | 0 | 0 | 0 | 0 | 0 | 0 | 0 |
| -++-- | 1 | 0 | 0 | 0 | 0 | 0 | 0 | 0 | 1 | 0 | 1 | 0 | 0 | 0 | 1 | 1 |
| +-+-- | 0 | 1 | 0 | 0 | 0 | 0 | 0 | 0 | 0 | 0 | 0 | 0 | 0 | 0 | 0 | 0 |
| --+-- | 0 | 0 | 0 | 0 | 0 | 1 | 0 | 1 | 0 | 0 | 0 | 0 | 0 | 0 | 0 | 0 |
| ++--- | 0 | 0 | 1 | 0 | 0 | 0 | 0 | 0 | 0 | 0 | 0 | 0 | 0 | 0 | 0 | 0 |
| -+--- | 1 | 0 | 0 | 0 | 0 | 0 | 0 | 1 | 0 | 0 | 0 | 0 | 0 | 0 | 0 | 0 |
| +---- | 1 | 0 | 1 | 0 | 0 | 0 | 0 | 0 | 0 | 0 | 1 | 0 | 0 | 0 | 0 | 0 |
| ----- | 0 | 0 | 0 | 0 | 0 | 0 | 0 | 0 | 0 | 0 | 0 | 0 | 0 | 1 | 2 | 0 |

**TABLE S8.10: The experimentally measured raw data of a complete set of the 1024 possible combinations of ten-photon coincidence events with each photon measured in the basis of $(|H\rangle \pm e^{i8\pi/10}|V\rangle)/\sqrt{2}$ (indicated by the $\pm$ label in the first column and the first line) to obtain $\langle M_{8\pi/10}^{\otimes 10}\rangle$.** The five + or – label in the first column represent the measurement base choice for the five $o$ photons, while the five + or – label in the first line represent the measurement base choice for the five $e$ photons.

| | +++ ++ | - +++ + | +- +++ | -- +++ | ++- ++ | -+- ++ | +-- ++ | --- ++ | +++ -+ | -+- -+ | +-+- + | --+- + | ++-- + | -+-- + | +--- + | ----+ |
|---|---|---|---|---|---|---|---|---|---|---|---|---|---|---|---|---|
| +++++ | 0 | 0 | 0 | 1 | 0 | 0 | 0 | 0 | 0 | 0 | 0 | 0 | 0 | 0 | 0 | 0 |
| -++++ | 0 | 0 | 0 | 0 | 0 | 0 | 0 | 2 | 0 | 0 | 0 | 0 | 0 | 0 | 0 | 0 |
| +-+++ | 0 | 0 | 0 | 0 | 2 | 0 | 0 | 0 | 0 | 1 | 0 | 0 | 0 | 1 | 0 | 0 |
| --+++ | 0 | 0 | 0 | 0 | 0 | 1 | 0 | 0 | 0 | 0 | 0 | 0 | 0 | 0 | 0 | 0 |
| ++-++ | 0 | 0 | 0 | 0 | 0 | 0 | 0 | 0 | 0 | 0 | 0 | 0 | 0 | 0 | 0 | 0 |
| -+-++ | 0 | 0 | 0 | 0 | 0 | 0 | 0 | 0 | 0 | 0 | 0 | 0 | 0 | 0 | 0 | 0 |
| +--++ | 0 | 0 | 0 | 0 | 0 | 1 | 0 | 0 | 0 | 0 | 0 | 0 | 0 | 0 | 2 | 0 |
| ---++ | 0 | 0 | 0 | 0 | 1 | 0 | 0 | 1 | 0 | 0 | 0 | 0 | 0 | 0 | 0 | 0 |
| +++-+ | 0 | 0 | 0 | 0 | 0 | 0 | 0 | 1 | 1 | 0 | 0 | 0 | 0 | 0 | 0 | 0 |
| -++-+ | 0 | 0 | 0 | 0 | 0 | 1 | 0 | 0 | 0 | 0 | 0 | 0 | 0 | 1 | 0 | 0 |
| +-+-+ | 0 | 0 | 0 | 0 | 0 | 0 | 2 | 0 | 0 | 0 | 0 | 0 | 0 | 0 | 0 | 0 |
| --+-+ | 0 | 0 | 0 | 0 | 0 | 0 | 0 | 0 | 0 | 1 | 0 | 0 | 0 | 0 | 0 | 0 |
| ++--+ | 0 | 0 | 0 | 0 | 0 | 0 | 0 | 0 | 0 | 0 | 0 | 1 | 0 | 0 | 0 | 0 |
| -+--+ | 0 | 0 | 0 | 0 | 0 | 0 | 0 | 0 | 0 | 0 | 0 | 0 | 0 | 0 | 0 | 0 |
| +---+ | 0 | 0 | 0 | 0 | 0 | 0 | 0 | 0 | 0 | 0 | 0 | 0 | 0 | 0 | 0 | 0 |

| | ++++- | -+++- | +-++- | --++- | ++-+- | -+-+- | +--+- | ---+- | +++-- | -++-- | +-+-- | --+-- | ++--- | -+--- | +---- | ----- |
|---|---|---|---|---|---|---|---|---|---|---|---|---|---|---|---|---|
| ----+ | 0 | 0 | 0 | 0 | 0 | 0 | 0 | 0 | 0 | 0 | 0 | 0 | 0 | 0 | 0 | 0 |
| ++++- | 0 | 1 | 1 | 0 | 0 | 0 | 0 | 0 | 0 | 0 | 1 | 0 | 0 | 0 | 0 | 0 |
| -+++- | 0 | 0 | 0 | 0 | 0 | 0 | 0 | 0 | 0 | 0 | 0 | 0 | 0 | 0 | 0 | 0 |
| +-++- | 0 | 0 | 0 | 0 | 0 | 0 | 0 | 0 | 0 | 0 | 0 | 0 | 0 | 0 | 0 | 0 |
| --++- | 0 | 0 | 0 | 0 | 0 | 0 | 0 | 0 | 0 | 0 | 0 | 0 | 0 | 0 | 0 | 0 |
| ++-+- | 0 | 0 | 0 | 0 | 0 | 0 | 0 | 0 | 0 | 2 | 0 | 0 | 0 | 0 | 0 | 0 |
| -+-+- | 0 | 1 | 0 | 0 | 0 | 0 | 0 | 0 | 0 | 0 | 0 | 0 | 0 | 0 | 0 | 0 |
| +--+- | 0 | 0 | 0 | 0 | 0 | 0 | 0 | 0 | 0 | 0 | 0 | 0 | 0 | 0 | 0 | 0 |
| ---+- | 0 | 0 | 0 | 0 | 0 | 0 | 0 | 1 | 0 | 0 | 1 | 0 | 0 | 0 | 0 | 0 |
| +++-- | 0 | 0 | 0 | 0 | 0 | 0 | 0 | 0 | 0 | 0 | 0 | 0 | 0 | 0 | 0 | 0 |
| -++-- | 0 | 0 | 0 | 0 | 0 | 0 | 0 | 1 | 0 | 0 | 0 | 0 | 0 | 0 | 1 | 0 |
| +-+-- | 0 | 0 | 0 | 0 | 0 | 0 | 0 | 0 | 0 | 1 | 0 | 0 | 1 | 0 | 1 | 0 |
| --+-- | 0 | 0 | 0 | 0 | 0 | 0 | 0 | 0 | 0 | 0 | 0 | 0 | 0 | 0 | 0 | 1 |
| ++--- | 0 | 0 | 0 | 0 | 0 | 0 | 1 | 0 | 1 | 0 | 0 | 0 | 0 | 0 | 0 | 0 |
| -+--- | 0 | 0 | 0 | 0 | 0 | 0 | 0 | 0 | 0 | 0 | 0 | 0 | 0 | 0 | 0 | 0 |
| +---- | 0 | 0 | 0 | 0 | 0 | 1 | 1 | 0 | 0 | 0 | 0 | 0 | 0 | 0 | 0 | 0 |
| ----- | 0 | 0 | 0 | 0 | 0 | 0 | 0 | 0 | 0 | 0 | 0 | 0 | 0 | 0 | 0 | 0 |

| | ++++- | -+++- | +-++- | --++- | ++-+- | -+-+- | +--+- | ---+- | +++-- | -++-- | +-+-- | --+-- | ++--- | -+--- | +---- | ----- |
|---|---|---|---|---|---|---|---|---|---|---|---|---|---|---|---|---|
| +++++ | 0 | 0 | 0 | 0 | 0 | 0 | 0 | 0 | 0 | 0 | 0 | 0 | 0 | 0 | 0 | 0 |
| -++++ | 0 | 0 | 0 | 0 | 0 | 1 | 1 | 0 | 0 | 0 | 0 | 0 | 0 | 0 | 0 | 1 |
| +-+++ | 0 | 0 | 0 | 0 | 0 | 0 | 1 | 0 | 0 | 0 | 0 | 0 | 0 | 0 | 0 | 2 |
| --+++ | 0 | 0 | 0 | 0 | 0 | 0 | 0 | 0 | 0 | 0 | 0 | 0 | 0 | 0 | 1 | 0 |
| ++-++ | 0 | 0 | 0 | 0 | 0 | 1 | 0 | 1 | 0 | 0 | 0 | 0 | 0 | 1 | 0 | 0 |
| -+-++ | 1 | 0 | 0 | 0 | 0 | 0 | 0 | 0 | 0 | 0 | 0 | 0 | 0 | 1 | 0 | 0 |
| +--++ | 0 | 0 | 0 | 0 | 0 | 0 | 0 | 0 | 1 | 0 | 0 | 0 | 0 | 0 | 1 | 0 |
| ---++ | 0 | 0 | 1 | 0 | 0 | 0 | 0 | 0 | 0 | 0 | 0 | 0 | 0 | 0 | 0 | 0 |
| +++-+ | 0 | 0 | 0 | 1 | 0 | 0 | 0 | 0 | 0 | 1 | 0 | 0 | 0 | 0 | 0 | 0 |
| -++-+ | 0 | 0 | 0 | 0 | 0 | 0 | 0 | 0 | 0 | 1 | 0 | 1 | 0 | 0 | 0 | 0 |
| +-+-+ | 0 | 0 | 0 | 0 | 0 | 0 | 0 | 0 | 0 | 0 | 0 | 0 | 0 | 0 | 0 | 0 |
| --+-+ | 0 | 0 | 0 | 0 | 0 | 0 | 0 | 1 | 0 | 0 | 0 | 0 | 0 | 0 | 0 | 0 |
| ++--+ | 0 | 0 | 0 | 1 | 0 | 0 | 0 | 0 | 0 | 0 | 0 | 0 | 0 | 0 | 0 | 0 |
| -+--+ | 0 | 0 | 0 | 0 | 0 | 0 | 0 | 0 | 0 | 0 | 0 | 0 | 0 | 0 | 0 | 0 |
| +---+ | 0 | 0 | 0 | 0 | 0 | 0 | 0 | 0 | 0 | 1 | 0 | 0 | 0 | 0 | 0 | 0 |
| ----+ | 0 | 0 | 0 | 0 | 0 | 0 | 0 | 0 | 0 | 0 | 0 | 0 | 0 | 0 | 0 | 0 |
| ++++- | 1 | 0 | 0 | 0 | 0 | 1 | 0 | 0 | 0 | 0 | 0 | 0 | 0 | 0 | 0 | 1 |
| -+++- | 1 | 1 | 0 | 0 | 0 | 0 | 0 | 0 | 0 | 0 | 0 | 0 | 0 | 0 | 0 | 0 |
| +-++- | 0 | 0 | 0 | 0 | 0 | 0 | 0 | 0 | 0 | 0 | 0 | 1 | 1 | 0 | 0 | 0 |
| --++- | 0 | 0 | 0 | 0 | 0 | 0 | 0 | 0 | 0 | 0 | 0 | 0 | 0 | 0 | 0 | 0 |
| ++-+- | 0 | 0 | 0 | 0 | 0 | 0 | 0 | 1 | 0 | 0 | 0 | 0 | 0 | 0 | 0 | 0 |
| -+-+- | 0 | 0 | 0 | 0 | 0 | 0 | 1 | 0 | 0 | 0 | 0 | 0 | 0 | 0 | 0 | 0 |
| +--+- | 1 | 0 | 0 | 0 | 0 | 0 | 0 | 1 | 0 | 0 | 1 | 0 | 0 | 0 | 0 | 0 |
| ---+- | 0 | 0 | 0 | 0 | 0 | 0 | 1 | 0 | 0 | 0 | 0 | 0 | 0 | 0 | 0 | 0 |
| +++-- | 0 | 1 | 0 | 1 | 0 | 0 | 0 | 0 | 0 | 0 | 0 | 0 | 0 | 2 | 0 | 0 |
| -++-- | 0 | 0 | 0 | 0 | 0 | 0 | 0 | 0 | 0 | 0 | 0 | 0 | 0 | 0 | 1 | 0 |
| +-+-- | 0 | 0 | 0 | 0 | 0 | 0 | 0 | 1 | 0 | 0 | 0 | 1 | 1 | 0 | 0 | 0 |
| --+-- | 0 | 0 | 0 | 0 | 0 | 0 | 0 | 0 | 1 | 0 | 0 | 0 | 1 | 1 | 0 | 0 |
| ++--- | 0 | 0 | 0 | 0 | 0 | 0 | 0 | 0 | 0 | 0 | 0 | 0 | 0 | 0 | 1 | 0 |
| -+--- | 1 | 0 | 0 | 0 | 0 | 0 | 0 | 0 | 0 | 0 | 0 | 0 | 0 | 1 | 0 | 0 |
| +---- | 0 | 1 | 0 | 0 | 1 | 0 | 0 | 0 | 0 | 0 | 0 | 0 | 0 | 0 | 0 | 0 |
| ----- | 1 | 1 | 0 | 0 | 0 | 0 | 1 | 0 | 0 | 0 | 0 | 1 | 0 | 0 | 0 | 1 |

**TABLE S8.11: The experimentally measured raw data of a complete set of the 1024 possible combinations of ten-photon coincidence events with each photon measured in the basis of** $(|H\rangle \pm e^{i9\pi/10}|V\rangle)/\sqrt{2}$ **(indicated by the** $\pm$ **label in the first column and the first line) to obtain** $\left\langle M_{9\pi/10}^{\otimes 10} \right\rangle$**.** The five + or – label in the first column represent the measurement base choice for the five $o$ photons, while the five + or – label in the first line represent the measurement base choice for the five $e$ photons.

| | +++++ | -++++ | +-+++ | --+++ | ++-++ | -+-++ | +--++ | ---++ | +++-+ | -++-+ | +-+-+ | --+-+ | ++--+ | -+--+ | +---+ | ----+ |
|---|---|---|---|---|---|---|---|---|---|---|---|---|---|---|---|---|
| +++++ | 0 | 0 | 0 | 0 | 2 | 1 | 0 | 0 | 0 | 0 | 0 | 0 | 0 | 0 | 0 | 0 |
| -++++ | 0 | 0 | 0 | 0 | 0 | 1 | 0 | 0 | 0 | 0 | 0 | 0 | 0 | 0 | 0 | 0 |
| +-+++ | 0 | 0 | 0 | 0 | 0 | 0 | 0 | 1 | 0 | 0 | 0 | 0 | 0 | 0 | 0 | 0 |
| --+++ | 0 | 1 | 0 | 0 | 1 | 0 | 0 | 1 | 0 | 0 | 0 | 0 | 0 | 0 | 0 | 0 |
| ++-++ | 0 | 0 | 0 | 0 | 0 | 0 | 0 | 0 | 0 | 0 | 0 | 0 | 0 | 0 | 0 | 0 |
| -+-++ | 0 | 0 | 0 | 0 | 1 | 0 | 1 | 0 | 0 | 0 | 1 | 0 | 0 | 0 | 0 | 0 |
| +--++ | 0 | 0 | 0 | 0 | 0 | 0 | 0 | 0 | 0 | 0 | 0 | 0 | 0 | 0 | 0 | 0 |
| ---++ | 0 | 0 | 0 | 2 | 0 | 0 | 0 | 0 | 0 | 0 | 0 | 0 | 0 | 0 | 0 | 0 |
| +++-+ | 0 | 0 | 0 | 0 | 0 | 0 | 0 | 0 | 1 | 0 | 0 | 0 | 0 | 0 | 0 | 0 |
| -++-+ | 0 | 0 | 0 | 0 | 0 | 1 | 0 | 0 | 0 | 0 | 0 | 0 | 0 | 0 | 1 | 0 |
| +-+-+ | 0 | 0 | 0 | 0 | 1 | 0 | 0 | 0 | 0 | 0 | 0 | 1 | 0 | 0 | 0 | 0 |
| --+-+ | 1 | 0 | 0 | 0 | 0 | 1 | 0 | 0 | 0 | 0 | 0 | 0 | 1 | 1 | 0 | 0 |
| ++--+ | 0 | 0 | 0 | 0 | 0 | 0 | 0 | 0 | 0 | 1 | 0 | 0 | 1 | 0 | 0 | 0 |
| -+--+ | 0 | 1 | 0 | 0 | 0 | 0 | 0 | 0 | 0 | 0 | 0 | 0 | 0 | 0 | 0 | 0 |
| +---+ | 0 | 0 | 0 | 0 | 0 | 0 | 0 | 0 | 1 | 0 | 0 | 0 | 1 | 0 | 0 | 1 |
| ----+ | 0 | 0 | 0 | 0 | 0 | 0 | 0 | 0 | 0 | 0 | 0 | 0 | 0 | 0 | 0 | 0 |
| ++++- | 0 | 0 | 0 | 0 | 0 | 0 | 1 | 0 | 0 | 0 | 0 | 0 | 0 | 0 | 0 | 0 |
| -+++- | 0 | 0 | 0 | 0 | 0 | 0 | 0 | 0 | 0 | 0 | 0 | 0 | 0 | 0 | 0 | 0 |
| +-++- | 0 | 0 | 0 | 0 | 0 | 0 | 0 | 1 | 0 | 0 | 1 | 0 | 0 | 1 | 0 | 0 |
| --++- | 1 | 0 | 0 | 0 | 0 | 0 | 0 | 0 | 0 | 0 | 0 | 0 | 0 | 0 | 0 | 1 |
| ++-+- | 1 | 0 | 0 | 0 | 0 | 0 | 0 | 0 | 0 | 0 | 0 | 0 | 0 | 0 | 0 | 0 |
| -+-+- | 0 | 0 | 0 | 0 | 1 | 0 | 0 | 0 | 0 | 0 | 0 | 0 | 0 | 0 | 0 | 0 |
| +--+- | 0 | 0 | 0 | 0 | 0 | 0 | 0 | 0 | 1 | 0 | 0 | 0 | 0 | 0 | 0 | 0 |
| ---+- | 0 | 1 | 0 | 0 | 1 | 0 | 0 | 0 | 0 | 0 | 0 | 0 | 0 | 0 | 0 | 0 |
| +++-- | 0 | 0 | 0 | 0 | 0 | 0 | 0 | 0 | 0 | 1 | 0 | 0 | 0 | 0 | 0 | 1 |
| -++-- | 0 | 0 | 0 | 0 | 0 | 0 | 0 | 0 | 0 | 0 | 0 | 0 | 0 | 1 | 0 | 1 |
| +-+-- | 0 | 0 | 0 | 0 | 0 | 0 | 0 | 0 | 0 | 0 | 0 | 0 | 0 | 0 | 0 | 0 |
| --+-- | 0 | 0 | 0 | 0 | 0 | 1 | 1 | 0 | 0 | 0 | 0 | 0 | 0 | 1 | 0 | 0 |
| ++--- | 0 | 0 | 0 | 0 | 1 | 0 | 0 | 0 | 1 | 0 | 0 | 0 | 0 | 0 | 0 | 0 |
| -+--- | 0 | 0 | 1 | 0 | 0 | 0 | 0 | 0 | 0 | 0 | 1 | 0 | 0 | 0 | 0 | 0 |
| +---- | 0 | 0 | 0 | 0 | 1 | 0 | 0 | 0 | 0 | 0 | 0 | 0 | 0 | 1 | 0 | 1 |
| ----- | 1 | 0 | 0 | 0 | 0 | 0 | 0 | 0 | 0 | 0 | 0 | 0 | 1 | 0 | 0 | 0 |

| | ++++- | -+++- | +-++- | --++- | ++-+- | -+-+- | +--+- | ---+- | +++-- | -++-- | +-+-- | --+-- | ++--- | -+--- | +---- | ----- |
|---|---|---|---|---|---|---|---|---|---|---|---|---|---|---|---|---|
| +++++ | 0 | 0 | 0 | 0 | 0 | 0 | 0 | 0 | 0 | 0 | 0 | 0 | 0 | 0 | 0 | 0 |
| -++++ | 0 | 1 | 0 | 0 | 0 | 0 | 1 | 0 | 0 | 0 | 0 | 0 | 0 | 0 | 0 | 0 |
| +-+++ | 0 | 0 | 0 | 0 | 1 | 0 | 0 | 0 | 0 | 0 | 0 | 0 | 0 | 0 | 0 | 0 |
| --+++ | 0 | 0 | 0 | 0 | 0 | 0 | 0 | 0 | 0 | 0 | 0 | 0 | 0 | 0 | 0 | 0 |
| ++-++ | 0 | 0 | 0 | 0 | 2 | 0 | 0 | 0 | 0 | 0 | 0 | 0 | 0 | 0 | 0 | 0 |
| -+-++ | 0 | 0 | 0 | 0 | 0 | 0 | 0 | 0 | 0 | 0 | 1 | 0 | 0 | 0 | 0 | 0 |
| +--++ | 0 | 0 | 0 | 0 | 1 | 0 | 0 | 0 | 0 | 0 | 0 | 0 | 0 | 0 | 0 | 0 |
| ---++ | 0 | 0 | 0 | 0 | 2 | 0 | 0 | 0 | 1 | 0 | 0 | 0 | 0 | 0 | 0 | 1 |
| +++-+ | 1 | 1 | 0 | 0 | 0 | 0 | 0 | 0 | 0 | 0 | 0 | 0 | 1 | 2 | 0 | 0 |
| -++-+ | 0 | 0 | 0 | 0 | 0 | 0 | 0 | 0 | 0 | 0 | 0 | 0 | 0 | 0 | 0 | 0 |
| +-+-+ | 1 | 0 | 0 | 0 | 0 | 1 | 0 | 0 | 0 | 0 | 0 | 0 | 1 | 0 | 0 | 0 |
| --+-+ | 0 | 0 | 0 | 0 | 1 | 0 | 0 | 0 | 0 | 0 | 0 | 0 | 0 | 0 | 0 | 0 |
| ++--+ | 0 | 0 | 0 | 0 | 0 | 0 | 0 | 0 | 0 | 0 | 0 | 0 | 0 | 0 | 0 | 0 |
| -+--+ | 0 | 1 | 0 | 0 | 0 | 0 | 0 | 0 | 1 | 0 | 0 | 0 | 0 | 1 | 0 | 0 |
| +---+ | 0 | 0 | 0 | 0 | 0 | 0 | 0 | 0 | 0 | 1 | 0 | 0 | 0 | 0 | 0 | 0 |
| ----+ | 0 | 0 | 0 | 0 | 0 | 0 | 0 | 0 | 0 | 0 | 0 | 1 | 0 | 0 | 0 | 0 |
| ++++- | 0 | 0 | 0 | 0 | 0 | 0 | 0 | 0 | 0 | 0 | 0 | 0 | 0 | 0 | 0 | 0 |
| -+++- | 0 | 0 | 0 | 0 | 0 | 0 | 1 | 0 | 0 | 0 | 0 | 0 | 0 | 0 | 0 | 1 |
| +-++- | 0 | 0 | 0 | 1 | 1 | 1 | 0 | 0 | 0 | 0 | 1 | 0 | 0 | 0 | 0 | 0 |
| --++- | 0 | 0 | 0 | 1 | 0 | 0 | 0 | 0 | 0 | 0 | 0 | 0 | 0 | 0 | 0 | 0 |
| ++-+- | 0 | 0 | 0 | 1 | 0 | 0 | 0 | 0 | 0 | 0 | 1 | 0 | 1 | 0 | 0 | 0 |
| -+-+- | 0 | 0 | 0 | 0 | 0 | 0 | 0 | 0 | 1 | 0 | 0 | 0 | 0 | 0 | 0 | 0 |
| +--+- | 0 | 0 | 0 | 0 | 1 | 0 | 0 | 0 | 0 | 0 | 0 | 0 | 0 | 0 | 0 | 1 |
| ---+- | 0 | 0 | 0 | 0 | 0 | 0 | 0 | 0 | 0 | 1 | 0 | 0 | 0 | 0 | 0 | 0 |
| +++-- | 0 | 0 | 1 | 1 | 0 | 0 | 0 | 0 | 0 | 0 | 0 | 0 | 0 | 0 | 0 | 1 |
| -++-- | 0 | 0 | 0 | 0 | 0 | 0 | 0 | 0 | 0 | 1 | 0 | 0 | 0 | 0 | 1 | 0 |
| +-+-- | 1 | 0 | 0 | 0 | 0 | 1 | 0 | 0 | 0 | 0 | 0 | 0 | 0 | 0 | 0 | 0 |
| --+-- | 0 | 0 | 1 | 0 | 0 | 0 | 0 | 0 | 0 | 0 | 1 | 0 | 0 | 0 | 0 | 0 |
| ++--- | 0 | 0 | 0 | 0 | 0 | 0 | 0 | 0 | 1 | 0 | 0 | 0 | 0 | 0 | 1 | 0 |
| -+--- | 0 | 0 | 0 | 0 | 0 | 0 | 0 | 0 | 0 | 0 | 0 | 0 | 0 | 0 | 0 | 0 |

| | | | | | | | | | | | | | | | |
|---|---|---|---|---|---|---|---|---|---|---|---|---|---|---|---|
| +---- | 0 | 0 | 0 | 0 | 0 | 0 | 0 | 0 | 0 | 0 | 0 | 0 | 0 | 0 | 1 |
| ----- | 0 | 1 | 0 | 0 | 0 | 0 | 0 | 0 | 0 | 0 | 0 | 0 | 0 | 0 | 0 |

From table S8.1, we can summary that the counts for the desirred term of $(|H\rangle\langle H|)^{\otimes 10} + (|V\rangle\langle V|)^{\otimes 10}$ is 102 and the sum of the others is 42. Therfore, we can calculate the expectation of population is 0.708(38).

**TABLE S8.12: Summary of experimental results in table S8.2-S8.11 to calculate $\langle M_\theta^{\otimes 10} \rangle$**

| Phase $\theta$ (rad) | $0\pi/10$ | $1\pi/10$ | $2\pi/10$ | $3\pi/10$ | $4\pi/10$ | $5\pi/10$ | $6\pi/10$ | $7\pi/10$ | $8\pi/10$ | $9\pi/10$ |
|---|---|---|---|---|---|---|---|---|---|---|
| Eigen($M_\theta^{\otimes 10}$) = +1 | 73 | 39 | 84 | 29 | 69 | 29 | 69 | 37 | 68 | 33 |
| Eigen($M_\theta^{\otimes 10}$) = -1 | 30 | 98 | 33 | 79 | 23 | 80 | 28 | 90 | 27 | 76 |
| $\langle M_\theta^{\otimes 10} \rangle$ | 0.417(90) | -0.431(77) | 0.436(83) | -0.463(85) | 0.500(90) | -0.468(85) | 0.423(92) | -0.417(81) | 0.432(93) | -0.394(88) |

Then, through $F(\psi^{10}) = (\langle P^{10}\rangle + \langle C^{10}\rangle)/2 = [\langle P^{10}\rangle + (1/10)\sum_{k=0}^{k=9}(-1)^k \langle M_{k\pi/10}^{\otimes 10}\rangle]/2$, we can obtain the fideliy of the ten-photon entanglement is 0.573(23).

## S9. Distillable ten-photon entanglement under higher pump power

As discussed in the main text, by changing the pump laser power, we witness a transition from genuine ten-particle entanglement at a pump power of 0.57 W to distillable entanglement at a pump power of 0.7 W. In this supplementary information, we present the detailed data and analysis of the ten-photon distillable entanglement.

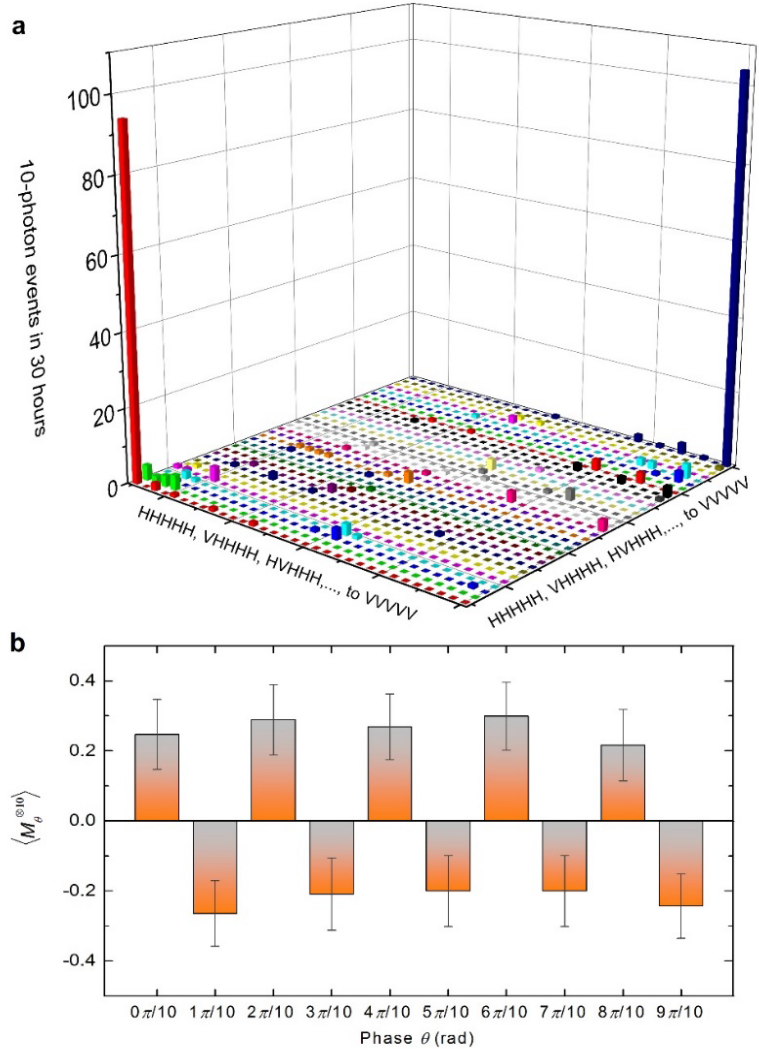

**Figure S1 | Experiment results for ten-photon distillable entanglement under pump power of 0.7 W. a**, Ten-photon coincidence counts in the $|H\rangle/|V\rangle$ basis accumulated for 30 hours. **b**, Expectation values of $M_{k\pi/10}^{\otimes10}$ (see main text for definition) obtained by the measurement in the basis of $(|H\rangle \pm e^{ik\pi/10}|V\rangle)/\sqrt{2}$. Each setting is measured for ~9 hours to accumulate ~100 events. The error bar represents one standard deviation, calculated from Poissonian counting statistics of the raw detection events.

Figure S1 shows the experimental results of the ten-photon state at the higher pump power of 0.7 W. A complete set of 1024 tenfold coincidence events in the $|H\rangle/|V\rangle$ basis are plotted in Fig. S1a. The measured expectation values of $M_{k\pi/10}^{\otimes10} = \cos(k\pi/10)\sigma_x + \sin(k\pi/10)\sigma_y$ ($k = 0,1,...9$) are shown in Fig. S1b, which give an average of $\langle C^{10}\rangle = 0.244(31)$, indicating the coherence of the ten-photon GHZ state. We calculate the state fidelity: $F = 0.429(21)$, which is below the threshold of 0.5. It has been theoretically shown that quantum states with fidelities below 0.5 can still be distillable entangled if they satisfy some specific criteria. We use the distillability criterion [S3] to test the multipartite distillability for the ten-photon state.

Denoting the horizontal ($H$) and vertical ($V$) polarizations of photons as $|0\rangle$ and $|1\rangle$, respectively. An arbitrary state $\rho$ is distillable if

$$\lambda_k < \frac{1}{2}(\langle 0|^{\otimes N}\rho\,|1\rangle^{\otimes N} + \langle 1|^{\otimes N}\rho\,|0\rangle^{\otimes N}) = \frac{1}{2}\langle C^N\rangle,\; for\,\forall\,|k\rangle \neq |0\rangle^{\otimes N} \qquad (1)$$

Where $\lambda_k = \frac{1}{2}(\langle k\,|\rho\,|k\rangle + \langle \bar{k}\,|\rho\,|\bar{k}\rangle)$, $k \in \{k_1 k_2 \ldots k_N | k_i = 0\, or\, 1\}$ is a chain of $N$ bits, and $\bar{k}_i = 1$ (or 0) if $k_i = 0$ (or 1). For the ten-photon state generated at 0.7 W pump power, we determine that $Max\{\lambda_k\} = 0.0093(5)$, $\langle C^{10}\rangle = 0.244(31)$. By using the distillability criterion, we can conclusively establish that it can be distillable with a high statistical significance (7.2 standard deviations)

The same as the calculation of the fidelity for the genuine ten-photon entanglement in S8, we can calculate the fidelity for the distillable ten-photon entanglement with the following meansured raw data as shown in table S9.1-S9.11.

**TABLE S9.1: The experimentally measured raw data of a complete set of the 1024 possible combinations of ten-photon coincidence events with each photon measured in the basis of $|H\rangle/|V\rangle$ (indicated by the H/V label in the first column and the first line) to obtain $\langle P^{10}\rangle$.**

The five H or V label in the first column represent the measurement base choice for the five $o$ photons, while the five H or V label in the first line represent the measurement base choice for the five $e$ photons.

| | HH HH H | VH HH H | HV HH H | VV HH H | HH VH H | VH VH H | HV VH H | VV VH H | HH HV H | VH HV H | HV HV H | VV HV H | HH VV H | VH VV H | HV VV H | VV VV H |
|---|---|---|---|---|---|---|---|---|---|---|---|---|---|---|---|---|
| HH HH H | 94 | 0 | 2 | 0 | 1 | 0 | 0 | 0 | 1 | 0 | 0 | 0 | 1 | 0 | 0 | 0 |
| VH HH H | 4 | 2 | 3 | 4 | 0 | 0 | 0 | 0 | 0 | 0 | 0 | 0 | 0 | 0 | 0 | 0 |
| HV HH H | 0 | 0 | 1 | 0 | 0 | 0 | 0 | 0 | 0 | 0 | 0 | 0 | 0 | 0 | 0 | 0 |
| VV HH H | 0 | 0 | 2 | 1 | 0 | 0 | 0 | 0 | 0 | 0 | 0 | 0 | 0 | 0 | 0 | 0 |
| HH VH H | 1 | 1 | 0 | 0 | 4 | 0 | 0 | 0 | 0 | 0 | 0 | 0 | 0 | 0 | 0 | 0 |
| VH VH H | 0 | 1 | 0 | 1 | 0 | 0 | 0 | 0 | 0 | 0 | 0 | 0 | 0 | 0 | 0 | 0 |
| HV VH H | 0 | 0 | 0 | 0 | 0 | 0 | 0 | 0 | 0 | 0 | 0 | 0 | 0 | 0 | 0 | 0 |

| | | | | | | | | | | | | | | | |
|---|---|---|---|---|---|---|---|---|---|---|---|---|---|---|---|
| VV VH H | 0 | 0 | 0 | 1 | 0 | 0 | 0 | 2 | 0 | 0 | 0 | 1 | 0 | 0 | 0 | 0 |
| HH HV H | 0 | 0 | 0 | 0 | 1 | 0 | 0 | 0 | 0 | 0 | 0 | 0 | 2 | 0 | 0 | 0 |
| VH HV H | 0 | 0 | 0 | 0 | 0 | 0 | 0 | 0 | 0 | 0 | 0 | 0 | 0 | 1 | 0 | 0 |
| HV HV H | 0 | 0 | 0 | 0 | 0 | 0 | 0 | 0 | 0 | 0 | 0 | 0 | 0 | 0 | 1 | 0 |
| VV HV H | 0 | 0 | 0 | 0 | 0 | 0 | 0 | 0 | 0 | 0 | 0 | 0 | 0 | 0 | 0 | 0 |
| HH VV H | 0 | 0 | 0 | 0 | 1 | 0 | 0 | 0 | 0 | 0 | 0 | 0 | 1 | 0 | 0 | 0 |
| VH VV H | 0 | 0 | 0 | 0 | 1 | 1 | 1 | 1 | 0 | 0 | 0 | 0 | 0 | 1 | 0 | 3 |
| HV VV H | 0 | 0 | 0 | 0 | 0 | 0 | 0 | 0 | 0 | 0 | 0 | 0 | 0 | 0 | 0 | 0 |
| VV VV H | 0 | 0 | 0 | 0 | 0 | 0 | 0 | 1 | 0 | 0 | 0 | 0 | 0 | 0 | 0 | 1 |
| HH HH V | 0 | 0 | 0 | 0 | 0 | 0 | 0 | 0 | 1 | 0 | 0 | 0 | 0 | 0 | 0 | 0 |
| VH HH V | 0 | 0 | 0 | 0 | 0 | 0 | 0 | 0 | 1 | 0 | 0 | 0 | 0 | 0 | 0 | 0 |
| HV HH V | 0 | 0 | 0 | 0 | 0 | 0 | 0 | 0 | 0 | 0 | 0 | 0 | 0 | 0 | 0 | 0 |
| VV HH V | 0 | 0 | 0 | 0 | 0 | 0 | 0 | 0 | 0 | 0 | 0 | 0 | 0 | 0 | 0 | 0 |
| HH VH V | 0 | 0 | 0 | 0 | 0 | 0 | 0 | 0 | 0 | 0 | 0 | 0 | 0 | 0 | 0 | 0 |
| VH VH V | 0 | 0 | 0 | 0 | 0 | 0 | 0 | 0 | 0 | 0 | 0 | 0 | 0 | 0 | 0 | 0 |
| HV VH V | 0 | 0 | 0 | 0 | 0 | 0 | 0 | 0 | 0 | 0 | 0 | 0 | 0 | 0 | 0 | 0 |
| VV VH V | 0 | 0 | 0 | 0 | 0 | 0 | 0 | 1 | 0 | 0 | 0 | 0 | 0 | 0 | 0 | 0 |
| HH HV V | 0 | 0 | 0 | 0 | 0 | 0 | 0 | 0 | 1 | 0 | 0 | 0 | 1 | 0 | 0 | 0 |
| VH HV V | 0 | 0 | 0 | 0 | 0 | 0 | 0 | 0 | 0 | 0 | 0 | 0 | 0 | 0 | 0 | 0 |
| HV HV V | 0 | 0 | 0 | 0 | 0 | 0 | 0 | 0 | 0 | 0 | 0 | 0 | 0 | 0 | 0 | 0 |
| VV HV V | 0 | 0 | 0 | 0 | 0 | 0 | 0 | 0 | 0 | 0 | 1 | 0 | 0 | 0 | 0 | 0 |
| HH VV V | 0 | 0 | 0 | 0 | 0 | 0 | 0 | 0 | 0 | 0 | 0 | 0 | 0 | 2 | 0 | 0 |
| VH VV V | 0 | 0 | 0 | 0 | 0 | 0 | 0 | 0 | 0 | 0 | 0 | 0 | 0 | 1 | 0 | 1 |
| HV VV V | 0 | 0 | 0 | 0 | 0 | 0 | 0 | 0 | 0 | 0 | 0 | 0 | 0 | 0 | 0 | 0 |
| VV VV V | 0 | 0 | 0 | 0 | 0 | 0 | 0 | 0 | 0 | 0 | 0 | 0 | 0 | 0 | 0 | 1 |

| | HH HH V | VH HH V | HV HH V | VV HH V | HH VH V | VH VH V | HV VH V | VV VH V | HH HV V | VH HV V | HV HV V | VV HV V | HH VV V | VH VV V | HV VV V | VV VV V |
|---|---|---|---|---|---|---|---|---|---|---|---|---|---|---|---|---|
| HH HH H | 0 | 0 | 0 | 0 | 0 | 0 | 0 | 0 | 0 | 0 | 0 | 0 | 0 | 0 | 0 | 0 |
| VH HH H | 0 | 0 | 0 | 0 | 0 | 0 | 0 | 0 | 0 | 0 | 0 | 0 | 0 | 0 | 0 | 0 |
| HV HH H | 1 | 0 | 3 | 0 | 0 | 0 | 0 | 0 | 0 | 0 | 0 | 0 | 0 | 0 | 1 | 0 |
| VV HH H | 0 | 0 | 3 | 1 | 0 | 0 | 0 | 0 | 0 | 0 | 0 | 0 | 0 | 0 | 0 | 0 |
| HH VH H | 0 | 0 | 0 | 0 | 0 | 0 | 0 | 0 | 0 | 0 | 0 | 0 | 0 | 0 | 0 | 0 |
| VH VH H | 0 | 0 | 0 | 0 | 0 | 0 | 0 | 0 | 0 | 0 | 0 | 0 | 0 | 0 | 0 | 0 |
| HV VH H | 0 | 0 | 0 | 0 | 0 | 0 | 0 | 0 | 0 | 0 | 0 | 0 | 0 | 0 | 0 | 0 |
| VV VH H | 0 | 0 | 0 | 0 | 0 | 0 | 0 | 1 | 0 | 0 | 0 | 0 | 0 | 0 | 0 | 0 |
| HH HV H | 0 | 0 | 0 | 0 | 0 | 0 | 0 | 0 | 0 | 0 | 0 | 0 | 0 | 0 | 0 | 0 |
| VH HV H | 0 | 0 | 0 | 0 | 0 | 0 | 0 | 0 | 0 | 0 | 0 | 0 | 0 | 0 | 0 | 0 |
| HV HV H | 0 | 0 | 0 | 0 | 0 | 0 | 0 | 0 | 0 | 0 | 0 | 0 | 0 | 0 | 0 | 0 |
| VV HV H | 0 | 0 | 0 | 0 | 0 | 0 | 0 | 0 | 0 | 0 | 0 | 0 | 0 | 0 | 0 | 0 |
| HH VV H | 0 | 0 | 0 | 0 | 0 | 0 | 0 | 0 | 0 | 0 | 0 | 0 | 0 | 0 | 0 | 0 |
| VH VV H | 0 | 0 | 0 | 0 | 0 | 0 | 0 | 0 | 0 | 0 | 0 | 0 | 0 | 0 | 0 | 0 |
| HV VV H | 0 | 0 | 0 | 0 | 0 | 0 | 0 | 0 | 0 | 0 | 0 | 0 | 0 | 0 | 0 | 0 |
| VV VV H | 0 | 0 | 0 | 0 | 0 | 0 | 0 | 3 | 0 | 0 | 0 | 0 | 0 | 0 | 0 | 3 |
| HH HH V | 1 | 0 | 0 | 0 | 0 | 0 | 0 | 0 | 0 | 0 | 0 | 0 | 0 | 0 | 0 | 0 |
| VH HH V | 0 | 0 | 0 | 1 | 0 | 0 | 0 | 0 | 0 | 0 | 0 | 0 | 0 | 0 | 0 | 0 |
| HV HH V | 0 | 0 | 1 | 0 | 0 | 0 | 0 | 0 | 1 | 0 | 3 | 0 | 0 | 0 | 0 | 0 |
| VV HH V | 0 | 0 | 3 | 0 | 0 | 0 | 0 | 0 | 0 | 0 | 0 | 0 | 0 | 0 | 0 | 0 |
| HH VH V | 0 | 0 | 0 | 0 | 0 | 0 | 0 | 0 | 0 | 0 | 0 | 0 | 0 | 0 | 0 | 0 |
| VH VH V | 0 | 0 | 0 | 0 | 0 | 1 | 0 | 0 | 0 | 0 | 0 | 0 | 0 | 0 | 0 | 0 |
| HV VH V | 0 | 0 | 0 | 0 | 0 | 0 | 0 | 0 | 0 | 0 | 0 | 0 | 0 | 0 | 0 | 1 |
| VV VH V | 0 | 0 | 0 | 0 | 0 | 0 | 0 | 2 | 0 | 0 | 0 | 2 | 0 | 0 | 0 | 3 |
| HH HV V | 0 | 0 | 0 | 0 | 0 | 0 | 0 | 0 | 3 | 0 | 0 | 0 | 3 | 0 | 0 | 0 |
| VH HV V | 0 | 0 | 0 | 0 | 0 | 0 | 0 | 0 | 0 | 0 | 0 | 0 | 0 | 0 | 0 | 0 |
| HV HV V | 0 | 0 | 0 | 0 | 0 | 0 | 0 | 0 | 0 | 0 | 0 | 0 | 1 | 0 | 3 | 0 |
| VV HV V | 0 | 0 | 0 | 0 | 0 | 0 | 0 | 0 | 0 | 0 | 2 | 2 | 0 | 0 | 4 | 0 |
| HH VV V | 0 | 0 | 0 | 0 | 0 | 0 | 0 | 0 | 0 | 0 | 0 | 0 | 0 | 1 | 0 | 0 |
| VH VV V | 0 | 0 | 0 | 0 | 0 | 0 | 0 | 0 | 0 | 0 | 0 | 0 | 0 | 0 | 0 | 0 |
| HV VV V | 0 | 0 | 0 | 0 | 0 | 0 | 0 | 0 | 0 | 0 | 0 | 0 | 0 | 0 | 0 | 1 |
| VV VV V | 0 | 0 | 0 | 0 | 0 | 0 | 0 | 2 | 0 | 1 | 0 | 3 | 0 | 1 | 0 | 104 |

**TABLE S9.2: The experimentally measured raw data of a complete set of the 1024 possible combinations of ten-photon coincidence events with each photon measured in the basis of** $(|H\rangle \pm e^{i0\pi/10}|V\rangle)/\sqrt{2}$ **(indicated by the** $\pm$ **label in the first column and the first line) to obtain** $\left\langle M_{0\pi/10}^{\otimes 10}\right\rangle$**.** The five + or – label in the first column represent the measurement base choice for the five $o$ photons, while the five + or – label in the first line represent the measurement base choice for the five $e$ photons.

| | +++++ | -++++ | +-+++ | --+++ | ++-++ | -+-++ | +--++ | ---++ | +++-+ | -++-+ | +-+-+ | --+-+ | ++--+ | -+--+ | +---+ | ----+ |
|---|---|---|---|---|---|---|---|---|---|---|---|---|---|---|---|---|
| +++++ | 0 | 0 | 0 | 0 | 0 | 0 | 0 | 0 | 0 | 0 | 0 | 0 | 0 | 0 | 0 | 0 |
| -++++ | 0 | 0 | 0 | 0 | 0 | 0 | 0 | 0 | 0 | 0 | 0 | 0 | 1 | 0 | 0 | 0 |
| +-+++ | 0 | 0 | 0 | 0 | 0 | 0 | 0 | 0 | 0 | 0 | 0 | 0 | 0 | 0 | 0 | 0 |
| --+++ | 0 | 0 | 0 | 0 | 0 | 0 | 0 | 0 | 1 | 1 | 0 | 0 | 0 | 0 | 0 | 0 |
| ++-++ | 0 | 0 | 0 | 0 | 0 | 0 | 1 | 0 | 0 | 1 | 1 | 0 | 0 | 0 | 0 | 0 |
| -+-++ | 0 | 0 | 0 | 0 | 0 | 0 | 0 | 0 | 0 | 0 | 0 | 0 | 0 | 0 | 0 | 0 |
| +--++ | 1 | 0 | 0 | 0 | 0 | 1 | 0 | 0 | 0 | 0 | 0 | 0 | 0 | 1 | 0 | 0 |
| ---++ | 1 | 0 | 0 | 0 | 0 | 0 | 0 | 0 | 0 | 0 | 0 | 0 | 0 | 0 | 0 | 0 |
| +++-+ | 0 | 0 | 0 | 0 | 0 | 0 | 0 | 0 | 0 | 0 | 1 | 0 | 0 | 0 | 0 | 0 |
| -++-+ | 0 | 1 | 0 | 0 | 0 | 0 | 0 | 0 | 0 | 0 | 0 | 1 | 0 | 0 | 0 | 0 |
| +-+-+ | 0 | 0 | 0 | 0 | 0 | 0 | 1 | 0 | 0 | 0 | 0 | 0 | 0 | 0 | 0 | 0 |
| --+-+ | 0 | 0 | 0 | 0 | 0 | 0 | 0 | 1 | 0 | 0 | 0 | 0 | 0 | 0 | 0 | 0 |
| ++--+ | 0 | 0 | 0 | 1 | 0 | 0 | 0 | 0 | 0 | 1 | 0 | 0 | 0 | 0 | 1 | 0 |
| -+--+ | 1 | 0 | 0 | 0 | 0 | 0 | 0 | 0 | 0 | 0 | 0 | 0 | 0 | 0 | 0 | 0 |
| +---+ | 0 | 0 | 0 | 0 | 0 | 0 | 0 | 0 | 0 | 0 | 1 | 0 | 0 | 0 | 0 | 0 |
| ----+ | 0 | 0 | 0 | 0 | 0 | 0 | 0 | 0 | 0 | 0 | 0 | 0 | 0 | 0 | 0 | 0 |
| ++++- | 0 | 0 | 0 | 0 | 0 | 0 | 0 | 0 | 0 | 0 | 0 | 0 | 0 | 0 | 0 | 0 |
| -+++- | 1 | 0 | 0 | 0 | 0 | 1 | 0 | 0 | 0 | 0 | 0 | 0 | 1 | 0 | 0 | 0 |
| +-++- | 0 | 0 | 0 | 0 | 0 | 0 | 0 | 0 | 0 | 0 | 0 | 0 | 0 | 0 | 0 | 1 |
| --++- | 0 | 0 | 0 | 0 | 1 | 0 | 0 | 0 | 0 | 0 | 0 | 0 | 0 | 0 | 0 | 0 |
| ++-+- | 0 | 0 | 0 | 0 | 0 | 0 | 0 | 0 | 0 | 0 | 0 | 0 | 0 | 0 | 0 | 0 |
| -+-+- | 1 | 0 | 0 | 0 | 0 | 0 | 0 | 0 | 1 | 0 | 0 | 0 | 0 | 0 | 0 | 0 |
| +--+- | 0 | 1 | 0 | 0 | 0 | 0 | 0 | 0 | 0 | 0 | 0 | 0 | 0 | 0 | 0 | 0 |
| ---+- | 0 | 0 | 0 | 0 | 0 | 0 | 0 | 0 | 0 | 0 | 0 | 0 | 0 | 0 | 0 | 0 |
| +++-- | 0 | 0 | 0 | 0 | 0 | 0 | 0 | 1 | 0 | 0 | 1 | 0 | 0 | 0 | 1 | 0 |
| -++-- | 0 | 0 | 0 | 0 | 0 | 0 | 0 | 0 | 0 | 0 | 0 | 0 | 0 | 0 | 0 | 0 |
| +-+-- | 0 | 0 | 0 | 0 | 0 | 1 | 0 | 0 | 0 | 0 | 1 | 0 | 0 | 0 | 0 | 0 |
| --+-- | 0 | 1 | 0 | 0 | 0 | 1 | 0 | 0 | 0 | 0 | 0 | 0 | 0 | 0 | 1 | 0 |
| ++--- | 0 | 0 | 0 | 0 | 0 | 0 | 0 | 0 | 0 | 0 | 0 | 0 | 0 | 0 | 0 | 0 |
| -+--- | 0 | 0 | 0 | 1 | 0 | 1 | 0 | 0 | 0 | 0 | 0 | 0 | 0 | 0 | 0 | 0 |
| +---- | 0 | 0 | 0 | 0 | 0 | 0 | 0 | 0 | 0 | 0 | 1 | 0 | 0 | 0 | 1 | 2 |
| ----- | 0 | 0 | 0 | 0 | 0 | 0 | 0 | 0 | 1 | 0 | 0 | 0 | 0 | 0 | 0 | 0 |

| | +++ +- | - +++ - | +- ++- | -- ++- | ++- +- | -+- +- | +-- +- | ---+- | +++ -- | -++- - | +-+- - | --+-- | ++-- - | -+--- | +---- | ----- |
|---|---|---|---|---|---|---|---|---|---|---|---|---|---|---|---|---|
| +++++ | 0 | 0 | 0 | 0 | 0 | 0 | 0 | 0 | 0 | 0 | 0 | 0 | 0 | 0 | 0 | 0 |
| -++++ | 0 | 0 | 0 | 0 | 0 | 0 | 0 | 0 | 0 | 0 | 0 | 0 | 0 | 0 | 1 | 0 |
| +-+++ | 0 | 0 | 0 | 0 | 0 | 0 | 0 | 0 | 0 | 0 | 0 | 0 | 0 | 0 | 0 | 0 |
| --+++ | 0 | 0 | 0 | 0 | 0 | 0 | 0 | 0 | 0 | 0 | 0 | 0 | 0 | 0 | 0 | 0 |
| ++-++ | 0 | 0 | 0 | 0 | 0 | 0 | 0 | 0 | 0 | 0 | 0 | 1 | 0 | 0 | 0 | 0 |
| -+-++ | 0 | 0 | 0 | 0 | 0 | 0 | 0 | 0 | 1 | 0 | 1 | 0 | 0 | 1 | 0 | 0 |
| +--++ | 0 | 1 | 1 | 0 | 0 | 0 | 0 | 0 | 0 | 0 | 0 | 1 | 0 | 0 | 0 | 0 |
| ---++ | 0 | 0 | 0 | 0 | 0 | 1 | 0 | 0 | 0 | 0 | 0 | 0 | 0 | 0 | 0 | 0 |
| +++-+ | 0 | 0 | 0 | 0 | 1 | 0 | 0 | 0 | 0 | 0 | 0 | 0 | 0 | 0 | 2 | 0 |
| -++-+ | 0 | 0 | 0 | 0 | 0 | 0 | 0 | 0 | 1 | 0 | 0 | 2 | 0 | 0 | 0 | 0 |
| +-+-+ | 0 | 0 | 0 | 0 | 0 | 0 | 0 | 0 | 0 | 2 | 0 | 0 | 0 | 0 | 0 | 1 |
| --+-+ | 0 | 0 | 0 | 0 | 0 | 0 | 1 | 0 | 0 | 0 | 0 | 0 | 0 | 0 | 0 | 0 |
| ++--+ | 0 | 0 | 0 | 0 | 1 | 0 | 0 | 0 | 0 | 0 | 0 | 1 | 0 | 0 | 0 | 0 |
| -+--+ | 0 | 0 | 0 | 0 | 0 | 1 | 0 | 0 | 0 | 0 | 0 | 0 | 0 | 1 | 0 | 1 |
| +---+ | 1 | 0 | 0 | 0 | 0 | 0 | 0 | 1 | 0 | 0 | 0 | 0 | 0 | 0 | 0 | 1 |
| ----+ | 0 | 0 | 0 | 0 | 0 | 0 | 0 | 0 | 1 | 0 | 0 | 0 | 1 | 0 | 0 | 0 |
| ++++- | 0 | 0 | 0 | 0 | 0 | 0 | 0 | 0 | 0 | 0 | 0 | 0 | 0 | 0 | 0 | 0 |
| -+++- | 0 | 0 | 0 | 0 | 0 | 0 | 0 | 0 | 0 | 0 | 0 | 0 | 0 | 0 | 0 | 0 |
| +-++- | 0 | 1 | 0 | 0 | 0 | 0 | 0 | 0 | 0 | 0 | 1 | 0 | 0 | 0 | 0 | 0 |
| --++- | 0 | 0 | 0 | 0 | 0 | 1 | 0 | 1 | 0 | 0 | 0 | 0 | 0 | 0 | 0 | 0 |
| ++-+- | 0 | 1 | 0 | 0 | 0 | 0 | 0 | 0 | 0 | 0 | 1 | 1 | 0 | 0 | 0 | 0 |
| -+-+- | 0 | 0 | 0 | 0 | 0 | 0 | 0 | 0 | 0 | 0 | 0 | 0 | 0 | 0 | 1 | 0 |
| +--+- | 0 | 0 | 0 | 0 | 0 | 0 | 0 | 0 | 0 | 0 | 0 | 0 | 0 | 0 | 0 | 0 |
| ---+- | 0 | 0 | 0 | 0 | 1 | 0 | 0 | 0 | 0 | 0 | 0 | 0 | 0 | 0 | 0 | 0 |
| +++-- | 0 | 0 | 0 | 0 | 0 | 0 | 0 | 0 | 0 | 0 | 0 | 0 | 0 | 0 | 1 | 0 |
| -++-- | 0 | 0 | 0 | 1 | 0 | 0 | 0 | 0 | 0 | 1 | 1 | 0 | 0 | 0 | 0 | 0 |
| +-+-- | 0 | 0 | 0 | 0 | 0 | 0 | 0 | 0 | 0 | 0 | 0 | 0 | 0 | 0 | 0 | 0 |
| --+-- | 0 | 0 | 0 | 0 | 0 | 0 | 0 | 0 | 0 | 0 | 0 | 0 | 0 | 0 | 0 | 0 |
| ++--- | 0 | 0 | 0 | 1 | 0 | 0 | 1 | 0 | 0 | 0 | 0 | 0 | 0 | 1 | 0 | 0 |
| -+--- | 0 | 0 | 0 | 0 | 0 | 0 | 0 | 0 | 0 | 0 | 1 | 0 | 0 | 1 | 0 | 0 |
| +---- | 0 | 0 | 0 | 0 | 0 | 0 | 0 | 0 | 0 | 0 | 0 | 0 | 0 | 0 | 0 | 0 |
| ----- | 0 | 1 | 0 | 0 | 0 | 1 | 0 | 0 | 0 | 0 | 0 | 0 | 0 | 0 | 0 | 0 |

**TABLE S9.3: The experimentally measured raw data of a complete set of the 1024 possible combinations of ten-photon coincidence events with each photon measured in the basis of $(|H\rangle \pm e^{i1\pi/10}|V\rangle)/\sqrt{2}$ (indicated by the $\pm$ label in the first column and the first line) to obtain $\langle M_{1\pi/10}^{\otimes 10}\rangle$. The five + or – label in the first column represent the measurement base choice for the five $o$ photons, while the five + or – label in the first line represent the measurement base choice for the five $e$ photons.**

| | +++ ++ | - +++ + | +- +++ | -- +++ | ++- ++ | -+- ++ | +-- ++ | --- ++ | +++ -+ | -++- + | +-+- + | --+-+ | ++-- + | -+-+ + | +--- + | ----+ |
|---|---|---|---|---|---|---|---|---|---|---|---|---|---|---|---|---|
| +++++ | 0 | 0 | 0 | 0 | 0 | 0 | 0 | 0 | 0 | 0 | 1 | 0 | 0 | 0 | 0 | 0 |
| -++++ | 0 | 0 | 0 | 0 | 0 | 0 | 0 | 0 | 0 | 1 | 0 | 0 | 0 | 0 | 0 | 0 |
| +-+++ | 0 | 0 | 0 | 0 | 0 | 0 | 1 | 0 | 0 | 0 | 1 | 0 | 0 | 0 | 0 | 0 |
| --+++ | 1 | 0 | 1 | 0 | 0 | 0 | 0 | 0 | 1 | 0 | 0 | 1 | 0 | 0 | 0 | 0 |
| ++-++ | 0 | 0 | 1 | 0 | 1 | 0 | 0 | 0 | 0 | 0 | 0 | 0 | 0 | 0 | 0 | 0 |
| -+-++ | 0 | 0 | 0 | 0 | 1 | 0 | 0 | 1 | 0 | 0 | 0 | 0 | 0 | 0 | 0 | 1 |
| +--++ | 0 | 0 | 0 | 0 | 0 | 0 | 0 | 0 | 0 | 0 | 0 | 0 | 0 | 0 | 0 | 0 |
| ---++ | 1 | 0 | 0 | 0 | 0 | 0 | 0 | 0 | 0 | 0 | 0 | 0 | 0 | 0 | 0 | 0 |
| +++-+ | 1 | 0 | 0 | 0 | 0 | 1 | 0 | 0 | 0 | 0 | 0 | 0 | 0 | 0 | 0 | 0 |
| -++++ | 0 | 0 | 0 | 0 | 0 | 1 | 0 | 0 | 1 | 0 | 1 | 0 | 0 | 0 | 0 | 0 |
| +-+++ | 0 | 0 | 0 | 0 | 0 | 0 | 0 | 0 | 0 | 0 | 0 | 0 | 0 | 0 | 0 | 1 |
| --+++ | 0 | 0 | 0 | 0 | 0 | 0 | 0 | 0 | 0 | 0 | 0 | 0 | 0 | 0 | 0 | 0 |
| ++-++ | 0 | 1 | 0 | 0 | 0 | 0 | 0 | 0 | 0 | 0 | 0 | 0 | 0 | 0 | 0 | 0 |
| -+-++ | 1 | 0 | 0 | 0 | 0 | 0 | 0 | 0 | 0 | 0 | 0 | 0 | 0 | 0 | 1 | 0 |
| +--++ | 0 | 0 | 0 | 0 | 0 | 0 | 0 | 0 | 1 | 0 | 0 | 0 | 0 | 0 | 1 | 0 |
| ---++ | 0 | 0 | 0 | 0 | 1 | 0 | 1 | 0 | 0 | 0 | 0 | 0 | 0 | 0 | 0 | 1 |
| ++++- | 0 | 0 | 0 | 0 | 1 | 0 | 0 | 0 | 0 | 2 | 0 | 0 | 1 | 1 | 0 | 0 |
| -+++- | 0 | 0 | 0 | 0 | 1 | 0 | 0 | 0 | 0 | 0 | 1 | 0 | 0 | 0 | 0 | 0 |
| +-++- | 0 | 0 | 0 | 1 | 0 | 0 | 0 | 0 | 0 | 0 | 0 | 0 | 0 | 0 | 1 | 0 |
| --++- | 0 | 0 | 0 | 0 | 0 | 0 | 1 | 0 | 0 | 0 | 0 | 0 | 0 | 0 | 0 | 0 |
| ++-+- | 0 | 0 | 0 | 0 | 0 | 0 | 0 | 0 | 0 | 0 | 0 | 0 | 0 | 0 | 1 | 0 |
| -+-+- | 0 | 0 | 0 | 0 | 0 | 0 | 0 | 0 | 0 | 0 | 0 | 0 | 0 | 0 | 0 | 0 |
| +--+- | 1 | 0 | 0 | 0 | 0 | 1 | 0 | 0 | 0 | 0 | 0 | 0 | 0 | 0 | 0 | 0 |
| ---+- | 0 | 0 | 0 | 0 | 0 | 0 | 0 | 0 | 1 | 0 | 0 | 0 | 0 | 0 | 0 | 0 |
| +++-- | 1 | 1 | 0 | 0 | 0 | 0 | 0 | 0 | 0 | 0 | 0 | 0 | 0 | 0 | 0 | 0 |
| -++-- | 1 | 1 | 0 | 0 | 0 | 0 | 0 | 0 | 0 | 0 | 0 | 0 | 0 | 0 | 0 | 0 |
| +-+-- | 1 | 0 | 0 | 0 | 0 | 0 | 0 | 0 | 0 | 0 | 0 | 0 | 0 | 0 | 0 | 0 |
| --+-- | 0 | 1 | 2 | 0 | 0 | 0 | 0 | 0 | 0 | 0 | 0 | 0 | 0 | 0 | 0 | 0 |
| ++--- | 0 | 1 | 0 | 0 | 0 | 0 | 0 | 0 | 1 | 0 | 0 | 0 | 0 | 0 | 0 | 0 |
| -+--- | 0 | 0 | 0 | 0 | 0 | 0 | 0 | 0 | 0 | 0 | 0 | 0 | 0 | 0 | 0 | 0 |
| +---- | 0 | 0 | 0 | 0 | 0 | 0 | 0 | 0 | 1 | 0 | 1 | 0 | 0 | 0 | 0 | 0 |
| ----- | 0 | 0 | 0 | 0 | 0 | 0 | 0 | 0 | 0 | 1 | 0 | 0 | 0 | 0 | 0 | 0 |

| | +++ +- | - +++ - | +- ++- | -- ++- | ++- +- | -+- +- | +-- +- | ---+- | +++ -- | -++ -- | +-+ -- | --+-- | ++- +- | -+- -- | +---- | ----- |
|---|---|---|---|---|---|---|---|---|---|---|---|---|---|---|---|---|
| +++++ | 0 | 0 | 0 | 0 | 0 | 1 | 0 | 0 | 0 | 1 | 0 | 0 | 0 | 0 | 0 | 1 |
| -++++ | 0 | 0 | 0 | 0 | 1 | 0 | 0 | 0 | 0 | 0 | 0 | 0 | 1 | 0 | 0 | 0 |
| +-+++ | 0 | 0 | 0 | 1 | 1 | 0 | 0 | 0 | 0 | 0 | 0 | 1 | 0 | 0 | 0 | 0 |
| --+++ | 0 | 0 | 0 | 0 | 0 | 0 | 0 | 0 | 0 | 0 | 0 | 0 | 1 | 0 | 0 | 0 |
| ++-++ | 0 | 0 | 0 | 0 | 0 | 0 | 0 | 0 | 0 | 0 | 0 | 0 | 0 | 0 | 0 | 0 |
| -+-++ | 0 | 1 | 0 | 0 | 0 | 0 | 0 | 0 | 0 | 0 | 1 | 0 | 0 | 0 | 0 | 0 |
| +--++ | 0 | 1 | 1 | 0 | 1 | 0 | 0 | 1 | 1 | 1 | 0 | 0 | 0 | 0 | 0 | 0 |
| ---++ | 1 | 1 | 0 | 0 | 0 | 0 | 0 | 0 | 0 | 0 | 0 | 0 | 0 | 0 | 0 | 0 |
| +++-+ | 0 | 0 | 1 | 0 | 0 | 0 | 0 | 0 | 0 | 0 | 0 | 0 | 0 | 0 | 0 | 0 |
| -++-+ | 0 | 0 | 0 | 0 | 0 | 0 | 0 | 0 | 0 | 0 | 0 | 0 | 0 | 0 | 0 | 1 |
| +-+-+ | 0 | 1 | 0 | 0 | 0 | 0 | 1 | 0 | 0 | 0 | 0 | 0 | 0 | 0 | 0 | 0 |
| --+-+ | 0 | 0 | 0 | 0 | 0 | 0 | 1 | 0 | 1 | 0 | 0 | 0 | 0 | 0 | 0 | 0 |
| ++--+ | 0 | 0 | 0 | 1 | 0 | 0 | 0 | 0 | 0 | 0 | 0 | 0 | 1 | 0 | 0 | 0 |
| -+--+ | 0 | 0 | 0 | 0 | 0 | 0 | 0 | 0 | 1 | 0 | 0 | 0 | 0 | 0 | 0 | 0 |
| +---+ | 0 | 0 | 0 | 0 | 0 | 0 | 0 | 0 | 0 | 0 | 0 | 0 | 0 | 0 | 0 | 0 |
| ----+ | 0 | 0 | 1 | 0 | 0 | 0 | 1 | 0 | 0 | 0 | 0 | 0 | 0 | 0 | 0 | 0 |
| ++++- | 0 | 0 | 0 | 0 | 0 | 0 | 0 | 0 | 0 | 0 | 0 | 0 | 0 | 0 | 1 | 0 |
| -+++- | 0 | 0 | 0 | 0 | 0 | 0 | 0 | 0 | 0 | 0 | 0 | 0 | 0 | 0 | 0 | 0 |
| +-++- | 0 | 0 | 0 | 0 | 0 | 1 | 1 | 0 | 0 | 0 | 0 | 0 | 0 | 0 | 0 | 0 |
| --++- | 0 | 0 | 1 | 0 | 0 | 0 | 0 | 0 | 0 | 0 | 0 | 0 | 0 | 0 | 0 | 0 |
| ++-+- | 0 | 1 | 0 | 1 | 0 | 0 | 0 | 0 | 0 | 0 | 0 | 0 | 0 | 0 | 0 | 0 |
| -+-+- | 0 | 0 | 0 | 0 | 0 | 0 | 0 | 0 | 0 | 0 | 0 | 0 | 0 | 0 | 0 | 0 |
| +--+- | 0 | 1 | 0 | 0 | 0 | 0 | 0 | 0 | 0 | 0 | 0 | 0 | 0 | 0 | 0 | 0 |
| ---+- | 0 | 0 | 0 | 0 | 0 | 1 | 0 | 0 | 0 | 1 | 0 | 0 | 0 | 1 | 1 | 0 |
| +++-- | 0 | 0 | 0 | 0 | 0 | 0 | 0 | 0 | 0 | 0 | 0 | 0 | 2 | 0 | 0 | 0 |
| -++-- | 0 | 0 | 0 | 0 | 0 | 0 | 0 | 0 | 0 | 0 | 0 | 0 | 0 | 0 | 0 | 0 |
| +-+-- | 0 | 0 | 0 | 0 | 0 | 0 | 0 | 0 | 0 | 0 | 0 | 0 | 0 | 0 | 0 | 0 |
| --+-- | 0 | 0 | 0 | 0 | 0 | 0 | 0 | 0 | 0 | 0 | 0 | 0 | 0 | 0 | 0 | 0 |
| ++--- | 1 | 0 | 0 | 0 | 0 | 0 | 0 | 0 | 0 | 0 | 0 | 0 | 0 | 0 | 1 | 0 |
| -+--- | 0 | 0 | 1 | 0 | 0 | 0 | 0 | 0 | 0 | 0 | 0 | 0 | 0 | 0 | 0 | 0 |
| +---- | 1 | 0 | 0 | 1 | 0 | 0 | 1 | 0 | 0 | 0 | 0 | 0 | 1 | 0 | 0 | 0 |
| ----- | 0 | 0 | 1 | 0 | 0 | 0 | 0 | 0 | 0 | 0 | 0 | 1 | 0 | 0 | 0 | 0 |

**TABLE S9.4: The experimentally measured raw data of a complete set of the 1024 possible combinations of ten-photon coincidence events with each photon measured in the basis of** $(|H\rangle \pm e^{i2\pi/10}|V\rangle)/\sqrt{2}$ **(indicated by the $\pm$ label in the first column and the first line) to**

obtain $\langle M_{2\pi/10}^{\otimes 10} \rangle$. The five + or – label in the first column represent the measurement base choice for the five $o$ photons, while the five + or – label in the first line represent the measurement base choice for the five $e$ photons.

| | +++ ++ | - +++ + | +- +++ | -- +++ | ++- ++ | -+- ++ | +-- ++ | --- ++ | +++ -+ | -++- + | +-+- + | --+- + | ++-- + | -+-- + | +--- + | ----+ |
|---|---|---|---|---|---|---|---|---|---|---|---|---|---|---|---|---|
| +++++ | 0 | 0 | 1 | 0 | 0 | 0 | 1 | 0 | 0 | 0 | 0 | 0 | 0 | 0 | 0 | 0 |
| -++++ | 0 | 0 | 0 | 0 | 0 | 0 | 0 | 0 | 0 | 0 | 0 | 0 | 0 | 0 | 0 | 0 |
| +-+++ | 0 | 0 | 0 | 0 | 0 | 0 | 0 | 0 | 0 | 0 | 1 | 0 | 0 | 0 | 0 | 0 |
| --+++ | 0 | 0 | 0 | 0 | 0 | 0 | 0 | 1 | 0 | 0 | 0 | 1 | 0 | 1 | 0 | 0 |
| ++-++ | 0 | 0 | 0 | 0 | 1 | 0 | 0 | 0 | 0 | 0 | 0 | 0 | 0 | 0 | 0 | 0 |
| -+-++ | 0 | 0 | 0 | 0 | 0 | 0 | 1 | 0 | 0 | 0 | 0 | 0 | 0 | 0 | 0 | 0 |
| +--++ | 0 | 0 | 1 | 0 | 0 | 0 | 0 | 0 | 0 | 0 | 0 | 0 | 0 | 0 | 1 | 0 |
| ---++ | 0 | 0 | 0 | 0 | 0 | 0 | 0 | 0 | 0 | 0 | 0 | 0 | 0 | 0 | 0 | 0 |
| +++-+ | 0 | 0 | 0 | 0 | 0 | 0 | 0 | 0 | 0 | 0 | 0 | 2 | 0 | 1 | 0 | 0 |
| -++-+ | 0 | 0 | 0 | 0 | 0 | 0 | 0 | 0 | 0 | 0 | 1 | 0 | 0 | 0 | 0 | 1 |
| +-+-+ | 0 | 0 | 0 | 0 | 0 | 1 | 0 | 0 | 0 | 1 | 0 | 0 | 0 | 0 | 0 | 0 |
| --+-+ | 0 | 0 | 1 | 0 | 0 | 0 | 0 | 1 | 0 | 0 | 0 | 0 | 0 | 0 | 0 | 0 |
| ++--+ | 0 | 0 | 0 | 0 | 0 | 1 | 0 | 0 | 0 | 0 | 0 | 0 | 0 | 0 | 0 | 0 |
| -+--+ | 0 | 0 | 1 | 1 | 0 | 0 | 0 | 0 | 0 | 0 | 0 | 0 | 1 | 0 | 0 | 0 |
| +---+ | 0 | 0 | 0 | 0 | 0 | 0 | 0 | 0 | 2 | 0 | 0 | 0 | 0 | 0 | 0 | 0 |
| ----+ | 0 | 0 | 0 | 0 | 0 | 0 | 0 | 0 | 0 | 0 | 0 | 0 | 0 | 0 | 0 | 0 |
| ++++- | 0 | 0 | 0 | 0 | 1 | 0 | 0 | 0 | 0 | 0 | 0 | 0 | 0 | 0 | 0 | 0 |
| -+++- | 0 | 0 | 0 | 0 | 0 | 1 | 0 | 0 | 0 | 0 | 0 | 0 | 0 | 0 | 0 | 0 |
| +-++- | 0 | 0 | 0 | 0 | 0 | 0 | 0 | 0 | 0 | 1 | 0 | 0 | 0 | 0 | 0 | 0 |
| --++- | 0 | 0 | 0 | 1 | 1 | 0 | 0 | 0 | 0 | 0 | 0 | 0 | 0 | 0 | 0 | 0 |
| ++-+- | 0 | 0 | 0 | 0 | 0 | 0 | 0 | 0 | 0 | 0 | 0 | 0 | 0 | 0 | 0 | 0 |
| -+-+- | 0 | 0 | 0 | 0 | 0 | 0 | 0 | 0 | 0 | 0 | 0 | 0 | 0 | 0 | 0 | 0 |
| +--+- | 0 | 0 | 0 | 0 | 0 | 1 | 0 | 0 | 0 | 0 | 0 | 0 | 0 | 0 | 0 | 0 |
| ---+- | 0 | 0 | 0 | 0 | 0 | 0 | 0 | 0 | 1 | 0 | 0 | 0 | 0 | 0 | 1 | 0 |
| +++-- | 0 | 0 | 0 | 0 | 0 | 0 | 0 | 0 | 0 | 0 | 0 | 0 | 0 | 0 | 0 | 0 |
| -++-- | 0 | 0 | 0 | 0 | 0 | 0 | 0 | 1 | 0 | 0 | 0 | 0 | 0 | 0 | 1 | 0 |
| +-+-- | 0 | 0 | 0 | 0 | 0 | 0 | 0 | 0 | 1 | 0 | 0 | 0 | 0 | 0 | 0 | 0 |
| --+-- | 0 | 0 | 0 | 0 | 1 | 0 | 0 | 0 | 0 | 0 | 0 | 1 | 0 | 0 | 2 | 1 |
| ++--- | 0 | 0 | 2 | 1 | 0 | 0 | 0 | 0 | 0 | 0 | 0 | 0 | 0 | 0 | 1 | 0 |
| -+--- | 0 | 0 | 0 | 0 | 0 | 1 | 0 | 0 | 0 | 0 | 0 | 0 | 0 | 0 | 0 | 0 |
| +---- | 0 | 0 | 0 | 1 | 0 | 0 | 0 | 1 | 0 | 0 | 0 | 0 | 0 | 0 | 0 | 1 |
| ----- | 0 | 0 | 1 | 0 | 0 | 0 | 0 | 0 | 0 | 0 | 0 | 0 | 0 | 0 | 0 | 0 |

| | +++ +- | - +++ - | +- ++- | -- ++- | ++- +- | -+- +- | +-- +- | ---+- | +++ -- | -++- - | +-+- - | --+- - | ++-- - | -+--- | +---- | ----- |
|---|---|---|---|---|---|---|---|---|---|---|---|---|---|---|---|---|
| +++++ | 0 | 0 | 0 | 0 | 0 | 0 | 0 | 0 | 0 | 0 | 0 | 0 | 0 | 0 | 0 | 0 |
| -++++ | 0 | 0 | 0 | 0 | 0 | 0 | 0 | 0 | 0 | 0 | 1 | 0 | 0 | 0 | 0 | 0 |
| +-+++ | 0 | 0 | 1 | 1 | 0 | 0 | 0 | 0 | 0 | 0 | 0 | 0 | 1 | 0 | 0 | 0 |
| --+++ | 0 | 0 | 0 | 0 | 0 | 0 | 0 | 0 | 0 | 0 | 0 | 0 | 0 | 0 | 0 | 0 |
| ++-++ | 0 | 0 | 0 | 1 | 1 | 0 | 0 | 0 | 0 | 0 | 0 | 1 | 0 | 0 | 0 | 0 |
| -+-++ | 0 | 0 | 0 | 0 | 0 | 1 | 0 | 2 | 0 | 0 | 0 | 0 | 0 | 0 | 1 | 1 |
| +--++ | 0 | 1 | 0 | 0 | 0 | 0 | 0 | 0 | 0 | 0 | 0 | 0 | 0 | 0 | 0 | 0 |
| ---++ | 0 | 1 | 0 | 0 | 0 | 0 | 0 | 0 | 0 | 0 | 0 | 0 | 0 | 0 | 0 | 0 |
| +++-+ | 0 | 0 | 0 | 0 | 0 | 0 | 0 | 0 | 0 | 0 | 0 | 0 | 0 | 0 | 0 | 0 |
| -++-+ | 0 | 0 | 0 | 0 | 0 | 0 | 0 | 0 | 0 | 0 | 0 | 0 | 0 | 0 | 0 | 0 |
| +-+-+ | 0 | 0 | 0 | 0 | 0 | 0 | 0 | 0 | 0 | 0 | 0 | 0 | 0 | 0 | 0 | 0 |
| --+-+ | 0 | 0 | 0 | 0 | 0 | 0 | 0 | 0 | 0 | 0 | 0 | 0 | 0 | 0 | 0 | 1 |
| ++--+ | 0 | 0 | 0 | 0 | 0 | 0 | 0 | 0 | 0 | 0 | 0 | 0 | 0 | 0 | 0 | 0 |
| -+--+ | 0 | 0 | 0 | 0 | 0 | 0 | 1 | 0 | 0 | 0 | 0 | 0 | 0 | 0 | 0 | 0 |
| +---+ | 0 | 0 | 0 | 0 | 0 | 1 | 0 | 0 | 0 | 0 | 0 | 0 | 0 | 0 | 0 | 0 |
| ----+ | 0 | 0 | 0 | 0 | 0 | 0 | 0 | 1 | 0 | 0 | 0 | 0 | 0 | 0 | 0 | 0 |
| ++++- | 0 | 0 | 0 | 0 | 0 | 0 | 1 | 0 | 0 | 0 | 0 | 0 | 0 | 0 | 1 | 0 |
| -+++- | 0 | 0 | 0 | 0 | 0 | 0 | 0 | 0 | 0 | 0 | 0 | 0 | 0 | 0 | 1 | 0 |
| +-++- | 0 | 0 | 0 | 1 | 0 | 0 | 0 | 0 | 0 | 0 | 0 | 0 | 0 | 0 | 0 | 0 |
| --++- | 0 | 0 | 0 | 1 | 0 | 0 | 0 | 0 | 0 | 0 | 1 | 0 | 0 | 0 | 0 | 0 |
| ++-+- | 0 | 0 | 0 | 0 | 0 | 0 | 0 | 0 | 0 | 0 | 0 | 0 | 0 | 0 | 1 | 0 |
| -+-+- | 0 | 0 | 0 | 0 | 0 | 1 | 0 | 0 | 0 | 0 | 1 | 0 | 1 | 0 | 0 | 0 |
| +--+- | 0 | 0 | 0 | 1 | 0 | 0 | 0 | 0 | 0 | 0 | 1 | 2 | 0 | 0 | 0 | 0 |
| ---+- | 0 | 0 | 0 | 0 | 0 | 0 | 0 | 1 | 0 | 0 | 0 | 1 | 0 | 0 | 0 | 0 |
| +++-- | 0 | 0 | 0 | 0 | 0 | 0 | 0 | 0 | 0 | 0 | 0 | 1 | 0 | 0 | 0 | 0 |
| -+++- | 1 | 0 | 1 | 0 | 0 | 0 | 0 | 0 | 0 | 0 | 0 | 0 | 0 | 0 | 0 | 0 |
| +-+-- | 0 | 0 | 0 | 0 | 0 | 0 | 0 | 0 | 0 | 0 | 0 | 0 | 0 | 0 | 0 | 0 |
| --+-- | 0 | 0 | 0 | 0 | 0 | 0 | 0 | 1 | 0 | 0 | 0 | 1 | 0 | 0 | 0 | 0 |
| ++--- | 0 | 0 | 0 | 0 | 0 | 0 | 0 | 0 | 0 | 0 | 0 | 0 | 0 | 0 | 0 | 0 |
| -+--- | 0 | 0 | 0 | 0 | 0 | 0 | 0 | 1 | 0 | 0 | 0 | 0 | 0 | 0 | 0 | 0 |
| +---- | 0 | 0 | 0 | 0 | 0 | 0 | 0 | 0 | 0 | 0 | 0 | 0 | 0 | 0 | 0 | 0 |
| ----- | 0 | 0 | 0 | 0 | 0 | 0 | 0 | 0 | 0 | 0 | 0 | 0 | 0 | 0 | 0 | 0 |

**TABLE S9.5: The experimentally measured raw data of a complete set of the 1024 possible combinations of ten-photon coincidence events with each photon measured in the basis of** $(|H\rangle \pm e^{i3\pi/10}|V\rangle)/\sqrt{2}$ **(indicated by the** $\pm$ **label in the first column and the first line) to obtain** $\langle M_{3\pi/10}^{\otimes 10}\rangle$. The five + or – label in the first column represent the measurement base choice for the five $o$ photons, while the five + or – label in the first line represent the measurement base choice for the five $e$ photons.

|  | +++ ++ | - +++ + | +- +++ | -- +++ | ++- ++ | -+- ++ | +-- ++ | --- ++ | +++ -+ | -++- + | +-+- + | --+- + | ++-- + | -+-- + | +--- + | ---- + |
|---|---|---|---|---|---|---|---|---|---|---|---|---|---|---|---|---|
| +++++ | 0 | 1 | 0 | 0 | 0 | 0 | 0 | 0 | 0 | 0 | 1 | 1 | 1 | 1 | 0 | 0 |
| -++++ | 0 | 0 | 0 | 0 | 0 | 1 | 0 | 0 | 0 | 0 | 0 | 0 | 0 | 0 | 0 | 0 |
| +-+++ | 0 | 0 | 0 | 0 | 0 | 0 | 0 | 0 | 0 | 0 | 0 | 0 | 0 | 0 | 0 | 0 |
| --+++ | 0 | 0 | 0 | 0 | 0 | 0 | 0 | 0 | 0 | 0 | 0 | 2 | 1 | 0 | 0 | 0 |
| ++-++ | 0 | 0 | 0 | 0 | 0 | 0 | 1 | 0 | 0 | 0 | 0 | 0 | 0 | 0 | 0 | 0 |
| -+-++ | 0 | 0 | 0 | 0 | 0 | 0 | 1 | 0 | 0 | 0 | 0 | 0 | 0 | 0 | 0 | 0 |
| +--++ | 0 | 0 | 0 | 0 | 0 | 0 | 0 | 0 | 0 | 0 | 0 | 0 | 0 | 0 | 0 | 0 |
| ---++ | 0 | 0 | 1 | 0 | 0 | 0 | 0 | 0 | 0 | 0 | 0 | 0 | 0 | 0 | 0 | 0 |
| +++-+ | 0 | 0 | 0 | 0 | 0 | 0 | 0 | 0 | 0 | 0 | 0 | 0 | 1 | 0 | 0 | 1 |
| -++-+ | 1 | 0 | 0 | 0 | 0 | 0 | 0 | 0 | 1 | 0 | 0 | 0 | 0 | 0 | 0 | 0 |
| +-+-+ | 0 | 0 | 0 | 0 | 0 | 0 | 0 | 0 | 1 | 0 | 0 | 0 | 0 | 0 | 0 | 0 |
| --+-+ | 0 | 0 | 0 | 0 | 0 | 0 | 0 | 0 | 0 | 0 | 0 | 0 | 0 | 0 | 0 | 0 |
| ++--+ | 0 | 1 | 0 | 0 | 0 | 0 | 0 | 0 | 1 | 0 | 0 | 0 | 0 | 0 | 0 | 0 |
| -+--+ | 0 | 0 | 0 | 0 | 0 | 0 | 0 | 0 | 0 | 0 | 0 | 0 | 0 | 0 | 0 | 0 |
| +--+ | 1 | 0 | 0 | 0 | 1 | 0 | 0 | 0 | 0 | 0 | 0 | 0 | 0 | 0 | 0 | 0 |
| ----+ | 0 | 0 | 0 | 0 | 0 | 0 | 0 | 0 | 0 | 0 | 0 | 0 | 0 | 1 | 0 | 0 |
| ++++- | 0 | 0 | 0 | 0 | 0 | 0 | 1 | 0 | 0 | 0 | 0 | 0 | 0 | 0 | 0 | 1 |
| -+++- | 0 | 1 | 0 | 0 | 0 | 0 | 0 | 0 | 0 | 0 | 0 | 0 | 0 | 0 | 1 | 0 |
| +-++- | 0 | 0 | 0 | 0 | 0 | 0 | 0 | 0 | 0 | 0 | 0 | 0 | 0 | 0 | 0 | 0 |
| --++- | 0 | 0 | 0 | 0 | 0 | 0 | 0 | 0 | 0 | 0 | 0 | 0 | 0 | 0 | 0 | 0 |
| ++-+- | 0 | 0 | 0 | 2 | 0 | 0 | 0 | 1 | 0 | 0 | 0 | 0 | 1 | 0 | 0 | 0 |
| -+-+- | 0 | 1 | 0 | 0 | 0 | 0 | 0 | 0 | 0 | 0 | 0 | 0 | 0 | 0 | 0 | 0 |
| +--+- | 0 | 0 | 0 | 0 | 0 | 0 | 0 | 1 | 0 | 1 | 0 | 0 | 0 | 0 | 0 | 0 |
| ---+- | 0 | 0 | 0 | 0 | 0 | 0 | 0 | 0 | 0 | 0 | 0 | 0 | 0 | 0 | 0 | 0 |
| +++-- | 0 | 0 | 0 | 0 | 0 | 0 | 0 | 0 | 0 | 0 | 0 | 0 | 0 | 0 | 0 | 0 |
| -++-- | 0 | 0 | 0 | 0 | 0 | 0 | 0 | 0 | 0 | 0 | 0 | 0 | 0 | 0 | 0 | 0 |
| +-+-- | 1 | 0 | 0 | 0 | 0 | 0 | 0 | 0 | 0 | 0 | 0 | 0 | 0 | 0 | 0 | 0 |
| --+-- | 0 | 0 | 0 | 0 | 0 | 0 | 0 | 0 | 0 | 0 | 0 | 0 | 0 | 0 | 0 | 0 |
| ++--- | 0 | 0 | 0 | 0 | 0 | 0 | 0 | 0 | 0 | 0 | 0 | 0 | 0 | 0 | 0 | 0 |
| -+--- | 0 | 0 | 0 | 0 | 0 | 1 | 0 | 1 | 0 | 0 | 1 | 0 | 0 | 0 | 0 | 0 |
| +---- | 0 | 0 | 0 | 0 | 1 | 0 | 0 | 1 | 0 | 0 | 0 | 0 | 0 | 0 | 0 | 0 |
| ----- | 0 | 0 | 0 | 0 | 1 | 0 | 0 | 0 | 0 | 0 | 0 | 0 | 0 | 0 | 0 | 0 |

|  | +++ +- | - +++ - | +- ++- | -- ++- | ++- +- | -+- +- | +-- +- | ---+- | +++ -- | -++- - | +-+- - | --+- | ++-- - | -+--- | +---- | ----- |
|---|---|---|---|---|---|---|---|---|---|---|---|---|---|---|---|---|
| +++++ | 0 | 0 | 0 | 0 | 0 | 0 | 0 | 0 | 1 | 0 | 0 | 0 | 0 | 0 | 0 | 0 |
| -++++ | 0 | 0 | 0 | 0 | 0 | 0 | 0 | 0 | 0 | 0 | 0 | 0 | 0 | 0 | 1 | 0 |
| +-+++ | 0 | 0 | 0 | 0 | 0 | 2 | 0 | 0 | 0 | 0 | 2 | 0 | 0 | 0 | 0 | 0 |
| --+++ | 0 | 0 | 0 | 0 | 0 | 3 | 0 | 0 | 1 | 0 | 0 | 0 | 0 | 0 | 0 | 1 |
| ++-++ | 0 | 0 | 0 | 3 | 0 | 0 | 0 | 0 | 0 | 0 | 0 | 0 | 0 | 0 | 0 | 1 |
| -+-++ | 0 | 0 | 0 | 0 | 0 | 0 | 1 | 0 | 1 | 0 | 1 | 0 | 0 | 0 | 0 | 1 |
| +--++ | 0 | 0 | 0 | 0 | 0 | 0 | 1 | 0 | 0 | 0 | 0 | 0 | 0 | 0 | 0 | 0 |

|  | | | | | | | | | | | | | | | |
|---|---|---|---|---|---|---|---|---|---|---|---|---|---|---|---|
| ---++ | 0 | 0 | 0 | 0 | 0 | 0 | 0 | 0 | 0 | 0 | 0 | 0 | 0 | 1 | 0 | 1 |
| +++-+ | 0 | 0 | 0 | 0 | 0 | 0 | 0 | 1 | 0 | 0 | 0 | 0 | 0 | 0 | 0 | 0 |
| -++-+ | 0 | 0 | 0 | 0 | 0 | 1 | 0 | 0 | 0 | 0 | 0 | 0 | 0 | 0 | 0 | 0 |
| +-+-+ | 0 | 0 | 0 | 0 | 0 | 0 | 0 | 0 | 0 | 0 | 0 | 0 | 0 | 0 | 0 | 0 |
| --+-+ | 0 | 0 | 0 | 0 | 0 | 0 | 0 | 1 | 1 | 0 | 0 | 0 | 1 | 0 | 0 | 0 |
| ++--+ | 0 | 0 | 0 | 0 | 0 | 0 | 0 | 0 | 0 | 0 | 0 | 0 | 0 | 0 | 0 | 0 |
| -+--+ | 1 | 0 | 0 | 0 | 0 | 0 | 0 | 0 | 0 | 0 | 0 | 0 | 0 | 0 | 0 | 0 |
| +---+ | 0 | 0 | 0 | 0 | 0 | 1 | 0 | 0 | 0 | 1 | 0 | 0 | 0 | 0 | 0 | 0 |
| ----+ | 0 | 1 | 0 | 0 | 0 | 0 | 0 | 0 | 0 | 0 | 0 | 0 | 0 | 0 | 0 | 0 |
| ++++- | 0 | 0 | 0 | 0 | 0 | 0 | 0 | 0 | 0 | 0 | 0 | 0 | 0 | 0 | 0 | 0 |
| -+++- | 0 | 0 | 0 | 1 | 0 | 0 | 0 | 1 | 0 | 0 | 0 | 0 | 0 | 0 | 0 | 0 |
| +-++- | 0 | 0 | 0 | 0 | 0 | 0 | 0 | 0 | 1 | 0 | 0 | 0 | 0 | 0 | 0 | 0 |
| --++- | 0 | 0 | 1 | 0 | 0 | 0 | 0 | 0 | 0 | 1 | 0 | 0 | 0 | 0 | 1 | 0 |
| ++-+- | 0 | 0 | 0 | 0 | 0 | 0 | 0 | 0 | 0 | 0 | 1 | 0 | 1 | 0 | 0 | 1 |
| -+-+- | 0 | 0 | 0 | 0 | 0 | 0 | 0 | 0 | 0 | 1 | 0 | 0 | 1 | 0 | 0 | 1 |
| +--+- | 0 | 0 | 0 | 0 | 1 | 0 | 0 | 0 | 0 | 0 | 0 | 0 | 0 | 0 | 0 | 0 |
| ---+- | 0 | 0 | 0 | 1 | 0 | 0 | 0 | 0 | 0 | 0 | 0 | 0 | 0 | 0 | 1 | 0 |
| +++-- | 0 | 0 | 0 | 0 | 0 | 0 | 0 | 0 | 0 | 1 | 0 | 0 | 0 | 0 | 0 | 0 |
| -++-- | 0 | 0 | 0 | 0 | 0 | 0 | 0 | 0 | 0 | 0 | 0 | 0 | 0 | 0 | 0 | 0 |
| +-+-- | 0 | 0 | 0 | 0 | 0 | 0 | 0 | 0 | 0 | 1 | 0 | 0 | 0 | 0 | 0 | 0 |
| --+-- | 0 | 0 | 0 | 0 | 0 | 0 | 0 | 0 | 0 | 0 | 1 | 0 | 0 | 0 | 0 | 0 |
| ++--- | 0 | 0 | 0 | 0 | 0 | 0 | 0 | 0 | 0 | 1 | 0 | 0 | 0 | 0 | 0 | 0 |
| -+--- | 0 | 0 | 0 | 1 | 0 | 1 | 0 | 0 | 0 | 0 | 0 | 0 | 0 | 0 | 0 | 0 |
| +---- | 0 | 0 | 0 | 0 | 0 | 0 | 0 | 0 | 0 | 0 | 0 | 0 | 0 | 0 | 0 | 0 |
| ----- | 0 | 0 | 0 | 0 | 0 | 0 | 0 | 0 | 0 | 0 | 0 | 0 | 0 | 0 | 0 | 0 |

**TABLE S9.6: The experimentally measured raw data of a complete set of the 1024 possible combinations of ten-photon coincidence events with each photon measured in the basis of $(|H\rangle \pm e^{i4\pi/10}|V\rangle)/\sqrt{2}$ (indicated by the $\pm$ label in the first column and the first line) to obtain $\langle M_{4\pi/10}^{\otimes 10}\rangle$.** The five + or − label in the first column represent the measurement base choice for the five $o$ photons, while the five + or − label in the first line represent the measurement base choice for the five $e$ photons.

|  | +++ ++ | - +++ + | +- +++ | -- +++ | ++- ++ | -+- ++ | +- ++ | --- ++ | +++ -+ | -++- + | +-+- + | --+- + | ++-- + | -+-- + | +--- + | ----+ |
|---|---|---|---|---|---|---|---|---|---|---|---|---|---|---|---|---|
| +++++ | 0 | 0 | 0 | 0 | 0 | 0 | 0 | 0 | 0 | 0 | 0 | 0 | 0 | 0 | 0 | 0 |
| -++++ | 1 | 0 | 0 | 0 | 0 | 0 | 0 | 0 | 0 | 0 | 0 | 0 | 0 | 0 | 0 | 0 |
| +-+++ | 0 | 0 | 0 | 0 | 0 | 0 | 0 | 0 | 0 | 0 | 0 | 0 | 0 | 1 | 0 | 0 |
| --+++ | 0 | 0 | 0 | 0 | 0 | 0 | 0 | 1 | 0 | 0 | 0 | 0 | 0 | 0 | 0 | 0 |
| ++-++ | 0 | 0 | 0 | 0 | 0 | 0 | 0 | 0 | 0 | 0 | 1 | 0 | 0 | 0 | 0 | 0 |
| -+-++ | 0 | 0 | 0 | 0 | 0 | 0 | 0 | 1 | 0 | 0 | 0 | 0 | 0 | 0 | 0 | 0 |
| +--++ | 0 | 0 | 0 | 0 | 0 | 0 | 0 | 0 | 1 | 0 | 0 | 0 | 0 | 0 | 0 | 0 |
| ---++ | 0 | 0 | 0 | 0 | 0 | 0 | 0 | 0 | 0 | 0 | 0 | 0 | 0 | 0 | 0 | 0 |
| +++-+ | 0 | 0 | 0 | 1 | 0 | 0 | 0 | 0 | 0 | 0 | 0 | 0 | 0 | 0 | 0 | 0 |
| -++-+ | 0 | 0 | 0 | 0 | 0 | 0 | 0 | 0 | 0 | 0 | 0 | 0 | 0 | 0 | 0 | 0 |
| +-+-+ | 0 | 0 | 0 | 0 | 0 | 0 | 0 | 1 | 0 | 2 | 0 | 0 | 0 | 0 | 0 | 0 |
| --+-+ | 1 | 1 | 0 | 0 | 0 | 0 | 0 | 1 | 0 | 1 | 1 | 1 | 0 | 0 | 0 | 0 |
| ++--+ | 0 | 0 | 0 | 0 | 0 | 0 | 0 | 0 | 0 | 0 | 0 | 1 | 0 | 0 | 0 | 0 |
| -+--+ | 1 | 0 | 0 | 0 | 0 | 0 | 0 | 0 | 0 | 0 | 0 | 0 | 0 | 0 | 1 | 0 |
| +---+ | 0 | 0 | 0 | 0 | 0 | 0 | 0 | 1 | 0 | 0 | 1 | 1 | 2 | 0 | 1 | 0 |
| ----+ | 1 | 0 | 0 | 1 | 0 | 0 | 0 | 0 | 0 | 0 | 1 | 0 | 0 | 0 | 0 | 1 |
| ++++- | 0 | 0 | 0 | 0 | 0 | 0 | 0 | 1 | 0 | 0 | 0 | 0 | 0 | 0 | 0 | 0 |
| -+++- | 1 | 0 | 0 | 0 | 0 | 0 | 0 | 0 | 0 | 0 | 0 | 1 | 1 | 0 | 0 | 0 |
| +-++- | 0 | 2 | 0 | 1 | 0 | 0 | 0 | 0 | 0 | 0 | 0 | 0 | 0 | 1 | 0 | 0 |
| --++- | 0 | 0 | 0 | 0 | 0 | 0 | 0 | 0 | 0 | 0 | 0 | 0 | 0 | 0 | 0 | 0 |
| ++-+- | 0 | 0 | 0 | 0 | 0 | 0 | 0 | 0 | 0 | 0 | 0 | 0 | 0 | 0 | 0 | 0 |
| -+-+- | 0 | 0 | 0 | 0 | 0 | 0 | 0 | 2 | 0 | 0 | 0 | 0 | 0 | 0 | 0 | 0 |
| +--+- | 0 | 0 | 0 | 0 | 1 | 1 | 0 | 0 | 0 | 0 | 0 | 0 | 0 | 0 | 0 | 0 |
| ---+- | 0 | 0 | 0 | 0 | 0 | 0 | 0 | 0 | 0 | 0 | 0 | 1 | 2 | 0 | 1 | 0 |
| +++-- | 0 | 0 | 0 | 0 | 0 | 0 | 0 | 0 | 0 | 0 | 0 | 1 | 0 | 0 | 0 | 0 |
| -++-- | 0 | 0 | 0 | 0 | 0 | 0 | 1 | 0 | 0 | 0 | 0 | 0 | 0 | 0 | 0 | 0 |
| +-+-- | 0 | 0 | 0 | 0 | 0 | 0 | 0 | 0 | 0 | 0 | 0 | 0 | 0 | 0 | 0 | 0 |

| | | | | | | | | | | | | | | | |
|---|---|---|---|---|---|---|---|---|---|---|---|---|---|---|---|
| --+-- | 0 | 0 | 0 | 0 | 0 | 0 | 0 | 0 | 0 | 0 | 0 | 0 | 1 | 0 | 0 | 1 |
| ++--- | 0 | 0 | 0 | 0 | 0 | 0 | 0 | 1 | 0 | 0 | 0 | 0 | 0 | 0 | 0 | 0 |
| -+--- | 0 | 0 | 0 | 0 | 0 | 1 | 0 | 0 | 0 | 0 | 0 | 0 | 1 | 0 | 0 | 0 |
| +---- | 0 | 0 | 0 | 0 | 0 | 2 | 0 | 0 | 0 | 0 | 0 | 0 | 0 | 0 | 0 | 0 |
| ----- | 0 | 1 | 1 | 0 | 0 | 0 | 0 | 0 | 0 | 0 | 0 | 0 | 0 | 0 | 0 | 0 |

| | ++++- | -+++- | +-++- | --++- | ++-+- | -+-+- | +--+- | ---+- | +++-- | -++-- | +-+-- | --+-- | ++--- | -+--- | +---- | ----- |
|---|---|---|---|---|---|---|---|---|---|---|---|---|---|---|---|---|
| +++++ | 0 | 0 | 0 | 0 | 0 | 0 | 0 | 0 | 1 | 0 | 0 | 0 | 0 | 0 | 0 | 0 |
| -++++ | 0 | 0 | 0 | 0 | 0 | 0 | 0 | 0 | 0 | 0 | 0 | 0 | 0 | 0 | 0 | 0 |
| +-+++ | 0 | 0 | 0 | 0 | 0 | 0 | 0 | 0 | 0 | 0 | 0 | 0 | 0 | 0 | 0 | 0 |
| --+++ | 0 | 0 | 0 | 0 | 0 | 0 | 0 | 0 | 0 | 0 | 0 | 0 | 0 | 0 | 0 | 0 |
| ++-++ | 0 | 0 | 0 | 0 | 0 | 0 | 0 | 0 | 1 | 0 | 0 | 0 | 0 | 0 | 0 | 0 |
| -+-++ | 0 | 0 | 0 | 0 | 0 | 0 | 0 | 0 | 0 | 1 | 0 | 0 | 0 | 0 | 0 | 0 |
| +--++ | 0 | 0 | 0 | 0 | 0 | 0 | 0 | 1 | 0 | 0 | 0 | 0 | 0 | 0 | 0 | 0 |
| ---++ | 0 | 0 | 0 | 0 | 0 | 0 | 0 | 1 | 0 | 0 | 0 | 0 | 0 | 0 | 1 | 0 |
| +++-+ | 0 | 0 | 0 | 0 | 0 | 0 | 0 | 0 | 0 | 0 | 0 | 0 | 1 | 0 | 0 | 0 |
| -++-+ | 0 | 1 | 0 | 0 | 1 | 0 | 0 | 0 | 0 | 0 | 0 | 0 | 0 | 0 | 0 | 0 |
| +-+-+ | 0 | 0 | 0 | 0 | 0 | 1 | 0 | 0 | 0 | 0 | 0 | 0 | 0 | 0 | 0 | 0 |
| --+-+ | 0 | 0 | 1 | 0 | 0 | 0 | 0 | 0 | 1 | 0 | 0 | 0 | 0 | 0 | 0 | 1 |
| ++--+ | 0 | 0 | 0 | 0 | 0 | 0 | 0 | 0 | 0 | 0 | 0 | 0 | 1 | 0 | 0 | 0 |
| -+--+ | 0 | 0 | 0 | 2 | 0 | 0 | 0 | 1 | 0 | 0 | 0 | 0 | 0 | 0 | 0 | 0 |
| +---+ | 0 | 0 | 0 | 0 | 0 | 0 | 0 | 0 | 0 | 0 | 0 | 0 | 1 | 0 | 1 | 0 |
| ----+ | 0 | 0 | 0 | 1 | 0 | 0 | 0 | 0 | 0 | 0 | 1 | 0 | 0 | 0 | 0 | 0 |
| ++++- | 0 | 0 | 0 | 0 | 0 | 0 | 0 | 0 | 0 | 0 | 0 | 0 | 0 | 0 | 0 | 0 |
| -+++- | 0 | 0 | 0 | 0 | 0 | 0 | 0 | 0 | 0 | 0 | 1 | 0 | 0 | 0 | 0 | 0 |
| +-++- | 0 | 0 | 0 | 0 | 0 | 0 | 0 | 0 | 0 | 0 | 0 | 0 | 0 | 1 | 1 | 0 |
| --++- | 1 | 0 | 0 | 1 | 0 | 0 | 0 | 1 | 0 | 0 | 0 | 0 | 0 | 0 | 0 | 0 |
| ++-+- | 0 | 0 | 0 | 0 | 0 | 0 | 0 | 0 | 0 | 0 | 0 | 1 | 0 | 0 | 0 | 0 |
| -+-+- | 0 | 0 | 0 | 0 | 0 | 0 | 0 | 0 | 2 | 1 | 0 | 0 | 0 | 0 | 0 | 0 |
| +--+- | 0 | 0 | 0 | 1 | 0 | 0 | 1 | 0 | 1 | 1 | 0 | 0 | 0 | 0 | 0 | 0 |
| ---+- | 0 | 0 | 0 | 0 | 0 | 1 | 0 | 0 | 0 | 0 | 0 | 1 | 0 | 0 | 0 | 0 |
| +++-- | 0 | 0 | 0 | 0 | 0 | 0 | 0 | 0 | 0 | 0 | 0 | 0 | 0 | 1 | 0 | 0 |
| -++-- | 0 | 0 | 0 | 0 | 0 | 0 | 0 | 0 | 0 | 0 | 0 | 1 | 0 | 0 | 0 | 0 |
| +-+-- | 0 | 0 | 0 | 0 | 0 | 0 | 0 | 0 | 0 | 0 | 0 | 0 | 1 | 0 | 1 | 0 |
| --+-- | 1 | 0 | 0 | 0 | 0 | 0 | 0 | 0 | 0 | 0 | 0 | 0 | 0 | 0 | 0 | 0 |
| ++--- | 0 | 0 | 0 | 0 | 0 | 0 | 0 | 0 | 0 | 0 | 0 | 0 | 0 | 0 | 0 | 0 |
| -+--- | 0 | 0 | 0 | 0 | 0 | 0 | 0 | 0 | 0 | 0 | 0 | 0 | 0 | 0 | 0 | 0 |
| +---- | 0 | 1 | 1 | 0 | 0 | 0 | 0 | 0 | 0 | 0 | 0 | 0 | 0 | 0 | 1 | 0 |
| ----- | 1 | 0 | 0 | 1 | 0 | 0 | 0 | 0 | 0 | 0 | 0 | 0 | 0 | 0 | 0 | 0 |

**TABLE S9.7: The experimentally measured raw data of a complete set of the 1024 possible combinations of ten-photon coincidence events with each photon measured in the basis of $(|H\rangle \pm e^{i5\pi/10}|V\rangle)/\sqrt{2}$ (indicated by the $\pm$ label in the first column and the first line) to obtain $\langle M_{5\pi/10}^{\otimes 10}\rangle$. The five + or – label in the first column represent the measurement base choice for the five $o$ photons, while the five + or – label in the first line represent the measurement base choice for the five $e$ photons.**

| | +++++ | -++++ | +-+++ | --+++ | ++-++ | -+-++ | +--++ | ---++ | +++-+ | -++-+ | +-+-+ | --+-+ | ++--+ | -+--+ | +---+ | ----+ |
|---|---|---|---|---|---|---|---|---|---|---|---|---|---|---|---|---|
| +++++ | 0 | 0 | 0 | 0 | 0 | 0 | 1 | 0 | 0 | 0 | 0 | 0 | 0 | 0 | 0 | 0 |
| -++++ | 0 | 0 | 0 | 0 | 0 | 1 | 0 | 0 | 0 | 0 | 0 | 0 | 1 | 0 | 1 | 0 |
| +-+++ | 0 | 0 | 0 | 0 | 0 | 0 | 0 | 0 | 0 | 0 | 0 | 0 | 0 | 0 | 0 | 0 |
| --+++ | 0 | 0 | 0 | 0 | 0 | 0 | 0 | 0 | 0 | 0 | 0 | 0 | 0 | 0 | 0 | 0 |
| ++-++ | 0 | 0 | 0 | 0 | 0 | 0 | 0 | 0 | 0 | 0 | 0 | 0 | 1 | 0 | 1 | 0 |
| -+-++ | 0 | 0 | 0 | 0 | 0 | 0 | 0 | 1 | 0 | 0 | 0 | 0 | 0 | 0 | 0 | 0 |
| +--++ | 0 | 0 | 0 | 0 | 0 | 0 | 0 | 0 | 0 | 0 | 0 | 0 | 0 | 0 | 0 | 0 |
| ---++ | 0 | 0 | 0 | 0 | 0 | 1 | 0 | 0 | 0 | 0 | 0 | 0 | 0 | 0 | 0 | 0 |
| +++-+ | 0 | 0 | 0 | 0 | 0 | 0 | 0 | 0 | 0 | 0 | 0 | 0 | 0 | 0 | 0 | 1 |
| -++-+ | 0 | 0 | 0 | 1 | 0 | 0 | 0 | 0 | 0 | 0 | 0 | 0 | 1 | 0 | 0 | 0 |

| | | | | | | | | | | | | | | | |
|---|---|---|---|---|---|---|---|---|---|---|---|---|---|---|---|
| +·+·+ | 0 | 1 | 0 | 1 | 0 | 0 | 0 | 1 | 0 | 0 | 0 | 0 | 0 | 0 | 0 | 0 |
| --·+·+ | 1 | 0 | 0 | 1 | 0 | 0 | 0 | 0 | 0 | 0 | 0 | 0 | 0 | 1 | 0 | 0 |
| ++--+ | 0 | 0 | 0 | 0 | 0 | 0 | 0 | 0 | 0 | 0 | 0 | 0 | 0 | 0 | 1 | 0 |
| -+--+ | 0 | 1 | 0 | 0 | 0 | 0 | 0 | 0 | 0 | 0 | 0 | 0 | 0 | 0 | 0 | 0 |
| +--+ | 1 | 0 | 0 | 0 | 0 | 0 | 0 | 0 | 0 | 0 | 1 | 0 | 0 | 0 | 1 | 1 |
| ----+ | 0 | 0 | 0 | 0 | 0 | 0 | 0 | 0 | 0 | 0 | 0 | 0 | 0 | 0 | 0 | 0 |
| ++++ | 0 | 0 | 0 | 0 | 0 | 0 | 0 | 0 | 0 | 0 | 0 | 0 | 0 | 0 | 0 | 0 |
| -+++- | 0 | 0 | 0 | 0 | 2 | 0 | 0 | 1 | 0 | 1 | 0 | 0 | 0 | 0 | 0 | 0 |
| +·+·+- | 0 | 0 | 0 | 0 | 0 | 0 | 0 | 0 | 0 | 0 | 0 | 0 | 0 | 0 | 0 | 0 |
| --·++ | 0 | 0 | 0 | 0 | 0 | 0 | 0 | 0 | 0 | 0 | 0 | 0 | 0 | 0 | 0 | 0 |
| ++·+· | 0 | 0 | 0 | 0 | 1 | 0 | 0 | 0 | 0 | 0 | 0 | 0 | 0 | 0 | 0 | 0 |
| -+·+· | 0 | 0 | 0 | 1 | 0 | 0 | 0 | 0 | 0 | 0 | 0 | 0 | 0 | 0 | 0 | 1 |
| +--+· | 0 | 0 | 0 | 0 | 0 | 0 | 0 | 0 | 0 | 0 | 0 | 0 | 0 | 0 | 0 | 0 |
| ---·+ | 0 | 0 | 0 | 0 | 0 | 0 | 0 | 0 | 0 | 0 | 0 | 0 | 0 | 0 | 0 | 0 |
| +++·- | 0 | 0 | 0 | 0 | 0 | 0 | 0 | 0 | 0 | 0 | 0 | 0 | 0 | 0 | 0 | 0 |
| -+·+-- | 0 | 0 | 0 | 1 | 0 | 0 | 0 | 0 | 0 | 0 | 0 | 0 | 0 | 0 | 0 | 0 |
| +·+·- | 0 | 0 | 0 | 0 | 0 | 0 | 0 | 0 | 0 | 0 | 1 | 0 | 0 | 0 | 0 | 0 |
| --·+-- | 0 | 0 | 0 | 0 | 0 | 0 | 0 | 1 | 0 | 0 | 0 | 0 | 0 | 1 | 0 | 0 |
| ++--- | 0 | 0 | 0 | 0 | 1 | 0 | 0 | 0 | 0 | 0 | 0 | 0 | 0 | 0 | 0 | 0 |
| -+--- | 0 | 0 | 0 | 0 | 0 | 0 | 0 | 1 | 0 | 0 | 1 | 0 | 0 | 0 | 0 | 0 |
| +---- | 0 | 0 | 0 | 0 | 1 | 0 | 0 | 0 | 0 | 0 | 0 | 0 | 0 | 0 | 0 | 0 |
| ----- | 0 | 0 | 0 | 0 | 0 | 0 | 0 | 0 | 0 | 0 | 0 | 0 | 0 | 0 | 0 | 0 |

| | +++ +- | - +++ - | +- ++- | -- ++- | ++- +- | -+- +- | +-- +- | ---+- | +++ -- | -++ -- | +-+- - | --+-- | ++-- - | -+--- | +---- | ----- |
|---|---|---|---|---|---|---|---|---|---|---|---|---|---|---|---|---|
| +++++ | 0 | 0 | 0 | 1 | 0 | 0 | 0 | 0 | 0 | 0 | 0 | 0 | 1 | 0 | 0 | 0 |
| -++++ | 0 | 0 | 0 | 1 | 0 | 0 | 0 | 0 | 0 | 0 | 0 | 0 | 0 | 0 | 0 | 0 |
| +·-+++ | 0 | 1 | 1 | 0 | 0 | 0 | 0 | 0 | 0 | 0 | 0 | 0 | 0 | 1 | 0 | 0 |
| --+++ | 0 | 0 | 0 | 1 | 0 | 0 | 0 | 0 | 0 | 0 | 0 | 0 | 0 | 0 | 0 | 0 |
| ++·-++ | 0 | 1 | 0 | 1 | 0 | 0 | 0 | 0 | 0 | 1 | 0 | 0 | 0 | 0 | 0 | 2 |
| -+·-++ | 0 | 0 | 0 | 0 | 0 | 0 | 0 | 0 | 0 | 1 | 0 | 0 | 0 | 0 | 0 | 1 |
| +--++ | 0 | 0 | 0 | 0 | 0 | 0 | 0 | 0 | 0 | 1 | 0 | 1 | 0 | 0 | 0 | 0 |
| ---++ | 0 | 1 | 0 | 0 | 0 | 1 | 0 | 0 | 0 | 0 | 0 | 0 | 0 | 0 | 0 | 0 |
| +++-+ | 0 | 0 | 0 | 1 | 0 | 0 | 0 | 0 | 0 | 0 | 1 | 0 | 0 | 0 | 0 | 0 |
| -++-+ | 0 | 1 | 0 | 0 | 0 | 0 | 0 | 0 | 0 | 0 | 0 | 0 | 0 | 0 | 0 | 0 |
| +·-+-+ | 0 | 0 | 0 | 0 | 0 | 0 | 0 | 0 | 0 | 1 | 0 | 0 | 0 | 1 | 0 | 0 |
| --+-+ | 0 | 0 | 0 | 0 | 1 | 1 | 0 | 0 | 0 | 0 | 0 | 0 | 0 | 0 | 0 | 0 |
| ++-·+ | 0 | 1 | 0 | 0 | 0 | 0 | 0 | 1 | 0 | 0 | 0 | 1 | 0 | 0 | 0 | 0 |
| -+--+ | 0 | 0 | 0 | 0 | 0 | 0 | 0 | 0 | 0 | 2 | 0 | 0 | 0 | 0 | 0 | 0 |
| +---+ | 0 | 0 | 0 | 0 | 0 | 1 | 0 | 0 | 0 | 0 | 0 | 0 | 0 | 0 | 0 | 0 |
| ----+ | 0 | 0 | 0 | 0 | 0 | 0 | 0 | 0 | 0 | 0 | 0 | 0 | 1 | 0 | 1 | 0 |
| ++++- | 0 | 0 | 0 | 0 | 0 | 0 | 0 | 0 | 1 | 1 | 0 | 0 | 0 | 0 | 0 | 0 |
| -+++- | 1 | 0 | 0 | 0 | 0 | 0 | 0 | 0 | 0 | 0 | 0 | 0 | 0 | 0 | 0 | 0 |
| +-++- | 0 | 0 | 0 | 0 | 0 | 0 | 0 | 0 | 0 | 0 | 0 | 0 | 0 | 1 | 0 | 0 |
| --++- | 0 | 0 | 1 | 0 | 0 | 1 | 0 | 0 | 0 | 0 | 0 | 0 | 0 | 1 | 0 | 0 |
| ++-+- | 0 | 0 | 0 | 1 | 0 | 1 | 0 | 0 | 0 | 0 | 0 | 0 | 0 | 0 | 0 | 1 |
| -+-+- | 0 | 0 | 0 | 0 | 0 | 0 | 0 | 1 | 0 | 0 | 0 | 0 | 0 | 0 | 0 | 0 |
| +--+- | 0 | 0 | 0 | 0 | 0 | 0 | 0 | 0 | 0 | 0 | 0 | 0 | 0 | 0 | 0 | 0 |
| ---+- | 0 | 0 | 0 | 0 | 0 | 0 | 0 | 0 | 0 | 0 | 0 | 0 | 0 | 0 | 0 | 0 |
| +++-- | 0 | 0 | 0 | 0 | 0 | 0 | 0 | 0 | 0 | 0 | 0 | 0 | 0 | 0 | 0 | 0 |
| -++-- | 0 | 0 | 0 | 0 | 0 | 0 | 0 | 1 | 1 | 0 | 0 | 1 | 0 | 0 | 0 | 0 |
| +-+-- | 0 | 0 | 0 | 0 | 0 | 0 | 0 | 1 | 1 | 0 | 0 | 0 | 0 | 0 | 1 | 0 |
| --+-- | 0 | 0 | 0 | 1 | 0 | 0 | 0 | 0 | 0 | 0 | 0 | 0 | 0 | 0 | 0 | 0 |
| ++--- | 0 | 0 | 0 | 0 | 0 | 0 | 0 | 1 | 1 | 0 | 0 | 0 | 0 | 0 | 0 | 0 |
| -+--- | 0 | 1 | 0 | 0 | 0 | 0 | 0 | 0 | 0 | 1 | 0 | 1 | 1 | 1 | 0 | 0 |
| +---- | 0 | 0 | 0 | 0 | 0 | 0 | 0 | 0 | 0 | 0 | 0 | 1 | 1 | 0 | 0 | 0 |
| ----- | 0 | 0 | 0 | 0 | 0 | 0 | 0 | 0 | 0 | 0 | 0 | 0 | 0 | 0 | 0 | 0 |

**TABLE S9.8: The experimentally measured raw data of a complete set of the 1024 possible combinations of ten-photon coincidence events with each photon measured in the basis of** $(|H\rangle \pm e^{i6\pi/10}|V\rangle)/\sqrt{2}$ **(indicated by the $\pm$ label in the first column and the first line) to obtain** $\langle M_{6\pi/10}^{\otimes 10}\rangle$**. The five + or − label in the first column represent the measurement base**

choice for the five *o* photons, while the five + or – label in the first line represent the measurement base choice for the five *e* photons.

| | +++ ++ | - +++ + | +- +++ | -- +++ | ++- ++ | -+- ++ | +-- ++ | --- ++ | +++ -+ | -++- + | +-+- + | --+- + | ++-- + | -+-- + | +--- + | ----+ |
|---|---|---|---|---|---|---|---|---|---|---|---|---|---|---|---|---|
| +++++ | 0 | 0 | 0 | 0 | 0 | 0 | 0 | 0 | 0 | 0 | 0 | 0 | 0 | 0 | 0 | 0 |
| -++++ | 0 | 0 | 0 | 0 | 0 | 0 | 0 | 1 | 0 | 0 | 0 | 0 | 1 | 0 | 0 | 0 |
| +-+++ | 0 | 0 | 0 | 0 | 0 | 0 | 0 | 0 | 0 | 0 | 0 | 0 | 0 | 0 | 1 | 0 |
| --+++ | 0 | 0 | 0 | 0 | 0 | 0 | 0 | 0 | 0 | 0 | 0 | 0 | 0 | 0 | 0 | 0 |
| ++-++ | 0 | 1 | 0 | 0 | 1 | 0 | 0 | 0 | 0 | 0 | 0 | 0 | 0 | 0 | 0 | 0 |
| -+-++ | 0 | 0 | 0 | 0 | 0 | 0 | 0 | 0 | 0 | 0 | 0 | 0 | 0 | 0 | 0 | 1 |
| +--++ | 0 | 0 | 0 | 0 | 0 | 0 | 0 | 0 | 0 | 0 | 0 | 0 | 0 | 0 | 0 | 0 |
| ---++ | 0 | 0 | 0 | 0 | 0 | 0 | 0 | 0 | 1 | 0 | 0 | 0 | 0 | 0 | 0 | 0 |
| +++-+ | 0 | 0 | 0 | 0 | 1 | 0 | 0 | 0 | 1 | 0 | 0 | 0 | 0 | 0 | 0 | 0 |
| -++-+ | 0 | 0 | 0 | 0 | 0 | 0 | 0 | 0 | 0 | 0 | 0 | 0 | 0 | 2 | 0 | 1 |
| +-+-+ | 0 | 0 | 0 | 0 | 0 | 0 | 1 | 0 | 0 | 0 | 0 | 0 | 1 | 0 | 0 | 0 |
| --+-+ | 0 | 2 | 0 | 0 | 0 | 0 | 0 | 0 | 0 | 0 | 0 | 0 | 0 | 0 | 0 | 0 |
| ++--+ | 0 | 0 | 0 | 0 | 0 | 0 | 0 | 0 | 0 | 0 | 0 | 0 | 0 | 0 | 0 | 0 |
| -+--+ | 0 | 0 | 0 | 0 | 0 | 0 | 0 | 0 | 0 | 0 | 0 | 0 | 1 | 0 | 0 | 0 |
| +---+ | 0 | 1 | 0 | 0 | 0 | 0 | 0 | 0 | 0 | 0 | 0 | 0 | 0 | 0 | 0 | 0 |
| ----+ | 0 | 0 | 0 | 0 | 0 | 0 | 0 | 0 | 0 | 0 | 0 | 0 | 0 | 0 | 0 | 0 |
| ++++- | 0 | 1 | 1 | 0 | 0 | 0 | 1 | 0 | 0 | 0 | 0 | 0 | 2 | 0 | 0 | 0 |
| -+++- | 0 | 1 | 0 | 0 | 0 | 0 | 0 | 0 | 0 | 1 | 1 | 0 | 0 | 0 | 0 | 0 |
| +-++- | 0 | 0 | 0 | 0 | 1 | 0 | 0 | 0 | 0 | 0 | 0 | 0 | 0 | 0 | 0 | 0 |
| --++- | 0 | 0 | 0 | 0 | 0 | 0 | 0 | 0 | 0 | 0 | 0 | 1 | 1 | 0 | 0 | 0 |
| ++-+- | 0 | 1 | 0 | 0 | 0 | 0 | 0 | 0 | 0 | 0 | 0 | 0 | 0 | 0 | 0 | 0 |
| -+-+- | 0 | 0 | 0 | 0 | 0 | 0 | 0 | 0 | 0 | 0 | 0 | 0 | 0 | 1 | 0 | 0 |
| +--+- | 0 | 0 | 1 | 0 | 2 | 0 | 0 | 0 | 0 | 0 | 0 | 0 | 0 | 0 | 0 | 0 |
| ---+- | 0 | 0 | 0 | 0 | 0 | 0 | 0 | 0 | 1 | 0 | 0 | 0 | 0 | 0 | 0 | 0 |
| +++-- | 0 | 0 | 0 | 0 | 0 | 0 | 0 | 1 | 0 | 0 | 0 | 0 | 0 | 0 | 0 | 0 |
| -++-- | 0 | 0 | 0 | 0 | 0 | 0 | 0 | 0 | 0 | 0 | 1 | 0 | 0 | 0 | 0 | 0 |
| +-+-- | 0 | 0 | 0 | 1 | 0 | 0 | 0 | 0 | 1 | 0 | 0 | 0 | 0 | 0 | 0 | 0 |
| --+-- | 1 | 0 | 0 | 0 | 0 | 0 | 0 | 0 | 0 | 0 | 0 | 0 | 0 | 0 | 0 | 1 |
| ++--- | 0 | 0 | 1 | 0 | 1 | 0 | 0 | 1 | 0 | 0 | 0 | 0 | 0 | 1 | 1 | 0 |
| -+--- | 0 | 0 | 0 | 0 | 0 | 0 | 0 | 0 | 0 | 0 | 0 | 0 | 0 | 0 | 0 | 0 |
| +---- | 0 | 0 | 1 | 0 | 0 | 1 | 0 | 0 | 0 | 0 | 0 | 0 | 0 | 0 | 0 | 0 |
| ----- | 0 | 0 | 0 | 1 | 0 | 0 | 0 | 0 | 0 | 0 | 0 | 0 | 0 | 0 | 0 | 0 |

| | +++ +- | - +++ - | +- ++- | -- ++- | ++- +- | -+- +- | +-- +- | ---+- | +++ -- | -++- - | +-+- - | --+-- | ++-- - | -+--- | +---- | ----- |
|---|---|---|---|---|---|---|---|---|---|---|---|---|---|---|---|---|
| +++++ | 0 | 0 | 1 | 0 | 0 | 0 | 0 | 0 | 0 | 0 | 0 | 0 | 0 | 1 | 0 | 0 |
| -++++ | 0 | 0 | 0 | 0 | 0 | 0 | 1 | 1 | 0 | 0 | 0 | 0 | 0 | 0 | 0 | 0 |
| +-+++ | 0 | 0 | 0 | 0 | 0 | 0 | 0 | 0 | 0 | 0 | 0 | 1 | 0 | 0 | 0 | 0 |
| --+++ | 0 | 0 | 0 | 0 | 0 | 0 | 0 | 0 | 0 | 0 | 0 | 0 | 0 | 0 | 0 | 0 |
| ++-++ | 0 | 0 | 0 | 0 | 0 | 0 | 0 | 0 | 0 | 0 | 0 | 0 | 0 | 0 | 0 | 0 |
| -+-++ | 0 | 0 | 0 | 0 | 0 | 0 | 0 | 0 | 0 | 1 | 0 | 0 | 1 | 0 | 1 | 0 |
| +--++ | 0 | 0 | 0 | 0 | 0 | 0 | 1 | 0 | 0 | 0 | 0 | 0 | 0 | 0 | 0 | 0 |
| ---++ | 0 | 0 | 0 | 1 | 1 | 0 | 0 | 0 | 0 | 0 | 0 | 0 | 0 | 0 | 0 | 0 |
| +++-+ | 0 | 0 | 1 | 1 | 0 | 0 | 0 | 0 | 0 | 0 | 0 | 0 | 0 | 0 | 0 | 0 |
| -++-+ | 0 | 0 | 0 | 0 | 0 | 0 | 0 | 0 | 0 | 0 | 0 | 0 | 0 | 0 | 0 | 0 |
| +-+-+ | 0 | 0 | 0 | 0 | 0 | 0 | 0 | 0 | 1 | 0 | 0 | 0 | 0 | 0 | 0 | 0 |
| --+-+ | 0 | 0 | 0 | 0 | 0 | 0 | 0 | 0 | 0 | 1 | 0 | 0 | 0 | 0 | 0 | 0 |
| ++--+ | 0 | 0 | 0 | 0 | 0 | 0 | 0 | 1 | 0 | 0 | 0 | 1 | 0 | 0 | 0 | 0 |
| -+--+ | 0 | 0 | 1 | 0 | 0 | 0 | 0 | 1 | 0 | 0 | 0 | 0 | 0 | 0 | 0 | 1 |
| +---+ | 0 | 0 | 0 | 0 | 0 | 0 | 1 | 0 | 0 | 0 | 0 | 0 | 0 | 0 | 0 | 0 |
| ----+ | 0 | 0 | 1 | 0 | 0 | 1 | 0 | 1 | 0 | 0 | 0 | 0 | 0 | 0 | 1 | 1 |
| ++++- | 0 | 0 | 0 | 0 | 0 | 0 | 0 | 0 | 1 | 0 | 0 | 0 | 0 | 0 | 1 | 0 |
| -+++- | 0 | 0 | 1 | 0 | 0 | 0 | 0 | 0 | 0 | 0 | 0 | 0 | 0 | 0 | 0 | 0 |
| +-++- | 0 | 0 | 1 | 0 | 0 | 0 | 0 | 0 | 0 | 0 | 0 | 0 | 1 | 1 | 0 | 0 |
| --++- | 0 | 1 | 0 | 0 | 0 | 0 | 0 | 1 | 0 | 0 | 0 | 0 | 0 | 0 | 1 | 0 |
| ++-+- | 0 | 0 | 1 | 0 | 0 | 0 | 0 | 0 | 0 | 0 | 0 | 0 | 0 | 0 | 0 | 0 |
| -+-+- | 0 | 0 | 0 | 0 | 0 | 0 | 0 | 0 | 0 | 0 | 0 | 0 | 0 | 0 | 0 | 0 |
| +--+- | 0 | 0 | 0 | 0 | 0 | 0 | 0 | 0 | 0 | 0 | 0 | 0 | 0 | 0 | 0 | 0 |
| ---+- | 0 | 0 | 0 | 0 | 0 | 0 | 0 | 1 | 0 | 1 | 0 | 0 | 0 | 0 | 0 | 0 |
| +++-- | 0 | 0 | 0 | 0 | 0 | 0 | 0 | 0 | 0 | 1 | 0 | 0 | 0 | 0 | 0 | 0 |
| -++-- | 0 | 0 | 0 | 0 | 1 | 0 | 1 | 0 | 0 | 0 | 0 | 0 | 0 | 0 | 0 | 0 |

| | | | | | | | | | | | | | | | |
|---|---|---|---|---|---|---|---|---|---|---|---|---|---|---|---|
| +-+-+-- | 0 | 0 | 0 | 0 | 0 | 0 | 0 | 0 | 0 | 0 | 1 | 0 | 0 | 0 | 0 | 0 |
| --+--- | 0 | 0 | 0 | 0 | 0 | 0 | 0 | 0 | 1 | 0 | 0 | 1 | 0 | 0 | 0 | 0 |
| ++--- | 0 | 0 | 0 | 1 | 0 | 0 | 1 | 0 | 0 | 0 | 0 | 0 | 0 | 0 | 0 | 0 |
| -+--- | 0 | 0 | 0 | 0 | 0 | 0 | 0 | 1 | 0 | 0 | 0 | 0 | 0 | 0 | 0 | 0 |
| +---- | 0 | 0 | 0 | 0 | 0 | 0 | 0 | 0 | 1 | 0 | 0 | 0 | 0 | 0 | 0 | 0 |
| ----- | 0 | 0 | 0 | 0 | 0 | 0 | 0 | 0 | 0 | 0 | 0 | 0 | 0 | 0 | 0 | 0 |

**TABLE S9.9: The experimentally measured raw data of a complete set of the 1024 possible combinations of ten-photon coincidence events with each photon measured in the basis of** $(|H\rangle \pm e^{i7\pi/10}|V\rangle)/\sqrt{2}$ **(indicated by the $\pm$ label in the first column and the first line) to obtain** $\langle M_{7\pi/10}^{\otimes 10}\rangle$. The five + or – label in the first column represent the measurement base choice for the five $o$ photons, while the five + or – label in the first line represent the measurement base choice for the five $e$ photons.

| | +++++ | -++++ | +-+++ | --+++ | ++-++ | -+-++ | +--++ | ---++ | +++-+ | -++-+ | +-+-+ | --+-+ | ++--+ | -+--+ | +---+ | ----+ |
|---|---|---|---|---|---|---|---|---|---|---|---|---|---|---|---|---|
| +++++ | 0 | 0 | 0 | 0 | 0 | 0 | 0 | 0 | 0 | 0 | 0 | 0 | 0 | 0 | 0 | 0 |
| -++++ | 1 | 0 | 1 | 0 | 0 | 0 | 0 | 0 | 0 | 0 | 0 | 0 | 1 | 1 | 0 | 0 |
| +-+++ | 1 | 0 | 0 | 0 | 0 | 0 | 0 | 0 | 0 | 0 | 0 | 0 | 0 | 0 | 0 | 0 |
| --+++ | 0 | 0 | 0 | 0 | 0 | 0 | 0 | 0 | 0 | 0 | 0 | 0 | 0 | 0 | 0 | 0 |
| ++-++ | 0 | 0 | 0 | 1 | 0 | 0 | 0 | 0 | 0 | 0 | 0 | 0 | 0 | 0 | 0 | 0 |
| -+-++ | 0 | 0 | 0 | 0 | 0 | 1 | 0 | 1 | 1 | 0 | 0 | 0 | 0 | 0 | 0 | 0 |
| +--++ | 0 | 0 | 0 | 0 | 0 | 0 | 0 | 0 | 0 | 0 | 0 | 0 | 0 | 0 | 0 | 0 |
| ---++ | 0 | 0 | 0 | 0 | 0 | 0 | 0 | 0 | 0 | 0 | 0 | 0 | 0 | 0 | 0 | 0 |
| +++-+ | 0 | 0 | 0 | 0 | 0 | 0 | 0 | 0 | 0 | 0 | 0 | 0 | 0 | 1 | 0 | 0 |
| -++-+ | 0 | 0 | 0 | 0 | 0 | 1 | 0 | 1 | 0 | 0 | 0 | 0 | 0 | 0 | 0 | 0 |
| +-+-+ | 0 | 0 | 0 | 0 | 0 | 0 | 0 | 0 | 0 | 0 | 0 | 0 | 0 | 0 | 0 | 0 |
| --+-+ | 0 | 0 | 0 | 0 | 0 | 0 | 0 | 0 | 0 | 0 | 0 | 0 | 0 | 0 | 0 | 0 |
| ++--+ | 0 | 0 | 0 | 0 | 0 | 0 | 0 | 0 | 0 | 0 | 0 | 0 | 0 | 0 | 0 | 0 |
| -+--+ | 0 | 0 | 1 | 0 | 0 | 0 | 0 | 0 | 0 | 0 | 0 | 0 | 0 | 0 | 0 | 0 |
| +---+ | 0 | 0 | 0 | 0 | 0 | 0 | 0 | 0 | 0 | 0 | 0 | 0 | 0 | 0 | 0 | 0 |
| ----+ | 0 | 0 | 0 | 0 | 1 | 0 | 0 | 0 | 0 | 0 | 0 | 1 | 0 | 0 | 0 | 0 |
| ++++- | 1 | 0 | 0 | 0 | 0 | 1 | 0 | 0 | 1 | 0 | 0 | 0 | 0 | 0 | 0 | 0 |
| -+++- | 0 | 0 | 0 | 0 | 0 | 0 | 0 | 0 | 0 | 0 | 0 | 0 | 1 | 1 | 0 | 1 |
| +-++- | 0 | 0 | 0 | 1 | 0 | 0 | 0 | 0 | 0 | 0 | 0 | 0 | 1 | 0 | 0 | 0 |
| --++- | 0 | 0 | 0 | 0 | 0 | 0 | 0 | 0 | 0 | 0 | 0 | 0 | 0 | 0 | 0 | 0 |
| ++-+- | 0 | 0 | 0 | 0 | 1 | 0 | 0 | 1 | 0 | 0 | 0 | 0 | 0 | 0 | 0 | 0 |
| -+-+- | 0 | 0 | 0 | 0 | 0 | 1 | 0 | 0 | 0 | 0 | 0 | 0 | 0 | 0 | 0 | 1 |
| +--+- | 0 | 0 | 0 | 0 | 0 | 0 | 1 | 0 | 0 | 0 | 0 | 0 | 0 | 0 | 0 | 0 |
| ---+- | 0 | 0 | 0 | 0 | 0 | 0 | 0 | 0 | 0 | 0 | 0 | 0 | 0 | 0 | 0 | 1 |
| +++-- | 0 | 0 | 0 | 0 | 0 | 0 | 0 | 1 | 0 | 0 | 0 | 0 | 0 | 1 | 0 | 0 |
| -++-- | 1 | 0 | 2 | 0 | 0 | 0 | 0 | 0 | 0 | 0 | 0 | 0 | 0 | 1 | 0 | 0 |
| +-+-- | 0 | 0 | 0 | 0 | 0 | 0 | 0 | 0 | 2 | 0 | 0 | 0 | 0 | 0 | 0 | 0 |
| --+-- | 0 | 0 | 0 | 0 | 0 | 0 | 0 | 1 | 0 | 0 | 0 | 0 | 0 | 0 | 0 | 0 |
| ++--- | 0 | 1 | 0 | 0 | 0 | 0 | 0 | 1 | 0 | 0 | 0 | 0 | 0 | 0 | 0 | 0 |
| -+--- | 0 | 0 | 0 | 0 | 0 | 0 | 0 | 1 | 0 | 0 | 0 | 0 | 0 | 0 | 0 | 0 |
| +---- | 0 | 0 | 0 | 0 | 0 | 0 | 0 | 0 | 0 | 0 | 0 | 0 | 0 | 0 | 0 | 0 |
| ----- | 0 | 0 | 0 | 0 | 0 | 0 | 1 | 0 | 0 | 1 | 0 | 0 | 0 | 0 | 0 | 0 |

| | ++++- | -+++- | +-++- | --++- | ++-+- | -+-+- | +--+- | ---+- | +++-- | -++-- | +-+-- | --+-- | ++--- | -+--- | +---- | ----- |
|---|---|---|---|---|---|---|---|---|---|---|---|---|---|---|---|---|
| +++++ | 0 | 0 | 0 | 0 | 0 | 0 | 0 | 0 | 0 | 1 | 0 | 0 | 1 | 0 | 1 | 0 |
| -++++ | 0 | 0 | 0 | 0 | 0 | 0 | 0 | 0 | 0 | 1 | 1 | 0 | 1 | 0 | 0 | 0 |
| +-+++ | 0 | 0 | 0 | 0 | 0 | 0 | 0 | 0 | 0 | 1 | 0 | 0 | 0 | 0 | 0 | 0 |
| --+++ | 0 | 0 | 1 | 0 | 0 | 0 | 0 | 0 | 0 | 1 | 0 | 0 | 0 | 0 | 0 | 0 |
| ++-++ | 0 | 0 | 0 | 0 | 0 | 0 | 0 | 0 | 0 | 0 | 0 | 0 | 0 | 0 | 0 | 0 |
| -+-++ | 1 | 0 | 0 | 0 | 0 | 0 | 0 | 0 | 0 | 0 | 0 | 0 | 0 | 0 | 0 | 0 |
| +--++ | 0 | 0 | 0 | 0 | 0 | 0 | 0 | 0 | 0 | 1 | 0 | 0 | 0 | 0 | 0 | 0 |
| ---++ | 0 | 0 | 0 | 0 | 0 | 0 | 1 | 0 | 0 | 0 | 0 | 0 | 0 | 0 | 0 | 0 |
| +++-+ | 0 | 0 | 0 | 0 | 0 | 0 | 0 | 0 | 0 | 0 | 0 | 0 | 0 | 1 | 0 | 0 |
| -++++ | 0 | 0 | 0 | 1 | 0 | 0 | 0 | 0 | 0 | 0 | 0 | 0 | 0 | 0 | 1 | 0 |
| +-+++ | 0 | 0 | 0 | 0 | 0 | 0 | 0 | 0 | 0 | 0 | 0 | 0 | 1 | 0 | 0 | 0 |
| --+++ | 0 | 0 | 0 | 0 | 0 | 0 | 0 | 0 | 0 | 0 | 0 | 0 | 0 | 0 | 0 | 0 |
| ++-++ | 0 | 0 | 0 | 0 | 0 | 0 | 0 | 0 | 0 | 0 | 1 | 0 | 0 | 0 | 0 | 0 |
| -+-++ | 0 | 0 | 0 | 0 | 0 | 0 | 0 | 0 | 0 | 0 | 0 | 0 | 0 | 0 | 1 | 0 |
| +--++ | 1 | 0 | 1 | 0 | 0 | 0 | 1 | 0 | 2 | 0 | 0 | 0 | 1 | 0 | 0 | 0 |
| ----+ | 0 | 0 | 0 | 0 | 0 | 0 | 0 | 0 | 0 | 1 | 0 | 0 | 0 | 0 | 0 | 0 |
| ++++- | 0 | 0 | 1 | 0 | 0 | 0 | 0 | 0 | 1 | 0 | 0 | 0 | 0 | 0 | 0 | 0 |
| -+++- | 0 | 0 | 0 | 0 | 0 | 0 | 0 | 0 | 0 | 0 | 0 | 0 | 0 | 0 | 1 | 0 |
| +-++- | 0 | 0 | 0 | 0 | 0 | 0 | 0 | 0 | 2 | 0 | 0 | 0 | 0 | 0 | 0 | 0 |
| --++- | 0 | 1 | 0 | 0 | 0 | 0 | 0 | 0 | 0 | 0 | 0 | 0 | 0 | 0 | 0 | 0 |
| ++-+- | 0 | 0 | 0 | 0 | 0 | 0 | 0 | 0 | 0 | 0 | 1 | 1 | 0 | 0 | 0 | 0 |
| -+-+- | 0 | 1 | 0 | 0 | 0 | 1 | 0 | 0 | 0 | 0 | 1 | 0 | 0 | 0 | 0 | 0 |
| +--+- | 0 | 0 | 0 | 0 | 0 | 0 | 0 | 0 | 0 | 1 | 0 | 0 | 0 | 0 | 0 | 0 |
| ---+- | 0 | 0 | 0 | 0 | 0 | 0 | 0 | 0 | 0 | 0 | 0 | 0 | 0 | 0 | 0 | 0 |
| +++-- | 0 | 0 | 0 | 0 | 0 | 0 | 0 | 0 | 1 | 0 | 0 | 0 | 0 | 0 | 0 | 2 |
| -++-- | 0 | 0 | 0 | 1 | 0 | 0 | 0 | 0 | 0 | 0 | 1 | 0 | 0 | 0 | 0 | 0 |
| +-+-- | 0 | 0 | 0 | 0 | 0 | 0 | 0 | 0 | 0 | 0 | 0 | 0 | 0 | 0 | 0 | 0 |
| --+-- | 0 | 0 | 0 | 1 | 0 | 1 | 0 | 0 | 0 | 0 | 0 | 0 | 0 | 0 | 0 | 0 |
| ++--- | 0 | 0 | 0 | 0 | 0 | 1 | 0 | 1 | 0 | 0 | 0 | 0 | 0 | 0 | 2 | 0 |
| -+--- | 0 | 0 | 0 | 0 | 0 | 0 | 0 | 0 | 0 | 0 | 0 | 1 | 0 | 0 | 0 | 0 |
| +---- | 0 | 0 | 0 | 0 | 0 | 0 | 0 | 1 | 0 | 0 | 0 | 0 | 1 | 0 | 0 | 0 |
| ----- | 0 | 0 | 0 | 0 | 0 | 0 | 0 | 0 | 0 | 0 | 1 | 0 | 0 | 0 | 0 | 0 |

**TABLE S9.10: The experimentally measured raw data of a complete set of the 1024 possible combinations of ten-photon coincidence events with each photon measured in the basis of** $(|H\rangle \pm e^{i8\pi/10}|V\rangle)/\sqrt{2}$ **(indicated by the** $\pm$ **label in the first column and the first line) to obtain** $\langle M_{8\pi/10}^{\otimes 10}\rangle$**.** The five + or – label in the first column represent the measurement base choice for the five $o$ photons, while the five + or – label in the first line represent the measurement base choice for the five $e$ photons.

| | +++++ | -++++ | +-+++ | --+++ | ++-++ | -+-++ | +--++ | ---++ | +++-+ | -++-+ | +-+-+ | --+-+ | ++--+ | -+--+ | +---+ | ----+ |
|---|---|---|---|---|---|---|---|---|---|---|---|---|---|---|---|---|
| +++++ | 0 | 0 | 0 | 1 | 0 | 1 | 0 | 0 | 0 | 0 | 0 | 0 | 0 | 0 | 0 | 0 |
| -++++ | 0 | 0 | 1 | 0 | 0 | 0 | 0 | 0 | 0 | 0 | 0 | 0 | 0 | 1 | 0 | 0 |
| +-+++ | 0 | 0 | 0 | 0 | 0 | 0 | 0 | 0 | 0 | 0 | 0 | 0 | 0 | 0 | 0 | 0 |
| --+++ | 0 | 0 | 0 | 0 | 0 | 0 | 1 | 0 | 0 | 0 | 0 | 1 | 0 | 0 | 0 | 0 |
| ++-++ | 0 | 0 | 0 | 0 | 0 | 0 | 0 | 0 | 0 | 0 | 0 | 0 | 0 | 0 | 0 | 0 |
| -+-++ | 0 | 0 | 0 | 0 | 0 | 0 | 0 | 0 | 0 | 0 | 0 | 0 | 0 | 0 | 0 | 0 |
| +--++ | 0 | 0 | 0 | 0 | 0 | 0 | 0 | 0 | 0 | 0 | 0 | 0 | 0 | 0 | 0 | 0 |
| ---++ | 1 | 0 | 0 | 0 | 0 | 1 | 0 | 0 | 1 | 0 | 0 | 0 | 0 | 0 | 0 | 0 |
| +++-+ | 0 | 0 | 1 | 1 | 0 | 0 | 0 | 0 | 0 | 0 | 0 | 1 | 0 | 0 | 0 | 0 |
| -++-+ | 0 | 0 | 0 | 0 | 0 | 0 | 0 | 0 | 0 | 0 | 0 | 0 | 0 | 1 | 0 | 0 |
| +-+-+ | 1 | 0 | 0 | 1 | 0 | 0 | 0 | 0 | 0 | 0 | 0 | 0 | 0 | 0 | 0 | 0 |
| --+-+ | 0 | 2 | 0 | 0 | 0 | 0 | 0 | 0 | 1 | 0 | 0 | 0 | 0 | 0 | 0 | 0 |
| ++--+ | 0 | 0 | 0 | 0 | 0 | 0 | 0 | 0 | 1 | 0 | 0 | 0 | 0 | 0 | 0 | 0 |
| -+--+ | 0 | 0 | 0 | 0 | 0 | 0 | 2 | 0 | 0 | 0 | 0 | 0 | 0 | 0 | 0 | 0 |
| +---+ | 0 | 0 | 0 | 0 | 0 | 0 | 0 | 0 | 0 | 0 | 0 | 1 | 0 | 0 | 1 | 0 |
| ----+ | 0 | 0 | 0 | 0 | 0 | 0 | 0 | 0 | 0 | 0 | 0 | 0 | 0 | 0 | 0 | 0 |
| ++++- | 0 | 0 | 0 | 0 | 0 | 0 | 0 | 0 | 0 | 0 | 0 | 0 | 0 | 1 | 0 | 0 |
| -+++- | 0 | 0 | 0 | 0 | 0 | 2 | 0 | 0 | 0 | 0 | 0 | 0 | 0 | 0 | 0 | 1 |
| +-++- | 0 | 0 | 0 | 0 | 0 | 0 | 0 | 0 | 0 | 0 | 0 | 0 | 0 | 0 | 0 | 0 |
| --++- | 0 | 1 | 0 | 0 | 0 | 0 | 0 | 0 | 1 | 0 | 0 | 0 | 0 | 0 | 0 | 0 |
| ++-+- | 1 | 0 | 0 | 0 | 0 | 0 | 0 | 0 | 0 | 0 | 0 | 0 | 0 | 0 | 0 | 0 |
| -+-+- | 0 | 1 | 0 | 0 | 0 | 0 | 0 | 0 | 0 | 0 | 0 | 0 | 0 | 0 | 0 | 0 |
| +--+- | 0 | 0 | 1 | 0 | 0 | 0 | 0 | 1 | 0 | 0 | 0 | 0 | 0 | 0 | 0 | 0 |
| ---+- | 0 | 0 | 0 | 0 | 0 | 0 | 0 | 0 | 0 | 0 | 0 | 1 | 0 | 0 | 0 | 0 |
| +++-- | 0 | 0 | 0 | 0 | 0 | 0 | 0 | 0 | 0 | 0 | 0 | 0 | 0 | 0 | 0 | 0 |
| -++-- | 0 | 0 | 0 | 0 | 0 | 0 | 0 | 0 | 0 | 0 | 0 | 0 | 0 | 0 | 0 | 0 |
| +-+-- | 0 | 0 | 0 | 0 | 0 | 1 | 0 | 1 | 0 | 0 | 0 | 0 | 1 | 0 | 0 | 0 |
| --+-- | 0 | 0 | 0 | 0 | 0 | 0 | 0 | 0 | 0 | 0 | 0 | 0 | 0 | 0 | 0 | 0 |
| ++--- | 0 | 0 | 0 | 0 | 0 | 0 | 0 | 0 | 0 | 0 | 0 | 0 | 0 | 0 | 0 | 0 |
| -+--- | 0 | 0 | 0 | 0 | 0 | 0 | 0 | 0 | 0 | 0 | 0 | 0 | 0 | 0 | 0 | 0 |
| +---- | 0 | 0 | 0 | 0 | 0 | 0 | 0 | 1 | 1 | 0 | 0 | 0 | 0 | 0 | 0 | 0 |
| ----- | 0 | 0 | 0 | 1 | 0 | 0 | 0 | 0 | 0 | 0 | 0 | 0 | 0 | 0 | 2 | 0 |

| | +++ +- | -+++- | +- ++- | -- ++- | ++- +- | -+- +- | +-- +- | ---+- | +++ -- | -++- - | +-+- - | --+- - | ++-- - | -+--- | +---- | ----- |
|---|---|---|---|---|---|---|---|---|---|---|---|---|---|---|---|---|
| +++++ | 0 | 0 | 0 | 0 | 0 | 0 | 0 | 2 | 0 | 0 | 0 | 0 | 0 | 0 | 1 | 0 |
| -++++ | 0 | 0 | 0 | 0 | 0 | 0 | 0 | 0 | 0 | 0 | 0 | 0 | 0 | 0 | 0 | 0 |
| +-+++ | 0 | 0 | 0 | 0 | 0 | 0 | 0 | 0 | 0 | 0 | 0 | 0 | 0 | 1 | 0 | 0 |
| --+++ | 0 | 0 | 0 | 0 | 1 | 0 | 0 | 0 | 0 | 0 | 0 | 0 | 0 | 0 | 0 | 0 |
| ++-++ | 1 | 0 | 0 | 1 | 0 | 0 | 0 | 0 | 0 | 1 | 0 | 0 | 0 | 0 | 0 | 0 |
| -+-++ | 0 | 0 | 0 | 0 | 0 | 0 | 0 | 0 | 0 | 1 | 0 | 0 | 0 | 0 | 0 | 0 |
| +--++ | 0 | 0 | 0 | 0 | 0 | 0 | 0 | 0 | 0 | 0 | 0 | 0 | 1 | 0 | 0 | 0 |
| ---++ | 0 | 0 | 1 | 0 | 0 | 0 | 0 | 0 | 2 | 1 | 0 | 0 | 0 | 0 | 0 | 0 |
| +++-+ | 0 | 0 | 0 | 0 | 1 | 0 | 0 | 0 | 0 | 0 | 0 | 1 | 0 | 0 | 0 | 0 |
| -++-+ | 0 | 0 | 0 | 0 | 1 | 0 | 0 | 0 | 0 | 0 | 0 | 1 | 0 | 0 | 0 | 0 |
| +-+-+ | 0 | 0 | 0 | 0 | 0 | 0 | 0 | 0 | 0 | 0 | 0 | 0 | 0 | 0 | 0 | 0 |
| --+-+ | 0 | 0 | 0 | 0 | 0 | 0 | 0 | 0 | 0 | 0 | 0 | 0 | 0 | 0 | 0 | 0 |
| ++--+ | 0 | 1 | 0 | 0 | 0 | 0 | 0 | 0 | 0 | 1 | 0 | 0 | 0 | 0 | 0 | 1 |
| -+--+ | 0 | 0 | 0 | 0 | 0 | 0 | 0 | 0 | 0 | 0 | 0 | 0 | 0 | 0 | 0 | 0 |
| +---+ | 0 | 0 | 1 | 0 | 0 | 0 | 0 | 0 | 0 | 0 | 0 | 1 | 0 | 0 | 0 | 0 |
| ----+ | 0 | 0 | 0 | 0 | 0 | 0 | 0 | 0 | 0 | 0 | 0 | 0 | 0 | 0 | 0 | 1 |
| ++++- | 0 | 0 | 0 | 1 | 1 | 0 | 1 | 0 | 0 | 0 | 0 | 0 | 0 | 1 | 0 | 0 |
| -+++- | 0 | 0 | 0 | 0 | 0 | 0 | 0 | 0 | 0 | 1 | 0 | 0 | 0 | 1 | 0 | 0 |
| +-++- | 0 | 1 | 0 | 0 | 0 | 0 | 0 | 0 | 0 | 0 | 0 | 0 | 0 | 0 | 0 | 0 |
| --++- | 0 | 0 | 0 | 0 | 0 | 0 | 0 | 0 | 0 | 0 | 0 | 0 | 0 | 0 | 0 | 0 |
| ++-+- | 0 | 0 | 0 | 0 | 1 | 0 | 0 | 0 | 0 | 0 | 0 | 0 | 0 | 0 | 0 | 0 |
| -+-+- | 0 | 0 | 0 | 0 | 0 | 0 | 0 | 1 | 0 | 1 | 0 | 0 | 0 | 0 | 1 | 1 |
| +--+- | 0 | 0 | 0 | 0 | 0 | 0 | 0 | 0 | 0 | 0 | 0 | 0 | 0 | 0 | 0 | 0 |
| ---+- | 0 | 0 | 1 | 0 | 0 | 0 | 0 | 1 | 0 | 0 | 0 | 0 | 0 | 0 | 0 | 0 |
| +++-- | 0 | 1 | 0 | 0 | 0 | 0 | 0 | 0 | 0 | 0 | 0 | 0 | 0 | 0 | 0 | 0 |
| -++-- | 0 | 1 | 0 | 0 | 1 | 0 | 0 | 0 | 0 | 0 | 0 | 0 | 0 | 0 | 0 | 0 |
| +-+-- | 0 | 0 | 0 | 0 | 0 | 0 | 0 | 0 | 0 | 0 | 0 | 0 | 0 | 0 | 0 | 0 |
| --+-- | 0 | 0 | 0 | 0 | 0 | 0 | 0 | 0 | 1 | 0 | 0 | 0 | 0 | 0 | 0 | 1 |
| ++--- | 0 | 0 | 0 | 1 | 0 | 0 | 0 | 1 | 0 | 0 | 0 | 0 | 0 | 0 | 0 | 0 |
| -+--- | 0 | 0 | 0 | 0 | 0 | 0 | 0 | 0 | 0 | 0 | 0 | 1 | 0 | 0 | 0 | 0 |
| +---- | 0 | 0 | 0 | 0 | 0 | 0 | 0 | 0 | 0 | 0 | 0 | 0 | 0 | 0 | 0 | 0 |
| ----- | 0 | 0 | 1 | 0 | 0 | 0 | 0 | 0 | 1 | 2 | 0 | 0 | 0 | 0 | 0 | 0 |

**TABLE S9.11: The experimentally measured raw data of a complete set of the 1024 possible combinations of ten-photon coincidence events with each photon measured in the basis of** $(|H\rangle \pm e^{i9\pi/10}|V\rangle)/\sqrt{2}$ **(indicated by the $\pm$ label in the first column and the first line) to obtain** $\langle M_{9\pi/10}^{\otimes 10}\rangle$. The five + or − label in the first column represent the measurement base choice for the five $o$ photons, while the five + or − label in the first line represent the measurement base choice for the five $e$ photons.

| | +++ ++ | -+++ + | +- +++ | -- +++ | ++- ++ | -+- ++ | +-- ++ | --- ++ | +++ -+ | -++- + | +-+- + | --+- + | ++-- + | -+-- + | +--- + | ----+ |
|---|---|---|---|---|---|---|---|---|---|---|---|---|---|---|---|---|
| +++++ | 0 | 0 | 0 | 0 | 0 | 0 | 0 | 0 | 0 | 0 | 0 | 0 | 0 | 0 | 0 | 1 |
| -++++ | 1 | 0 | 0 | 0 | 0 | 0 | 0 | 0 | 1 | 0 | 0 | 0 | 0 | 0 | 0 | 0 |
| +-+++ | 0 | 0 | 0 | 0 | 0 | 1 | 0 | 0 | 0 | 0 | 0 | 0 | 0 | 0 | 0 | 0 |
| --+++ | 0 | 0 | 0 | 0 | 0 | 0 | 0 | 0 | 1 | 0 | 0 | 0 | 0 | 0 | 0 | 0 |
| ++-++ | 0 | 0 | 0 | 0 | 0 | 0 | 0 | 0 | 1 | 0 | 0 | 0 | 0 | 0 | 0 | 0 |
| -+-++ | 0 | 0 | 0 | 0 | 0 | 0 | 0 | 0 | 0 | 0 | 0 | 0 | 0 | 1 | 1 | 0 |
| +--++ | 0 | 0 | 0 | 0 | 1 | 0 | 0 | 1 | 0 | 0 | 0 | 0 | 0 | 1 | 0 | 0 |
| ---++ | 0 | 0 | 0 | 0 | 0 | 0 | 0 | 0 | 0 | 0 | 0 | 0 | 0 | 0 | 0 | 0 |
| +++-+ | 0 | 0 | 0 | 0 | 0 | 0 | 1 | 0 | 0 | 0 | 0 | 0 | 0 | 0 | 0 | 0 |
| -++-+ | 0 | 0 | 0 | 0 | 0 | 0 | 0 | 0 | 0 | 0 | 0 | 0 | 0 | 0 | 0 | 0 |
| +-+-+ | 0 | 0 | 0 | 0 | 1 | 0 | 0 | 0 | 0 | 1 | 0 | 0 | 0 | 1 | 0 | 2 |
| --+-+ | 0 | 0 | 0 | 0 | 0 | 0 | 1 | 0 | 0 | 0 | 0 | 0 | 0 | 0 | 0 | 0 |

| ++--+ | 0 | 0 | 0 | 0 | 0 | 0 | 0 | 0 | 0 | 0 | 0 | 0 | 0 | 1 | 1 | 0 |
|---|---|---|---|---|---|---|---|---|---|---|---|---|---|---|---|---|
| -+--+ | 0 | 0 | 1 | 0 | 0 | 1 | 0 | 0 | 0 | 0 | 0 | 0 | 1 | 0 | 0 | 0 |
| +--+ | 1 | 0 | 0 | 0 | 0 | 0 | 0 | 0 | 0 | 0 | 0 | 1 | 1 | 0 | 0 | 0 |
| ---+ | 0 | 0 | 0 | 0 | 0 | 0 | 0 | 0 | 0 | 0 | 0 | 0 | 0 | 0 | 1 | 0 |
| +++- | 1 | 0 | 0 | 0 | 0 | 0 | 0 | 0 | 0 | 0 | 0 | 0 | 0 | 0 | 0 | 0 |
| -++- | 0 | 0 | 0 | 0 | 1 | 1 | 0 | 0 | 1 | 0 | 0 | 0 | 0 | 0 | 0 | 0 |
| +-++- | 0 | 0 | 0 | 0 | 1 | 0 | 0 | 0 | 0 | 0 | 0 | 0 | 0 | 0 | 1 | 0 |
| --++- | 0 | 0 | 0 | 1 | 0 | 1 | 0 | 0 | 0 | 0 | 0 | 0 | 1 | 0 | 0 | 0 |
| ++-+- | 0 | 0 | 1 | 0 | 0 | 0 | 1 | 0 | 0 | 0 | 0 | 1 | 0 | 0 | 0 | 2 |
| -+-+- | 0 | 0 | 0 | 0 | 0 | 0 | 1 | 0 | 2 | 1 | 0 | 0 | 0 | 0 | 0 | 0 |
| +--+- | 0 | 0 | 0 | 0 | 0 | 0 | 0 | 0 | 0 | 0 | 0 | 0 | 0 | 2 | 0 | 0 |
| ---+- | 0 | 0 | 0 | 0 | 1 | 1 | 0 | 0 | 0 | 0 | 0 | 0 | 1 | 0 | 0 | 0 |
| +++-- | 0 | 0 | 1 | 1 | 0 | 0 | 0 | 0 | 0 | 0 | 0 | 0 | 0 | 0 | 0 | 0 |
| -++-- | 0 | 0 | 0 | 0 | 0 | 0 | 0 | 0 | 0 | 1 | 0 | 0 | 0 | 0 | 0 | 0 |
| +-+-- | 0 | 0 | 0 | 0 | 0 | 0 | 0 | 0 | 0 | 0 | 0 | 1 | 0 | 0 | 0 | 0 |
| --+-- | 0 | 0 | 0 | 0 | 1 | 0 | 0 | 1 | 0 | 0 | 0 | 1 | 0 | 0 | 0 | 0 |
| ++--- | 1 | 0 | 0 | 0 | 0 | 0 | 0 | 0 | 0 | 0 | 0 | 0 | 0 | 0 | 0 | 0 |
| -+--- | 0 | 0 | 1 | 0 | 0 | 0 | 0 | 0 | 0 | 0 | 0 | 0 | 0 | 0 | 0 | 0 |
| +---- | 0 | 0 | 0 | 0 | 0 | 0 | 0 | 0 | 0 | 0 | 0 | 0 | 0 | 0 | 0 | 0 |
| ----- | 0 | 0 | 0 | 0 | 0 | 0 | 0 | 0 | 1 | 0 | 1 | 0 | 0 | 0 | 0 | 0 |

| | +++ +- | - +++ - | +- ++- | -- ++- | ++- +- | -+- +- | +-- +- | ---+- | +++ -- | -++- - | +-+- - | --+-- | +-- - | -+--- | +---- | ----- |
|---|---|---|---|---|---|---|---|---|---|---|---|---|---|---|---|---|
| +++++ | 0 | 0 | 1 | 1 | 0 | 0 | 0 | 0 | 0 | 0 | 1 | 0 | 0 | 0 | 0 | 0 |
| -++++ | 0 | 0 | 0 | 0 | 1 | 0 | 1 | 0 | 0 | 0 | 0 | 0 | 0 | 0 | 0 | 0 |
| +-+++ | 0 | 0 | 0 | 0 | 0 | 0 | 0 | 0 | 0 | 0 | 0 | 0 | 0 | 0 | 0 | 1 |
| --+++ | 0 | 0 | 0 | 0 | 0 | 0 | 0 | 0 | 0 | 0 | 0 | 0 | 0 | 0 | 0 | 0 |
| ++-++ | 0 | 0 | 0 | 0 | 0 | 0 | 1 | 0 | 0 | 1 | 0 | 0 | 0 | 0 | 0 | 0 |
| -+-++ | 2 | 0 | 0 | 0 | 0 | 0 | 0 | 0 | 0 | 0 | 0 | 0 | 0 | 0 | 0 | 0 |
| +--++ | 0 | 0 | 0 | 0 | 0 | 1 | 0 | 0 | 0 | 0 | 0 | 1 | 0 | 0 | 0 | 0 |
| ---++ | 0 | 1 | 0 | 0 | 0 | 0 | 0 | 0 | 0 | 0 | 0 | 0 | 0 | 0 | 0 | 0 |
| +++-+ | 0 | 0 | 0 | 0 | 0 | 0 | 0 | 0 | 0 | 0 | 0 | 0 | 0 | 0 | 0 | 0 |
| -++-+ | 1 | 1 | 0 | 0 | 0 | 0 | 1 | 0 | 0 | 0 | 0 | 0 | 0 | 0 | 0 | 0 |
| +-+-+ | 0 | 0 | 0 | 0 | 0 | 0 | 0 | 0 | 0 | 0 | 0 | 1 | 0 | 0 | 0 | 0 |
| --+-+ | 0 | 0 | 0 | 1 | 0 | 0 | 1 | 0 | 0 | 0 | 0 | 1 | 0 | 0 | 0 | 0 |
| ++--+ | 1 | 0 | 0 | 0 | 0 | 0 | 0 | 0 | 0 | 1 | 0 | 0 | 0 | 0 | 0 | 0 |
| -+--+ | 0 | 0 | 0 | 0 | 0 | 0 | 0 | 0 | 0 | 0 | 0 | 0 | 0 | 0 | 0 | 0 |
| +---+ | 0 | 0 | 0 | 0 | 0 | 0 | 0 | 0 | 0 | 0 | 0 | 0 | 0 | 0 | 0 | 0 |
| ----+ | 0 | 2 | 0 | 0 | 0 | 1 | 0 | 0 | 0 | 1 | 0 | 0 | 0 | 1 | 0 | 0 |
| ++++- | 0 | 1 | 0 | 0 | 0 | 1 | 0 | 0 | 0 | 1 | 1 | 0 | 0 | 0 | 0 | 0 |
| -+++- | 0 | 0 | 0 | 0 | 0 | 0 | 0 | 0 | 0 | 0 | 0 | 0 | 0 | 0 | 0 | 0 |
| +-++- | 0 | 1 | 0 | 0 | 1 | 0 | 0 | 0 | 0 | 3 | 0 | 0 | 0 | 0 | 0 | 0 |
| --++- | 0 | 0 | 0 | 0 | 0 | 0 | 0 | 0 | 0 | 0 | 0 | 0 | 0 | 0 | 0 | 0 |
| ++-+- | 0 | 0 | 0 | 0 | 0 | 0 | 1 | 0 | 0 | 0 | 0 | 0 | 0 | 0 | 0 | 0 |
| -+-+- | 0 | 1 | 0 | 0 | 0 | 0 | 0 | 0 | 0 | 0 | 0 | 0 | 1 | 0 | 0 | 0 |
| +--+- | 0 | 0 | 0 | 0 | 0 | 0 | 0 | 0 | 1 | 0 | 0 | 0 | 0 | 0 | 0 | 0 |
| ---+- | 0 | 0 | 0 | 0 | 0 | 0 | 0 | 1 | 0 | 0 | 0 | 0 | 0 | 0 | 0 | 0 |
| +++-- | 0 | 0 | 0 | 0 | 0 | 0 | 0 | 0 | 0 | 0 | 1 | 0 | 0 | 0 | 0 | 0 |
| -++-- | 0 | 0 | 1 | 0 | 0 | 0 | 0 | 0 | 0 | 0 | 0 | 0 | 0 | 0 | 0 | 0 |
| +-+-- | 0 | 0 | 0 | 0 | 1 | 0 | 0 | 0 | 0 | 0 | 0 | 0 | 0 | 0 | 0 | 0 |
| --+-- | 0 | 0 | 0 | 0 | 0 | 0 | 0 | 0 | 0 | 0 | 0 | 0 | 0 | 0 | 0 | 0 |
| ++--- | 0 | 0 | 0 | 1 | 0 | 0 | 0 | 0 | 0 | 0 | 0 | 0 | 0 | 0 | 0 | 1 |
| -+--- | 0 | 0 | 0 | 0 | 0 | 0 | 0 | 0 | 0 | 0 | 0 | 1 | 0 | 0 | 0 | 0 |
| +---- | 1 | 0 | 0 | 0 | 0 | 0 | 0 | 0 | 0 | 0 | 0 | 0 | 0 | 0 | 0 | 0 |
| ----- | 0 | 0 | 1 | 0 | 1 | 0 | 0 | 0 | 0 | 0 | 0 | 0 | 0 | 0 | 0 | 0 |

From table S9.1, we can summary that the counts for the desirred term of $(|H\rangle\langle H|)^{\otimes 10} + (|V\rangle\langle V|)^{\otimes 10}$ is 198 and the sum of the others is 124, therfore, we can calculate the expectation of population is 0.615(27).

**TABLE S9.12: Summary of experimental results in table S9.2-S9.11 to calculate $\left\langle M_\theta^{\otimes 10}\right\rangle$**

| Phase $\theta$ (rad) | $0\pi/10$ | $1\pi/10$ | $2\pi/10$ | $3\pi/10$ | $4\pi/10$ | $5\pi/10$ | $6\pi/10$ | $7\pi/10$ | $8\pi/10$ | $9\pi/10$ |
|---|---|---|---|---|---|---|---|---|---|---|
| Eigen($M_\theta^{\otimes 10}$) = +1 | 58 | 39 | 58 | 36 | 66 | 38 | 63 | 38 | 56 | 42 |
| Eigen($M_\theta^{\otimes 10}$) = -1 | 35 | 67 | 32 | 55 | 38 | 57 | 34 | 57 | 36 | 69 |
| $\left\langle M_\theta^{\otimes 10} \right\rangle$ | 0.247(100) | -0.264(94) | 0.289(101) | -0.209(103) | 0.269(94) | -0.200(101) | 0.299(97) | -0.2(101) | 0.217(102) | -0.243(92) |

Then, through $F(\psi^{10}) = (\langle P^{10} \rangle + \langle C^{10} \rangle)/2 = [\langle P^{10} \rangle + (1/10)\sum_{k=0}^{k=9}(-1)^k \left\langle M_{k\pi/10}^{\otimes 10} \right\rangle]/2$, we can

obtain the fideliy of the ten-photon entanglement is 0.429(21)

## S10. Experimentally measured raw data to calculate the fidelity of eight-photon state

Similar to the calculation method in S8, we can calculate the fidelity of eighter-photon entanglement through the following tables S10.1-S10.9.

**TABLE S10.1: The experimentally measured raw data of a complete set of the 256 possible combinations of eight-photon coincidence events with each photon measured in the basis of $|H\rangle/|V\rangle$ (indicated by the H/V label in the first column and the first line) to obtain $\langle P^8 \rangle$. The four H or V label in the first column represent the measurement base choice for the four $o$ photons, while the four H or V label in the first line represent the measurement base choice for the four $e$ photons.**

| | HH HH | VH HH | HV HH | VV HH | HH VH | VH VH | HV VH | VV VH | HH HV | VH HV | HV HV | VV HV | HH VV | VH VV | HV VV | VV VV |
|---|---|---|---|---|---|---|---|---|---|---|---|---|---|---|---|---|
| HH HH | 61 | 2 | 1 | 0 | 0 | 0 | 0 | 0 | 1 | 0 | 0 | 0 | 0 | 0 | 0 | 0 |
| VH HH | 0 | 2 | 0 | 0 | 0 | 0 | 0 | 0 | 3 | 0 | 0 | 0 | 0 | 0 | 0 | 0 |
| HV HH | 2 | 2 | 0 | 0 | 0 | 0 | 0 | 0 | 0 | 0 | 1 | 0 | 0 | 0 | 0 | 0 |
| VV HH | 0 | 1 | 0 | 0 | 0 | 0 | 0 | 0 | 0 | 0 | 0 | 1 | 0 | 0 | 0 | 0 |
| HH VH | 1 | 0 | 0 | 0 | 1 | 0 | 0 | 0 | 0 | 0 | 0 | 0 | 0 | 0 | 0 | 0 |
| VH VH | 0 | 0 | 0 | 0 | 0 | 0 | 0 | 0 | 0 | 0 | 0 | 1 | 0 | 0 | 0 | 0 |
| HV VH | 0 | 0 | 1 | 2 | 0 | 0 | 0 | 0 | 0 | 0 | 0 | 0 | 0 | 0 | 0 | 0 |
| VV VH | 0 | 0 | 0 | 0 | 0 | 0 | 0 | 0 | 0 | 0 | 0 | 0 | 0 | 0 | 0 | 0 |
| HH HV | 1 | 0 | 0 | 0 | 0 | 0 | 0 | 0 | 1 | 0 | 0 | 0 | 1 | 0 | 0 | 0 |
| VH HV | 0 | 0 | 0 | 0 | 0 | 0 | 0 | 0 | 1 | 0 | 0 | 0 | 1 | 0 | 0 | 0 |
| HV HV | 0 | 0 | 0 | 0 | 0 | 0 | 0 | 0 | 0 | 0 | 0 | 0 | 0 | 0 | 0 | 0 |
| VV HV | 0 | 0 | 0 | 0 | 0 | 0 | 0 | 0 | 0 | 0 | 0 | 1 | 0 | 0 | 0 | 0 |
| HH VV | 0 | 0 | 0 | 0 | 1 | 0 | 0 | 0 | 0 | 0 | 0 | 0 | 2 | 0 | 1 | 0 |
| VH VV | 0 | 0 | 0 | 0 | 0 | 0 | 0 | 0 | 0 | 0 | 0 | 0 | 3 | 1 | 1 | 0 |
| HV VV | 0 | 0 | 0 | 0 | 0 | 0 | 1 | 0 | 0 | 0 | 0 | 0 | 0 | 0 | 1 | 1 |

| VV VV | 0 | 0 | 0 | 0 | 0 | 0 | 0 | 0 | 0 | 0 | 0 | 0 | 0 | 2 | 0 | 65 |
|---|---|---|---|---|---|---|---|---|---|---|---|---|---|---|---|---|

| | ++<br>++ | -<br>++<br>+ | +-<br>++ | --<br>++ | ++<br>-+ | -+-<br>+ | +--<br>+ | ---<br>+ | ++<br>+- | -<br>++<br>- | +-<br>+- | --<br>+- | ++<br>-- | -+-<br>- | +--<br>- | ---- |
|---|---|---|---|---|---|---|---|---|---|---|---|---|---|---|---|---|
| ++++ | 1 | 0 | 0 | 0 | 1 | 0 | 0 | 1 | 2 | 2 | 0 | 0 | 1 | 0 | 0 | 0 |
| -+++ | 0 | 1 | 0 | 0 | 1 | 0 | 0 | 0 | 1 | 0 | 1 | 2 | 0 | 0 | 1 | 0 |
| +-++ | 0 | 0 | 0 | 0 | 1 | 1 | 0 | 2 | 0 | 0 | 0 | 0 | 0 | 0 | 0 | 0 |
| --++ | 0 | 0 | 0 | 0 | 0 | 1 | 1 | 0 | 0 | 0 | 0 | 0 | 2 | 0 | 0 | 2 |
| ++-+ | 0 | 1 | 1 | 0 | 1 | 0 | 0 | 1 | 1 | 0 | 0 | 1 | 0 | 1 | 0 | 0 |
| -+-+ | 0 | 0 | 0 | 1 | 0 | 0 | 0 | 1 | 0 | 0 | 0 | 0 | 1 | 0 | 0 | 0 |
| +--+ | 0 | 0 | 0 | 1 | 0 | 0 | 0 | 0 | 0 | 0 | 0 | 0 | 0 | 1 | 1 | 0 |
| ---+ | 0 | 0 | 0 | 0 | 0 | 0 | 0 | 2 | 2 | 0 | 0 | 0 | 1 | 0 | 0 | 0 |
| +++- | 0 | 2 | 1 | 0 | 0 | 0 | 0 | 1 | 1 | 0 | 0 | 0 | 0 | 0 | 0 | 0 |
| -++- | 2 | 0 | 0 | 1 | 0 | 0 | 1 | 0 | 0 | 0 | 1 | 0 | 0 | 0 | 0 | 0 |
| +-+- | 3 | 0 | 0 | 0 | 0 | 0 | 1 | 1 | 0 | 0 | 0 | 1 | 0 | 0 | 1 | 1 |
| --+- | 0 | 0 | 0 | 0 | 0 | 0 | 0 | 0 | 1 | 0 | 0 | 0 | 0 | 0 | 1 | 0 |
| ++-- | 1 | 0 | 1 | 0 | 0 | 2 | 1 | 0 | 0 | 1 | 2 | 0 | 0 | 1 | 0 | 2 |
| -+-- | 0 | 0 | 0 | 0 | 0 | 1 | 0 | 0 | 1 | 0 | 0 | 0 | 0 | 0 | 0 | 0 |
| +--- | 1 | 1 | 0 | 0 | 1 | 1 | 0 | 3 | 0 | 0 | 0 | 1 | 1 | 0 | 1 | 0 |
| ---- | 1 | 0 | 0 | 3 | 0 | 0 | 1 | 0 | 0 | 0 | 0 | 0 | 0 | 0 | 0 | 0 |

| | ++<br>++ | -<br>++<br>+ | +-<br>++ | --<br>++ | ++<br>-+ | -+-<br>+ | +--<br>+ | ---<br>+ | ++<br>+- | -<br>++<br>- | +-<br>+- | --<br>+- | ++<br>-- | -+-<br>- | +--<br>- | ---- |
|---|---|---|---|---|---|---|---|---|---|---|---|---|---|---|---|---|
| ++++ | 0 | 2 | 0 | 1 | 1 | 0 | 0 | 1 | 1 | 0 | 0 | 0 | 0 | 0 | 0 | 0 |
| -+++ | 2 | 0 | 0 | 1 | 0 | 0 | 0 | 0 | 0 | 1 | 0 | 1 | 1 | 1 | 0 | 0 |
| +-++ | 0 | 0 | 1 | 0 | 0 | 0 | 1 | 0 | 0 | 0 | 0 | 0 | 1 | 0 | 1 | 1 |
| --++ | 0 | 1 | 0 | 1 | 0 | 0 | 0 | 5 | 2 | 1 | 0 | 0 | 0 | 0 | 1 | 0 |

| | ++++ | -+++ | +-++ | --++ | ++-+ | -+-+ | +--+ | ---+ | +++- | -++- | +-+- | --+- | ++-- | -+-- | +--- | ---- |
|---|---|---|---|---|---|---|---|---|---|---|---|---|---|---|---|---|
| ++-+ | 0 | 0 | 0 | 1 | 0 | 0 | 0 | 0 | 0 | 1 | 1 | 0 | 1 | 0 | 1 | 2 |
| -+-+ | 0 | 2 | 0 | 0 | 1 | 0 | 0 | 2 | 0 | 1 | 0 | 1 | 0 | 0 | 0 | 0 |
| +--+ | 0 | 0 | 0 | 0 | 1 | 0 | 0 | 3 | 0 | 0 | 0 | 0 | 0 | 1 | 2 | 0 |
| ---+ | 0 | 1 | 0 | 1 | 0 | 0 | 2 | 0 | 0 | 0 | 0 | 1 | 1 | 1 | 1 | 1 |
| +++- | 0 | 0 | 0 | 1 | 0 | 0 | 0 | 1 | 0 | 1 | 0 | 0 | 0 | 1 | 0 | 0 |
| -++- | 0 | 0 | 0 | 0 | 1 | 1 | 0 | 2 | 0 | 0 | 0 | 0 | 0 | 0 | 1 | 0 |
| +-+- | 0 | 0 | 2 | 0 | 0 | 1 | 0 | 1 | 2 | 0 | 2 | 1 | 1 | 0 | 1 | 0 |
| --+- | 0 | 0 | 1 | 0 | 1 | 0 | 1 | 1 | 0 | 1 | 0 | 2 | 1 | 1 | 0 | 0 |
| ++-- | 0 | 1 | 0 | 0 | 0 | 0 | 0 | 1 | 0 | 0 | 0 | 0 | 0 | 0 | 1 | 0 |
| -+-- | 0 | 0 | 0 | 2 | 0 | 0 | 1 | 0 | 0 | 0 | 1 | 1 | 0 | 0 | 0 | 1 |
| +--- | 0 | 0 | 0 | 1 | 0 | 1 | 2 | 0 | 1 | 1 | 0 | 1 | 0 | 0 | 0 | 0 |
| ---- | 0 | 0 | 0 | 1 | 1 | 0 | 0 | 0 | 0 | 0 | 0 | 2 | 0 | 1 | 1 | 0 |

**TABLE S10.4: The experimentally measured raw data of a complete set of the 256 possible combinations of ten-photon coincidence events with each photon measured in the basis of** $(|H\rangle \pm e^{i2\pi/8}|V\rangle)/\sqrt{2}$ **(indicated by the $\pm$ label in the first column and the first line) to obtain** $\langle M_{2\pi/8}^{\otimes 8}\rangle$. The four + or – label in the first column represent the measurement base choice for the four $o$ photons, while the four + or – label in the first line represent the measurement base choice for the four $e$ photons.

| | ++ ++ | - ++ + | +- ++ | -- ++ | ++ -+ | -+- + | +-- + | --- + | ++ + | - ++ - | +- +- | -- +- | ++ -- | -+- - | +-- - | ---- |
|---|---|---|---|---|---|---|---|---|---|---|---|---|---|---|---|---|
| ++++ | 2 | 0 | 0 | 0 | 0 | 0 | 1 | 1 | 0 | 1 | 0 | 0 | 2 | 0 | 1 | 0 |
| -+++ | 0 | 0 | 0 | 1 | 1 | 0 | 0 | 0 | 0 | 0 | 0 | 1 | 0 | 0 | 0 | 1 |
| +-++ | 0 | 0 | 0 | 0 | 0 | 1 | 0 | 0 | 0 | 0 | 1 | 1 | 0 | 0 | 1 | 1 |
| --++ | 0 | 0 | 0 | 0 | 0 | 0 | 0 | 0 | 2 | 0 | 1 | 1 | 2 | 0 | 0 | 1 |
| ++-+ | 0 | 2 | 1 | 0 | 1 | 0 | 0 | 0 | 0 | 0 | 0 | 0 | 0 | 0 | 1 | 0 |
| -+-+ | 1 | 0 | 0 | 1 | 0 | 1 | 1 | 0 | 0 | 2 | 0 | 0 | 1 | 0 | 1 | 0 |
| +--+ | 2 | 0 | 0 | 0 | 0 | 0 | 0 | 1 | 2 | 0 | 0 | 1 | 1 | 0 | 1 | 1 |
| ---+ | 0 | 1 | 0 | 0 | 1 | 0 | 0 | 0 | 1 | 1 | 0 | 0 | 0 | 1 | 1 | 0 |
| +++- | 0 | 2 | 1 | 0 | 2 | 1 | 0 | 1 | 0 | 0 | 0 | 0 | 0 | 0 | 1 | 0 |
| -++- | 1 | 0 | 0 | 1 | 0 | 0 | 0 | 1 | 1 | 2 | 0 | 0 | 1 | 0 | 1 | 2 |
| +-+- | 0 | 0 | 1 | 0 | 0 | 2 | 0 | 1 | 0 | 1 | 0 | 0 | 1 | 0 | 0 | 1 |
| --+- | 1 | 0 | 0 | 0 | 1 | 0 | 0 | 0 | 0 | 0 | 0 | 1 | 0 | 2 | 0 | 0 |
| ++-- | 0 | 0 | 0 | 1 | 0 | 0 | 0 | 0 | 0 | 2 | 0 | 0 | 0 | 0 | 1 | 1 |
| -+-- | 0 | 0 | 1 | 0 | 1 | 0 | 1 | 1 | 2 | 0 | 0 | 1 | 0 | 0 | 0 | 1 |
| +--- | 0 | 2 | 2 | 0 | 0 | 0 | 0 | 1 | 3 | 0 | 0 | 0 | 1 | 0 | 0 | 0 |
| ---- | 2 | 0 | 0 | 1 | 0 | 1 | 2 | 0 | 0 | 0 | 0 | 0 | 1 | 0 | 1 | 1 |

**TABLE S10.5: The experimentally measured raw data of a complete set of the 256 possible combinations of ten-photon coincidence events with each photon measured in the basis of** $(|H\rangle \pm e^{i3\pi/8}|V\rangle)/\sqrt{2}$ **(indicated by the $\pm$ label in the first column and the first line) to**

obtain $\langle M_{3\pi/8}^{\otimes 8} \rangle$. The four + or – label in the first column represent the measurement base choice for the four $o$ photons, while the four + or – label in the first line represent the measurement base choice for the four $e$ photons.

| | ++ ++ | - ++ + | +- ++ | -- ++ | ++ -+ | -+- + | +-- + | --- + | ++ +- | - ++ - | +- +- | -- +- | ++ -- | -+- - | +-- - | ---- |
|---|---|---|---|---|---|---|---|---|---|---|---|---|---|---|---|---|
| ++++ | 0 | 2 | 0 | 1 | 0 | 0 | 0 | 0 | 0 | 0 | 0 | 1 | 0 | 1 | 0 | 0 |
| -+++ | 0 | 0 | 0 | 3 | 0 | 0 | 0 | 1 | 0 | 0 | 0 | 0 | 1 | 0 | 0 | 2 |
| +-++ | 0 | 2 | 0 | 2 | 1 | 1 | 2 | 0 | 1 | 1 | 0 | 0 | 1 | 1 | 0 | 0 |
| --++ | 1 | 1 | 0 | 0 | 0 | 0 | 0 | 1 | 1 | 0 | 0 | 1 | 0 | 0 | 0 | 0 |
| ++-+ | 0 | 1 | 1 | 0 | 0 | 1 | 1 | 0 | 0 | 1 | 1 | 0 | 0 | 1 | 1 | 0 |
| -+-+ | 0 | 0 | 0 | 0 | 1 | 0 | 0 | 1 | 0 | 0 | 0 | 0 | 0 | 0 | 0 | 0 |
| +--+ | 0 | 0 | 1 | 0 | 1 | 0 | 0 | 0 | 0 | 0 | 0 | 0 | 0 | 0 | 1 | 0 |
| ---+ | 3 | 0 | 0 | 0 | 0 | 0 | 0 | 1 | 0 | 0 | 0 | 0 | 1 | 0 | 1 | 1 |
| +++- | 0 | 0 | 0 | 0 | 0 | 0 | 2 | 0 | 0 | 0 | 0 | 0 | 0 | 0 | 0 | 0 |
| -++- | 0 | 0 | 0 | 0 | 1 | 0 | 0 | 0 | 0 | 0 | 0 | 1 | 0 | 2 | 0 | 0 |
| +-+- | 0 | 0 | 0 | 0 | 2 | 0 | 0 | 1 | 0 | 0 | 0 | 0 | 0 | 1 | 1 | 0 |
| --+- | 0 | 0 | 0 | 1 | 0 | 1 | 0 | 0 | 0 | 0 | 1 | 0 | 0 | 0 | 0 | 4 |
| ++-- | 0 | 1 | 1 | 0 | 0 | 0 | 1 | 1 | 0 | 0 | 0 | 1 | 0 | 2 | 1 | 0 |
| -+-- | 0 | 0 | 0 | 1 | 0 | 0 | 0 | 0 | 0 | 0 | 1 | 0 | 0 | 0 | 0 | 1 |
| +--- | 0 | 0 | 0 | 1 | 0 | 0 | 2 | 0 | 0 | 0 | 1 | 1 | 1 | 0 | 0 | 0 |
| ---- | 1 | 2 | 1 | 0 | 1 | 0 | 0 | 1 | 0 | 0 | 0 | 2 | 0 | 1 | 1 | 1 |

**TABLE S10.6: The experimentally measured raw data of a complete set of the 256 possible combinations of ten-photon coincidence events with each photon measured in the basis of** $(|H\rangle \pm e^{i4\pi/8}|V\rangle)/\sqrt{2}$ **(indicated by the $\pm$ label in the first column and the first line) to obtain** $\langle M_{4\pi/8}^{\otimes 8} \rangle$. The four + or – label in the first column represent the measurement base choice for the four $o$ photons, while the four + or – label in the first line represent the measurement base choice for the four $e$ photons.

| | ++ ++ | - ++ + | +- ++ | -- ++ | ++ -+ | -+- + | +-- + | --- + | ++ +- | - ++ - | +- +- | -- +- | ++ -- | -+- - | +-- - | ---- |
|---|---|---|---|---|---|---|---|---|---|---|---|---|---|---|---|---|---|
| ++++ | 1 | 0 | 0 | 1 | 1 | 0 | 0 | 0 | 0 | 0 | 1 | 0 | 0 | 0 | 1 | 0 |
| -+++ | 1 | 1 | 1 | 0 | 0 | 0 | 0 | 1 | 0 | 1 | 1 | 1 | 0 | 0 | 1 | 0 |
| +-++ | 1 | 1 | 0 | 1 | 0 | 0 | 0 | 0 | 1 | 0 | 0 | 0 | 1 | 0 | 1 | 0 |
| --++ | 1 | 0 | 0 | 0 | 0 | 1 | 0 | 0 | 0 | 0 | 1 | 1 | 1 | 0 | 0 | 2 |
| ++-+ | 0 | 0 | 2 | 0 | 0 | 0 | 0 | 0 | 1 | 0 | 0 | 0 | 0 | 0 | 1 | 0 |
| -+-+ | 0 | 0 | 0 | 0 | 0 | 1 | 1 | 0 | 0 | 0 | 1 | 0 | 1 | 1 | 1 | 0 |
| +--+ | 0 | 0 | 0 | 0 | 0 | 0 | 0 | 1 | 0 | 1 | 1 | 0 | 0 | 0 | 1 | 0 |
| ---+ | 0 | 1 | 1 | 0 | 0 | 0 | 0 | 2 | 0 | 0 | 0 | 2 | 0 | 0 | 2 | 0 |
| +++- | 0 | 0 | 1 | 0 | 1 | 0 | 1 | 2 | 0 | 1 | 1 | 5 | 0 | 0 | 0 | 0 |
| -++- | 1 | 0 | 0 | 0 | 0 | 0 | 0 | 0 | 0 | 0 | 1 | 0 | 0 | 0 | 0 | 0 |
| +-+- | 0 | 0 | 0 | 1 | 0 | 1 | 1 | 1 | 0 | 0 | 0 | 0 | 1 | 0 | 0 | 0 |

| | | | | | | | | | | | | | | | |
|---|---|---|---|---|---|---|---|---|---|---|---|---|---|---|---|
| --+- | 0 | 2 | 0 | 0 | 1 | 0 | 0 | 0 | 0 | 0 | 0 | 0 | 0 | 1 | 0 | 0 |
| ++-- | 1 | 0 | 0 | 0 | 0 | 1 | 3 | 1 | 0 | 0 | 0 | 0 | 0 | 0 | 0 | 1 |
| -+-- | 0 | 1 | 2 | 0 | 1 | 0 | 2 | 1 | 0 | 0 | 1 | 1 | 0 | 1 | 1 | 1 |
| +--- | 0 | 0 | 0 | 1 | 0 | 1 | 1 | 0 | 1 | 0 | 0 | 0 | 0 | 0 | 0 | 1 |
| ---- | 2 | 0 | 0 | 0 | 0 | 0 | 0 | 1 | 0 | 0 | 0 | 0 | 0 | 0 | 0 | 1 |

**TABLE S10.7: The experimentally measured raw data of a complete set of the 256 possible combinations of ten-photon coincidence events with each photon measured in the basis of** $(|H\rangle \pm e^{i5\pi/8}|V\rangle)/\sqrt{2}$ **(indicated by the $\pm$ label in the first column and the first line) to obtain** $\langle M_{5\pi/8}^{\otimes 8}\rangle$**.** The four + or – label in the first column represent the measurement base choice for the four $o$ photons, while the four + or – label in the first line represent the measurement base choice for the four $e$ photons.

| | ++ ++ | - ++ + | +- ++ | -- ++ | ++ -+ | -+- -+ | +-- + | --- + | ++ + | - ++ - | +- +- | -- +- | ++ -- | -+- - | +-- - | ---- |
|---|---|---|---|---|---|---|---|---|---|---|---|---|---|---|---|---|
| ++++ | 0 | 1 | 0 | 1 | 0 | 0 | 1 | 1 | 2 | 1 | 1 | 0 | 0 | 1 | 0 | 0 |
| -+++ | 0 | 0 | 1 | 2 | 0 | 1 | 2 | 0 | 1 | 1 | 0 | 0 | 1 | 0 | 0 | 1 |
| +-++ | 0 | 0 | 1 | 0 | 0 | 0 | 0 | 0 | 0 | 1 | 0 | 0 | 0 | 0 | 0 | 0 |
| --++ | 0 | 1 | 0 | 0 | 1 | 0 | 0 | 1 | 1 | 0 | 0 | 1 | 0 | 0 | 0 | 1 |
| ++-+ | 2 | 0 | 0 | 0 | 0 | 0 | 1 | 1 | 0 | 0 | 2 | 0 | 3 | 0 | 0 | 0 |
| -+-+ | 0 | 1 | 1 | 0 | 0 | 0 | 0 | 1 | 1 | 0 | 0 | 0 | 0 | 0 | 3 | 1 |
| +--+ | 0 | 0 | 0 | 0 | 1 | 0 | 0 | 1 | 2 | 0 | 0 | 1 | 0 | 0 | 0 | 0 |
| ---+ | 0 | 0 | 0 | 0 | 2 | 2 | 1 | 0 | 0 | 0 | 0 | 0 | 1 | 0 | 0 | 0 |
| +++- | 0 | 0 | 0 | 0 | 0 | 1 | 0 | 0 | 0 | 1 | 2 | 0 | 0 | 0 | 0 | 1 |
| -++- | 0 | 0 | 0 | 0 | 1 | 0 | 0 | 0 | 1 | 0 | 0 | 0 | 0 | 1 | 2 | 1 |
| +-+- | 0 | 1 | 4 | 0 | 1 | 0 | 0 | 2 | 0 | 0 | 1 | 1 | 0 | 0 | 1 | 0 |
| --+- | 1 | 0 | 0 | 1 | 0 | 0 | 3 | 0 | 1 | 1 | 0 | 0 | 0 | 0 | 1 | 0 |
| ++-- | 1 | 0 | 0 | 1 | 1 | 0 | 0 | 0 | 0 | 0 | 0 | 1 | 0 | 0 | 0 | 0 |
| -+-- | 1 | 1 | 0 | 0 | 0 | 0 | 1 | 0 | 0 | 0 | 1 | 0 | 0 | 0 | 0 | 1 |
| +--- | 1 | 1 | 0 | 0 | 0 | 0 | 0 | 0 | 0 | 2 | 0 | 0 | 1 | 1 | 0 | 1 |
| ---- | 0 | 1 | 0 | 0 | 0 | 0 | 0 | 0 | 0 | 0 | 1 | 1 | 0 | 2 | 0 | 0 |

**TABLE S10.8: The experimentally measured raw data of a complete set of the 256 possible combinations of ten-photon coincidence events with each photon measured in the basis of** $(|H\rangle \pm e^{i6\pi/8}|V\rangle)/\sqrt{2}$ **(indicated by the $\pm$ label in the first column and the first line) to obtain** $\langle M_{6\pi/8}^{\otimes 8}\rangle$**.** The four + or – label in the first column represent the measurement base choice for the four $o$ photons, while the four + or – label in the first line represent the measurement base choice for the four $e$ photons.

| | ++ ++ | - ++ + | +- ++ | -- ++ | ++ -+ | -+- -+ | +-- + | --- + | ++ + | - ++ - | +- +- | -- +- | ++ -- | -+- - | +-- - | ---- |
|---|---|---|---|---|---|---|---|---|---|---|---|---|---|---|---|---|

| | | | | | | | | | | | | | | | |
|---|---|---|---|---|---|---|---|---|---|---|---|---|---|---|---|
| ++++ | 1 | 0 | 1 | 0 | 0 | 1 | 1 | 0 | 0 | 0 | 0 | 0 | 0 | 0 | 0 | 0 |
| -+++ | 0 | 2 | 0 | 0 | 1 | 0 | 1 | 1 | 0 | 0 | 0 | 2 | 0 | 1 | 0 | 0 |
| +-++ | 0 | 1 | 0 | 1 | 0 | 0 | 0 | 0 | 1 | 0 | 0 | 0 | 1 | 1 | 2 | 0 |
| --++ | 2 | 0 | 0 | 0 | 0 | 0 | 1 | 0 | 0 | 1 | 1 | 0 | 1 | 1 | 0 | 2 |
| ++-+ | 0 | 1 | 1 | 0 | 0 | 0 | 0 | 1 | 0 | 0 | 0 | 3 | 0 | 0 | 1 | 0 |
| -+-+ | 1 | 0 | 0 | 0 | 0 | 0 | 0 | 0 | 0 | 0 | 3 | 1 | 1 | 1 | 0 | 2 |
| +--+ | 0 | 0 | 0 | 0 | 0 | 2 | 1 | 0 | 0 | 0 | 0 | 0 | 0 | 0 | 0 | 1 |
| ---+ | 1 | 2 | 0 | 0 | 0 | 1 | 0 | 0 | 0 | 0 | 0 | 0 | 0 | 2 | 0 | 0 |
| +++- | 0 | 1 | 2 | 0 | 0 | 0 | 1 | 0 | 0 | 0 | 0 | 1 | 1 | 0 | 2 | 0 |
| -++- | 1 | 0 | 0 | 1 | 0 | 0 | 1 | 0 | 1 | 3 | 0 | 0 | 1 | 0 | 0 | 2 |
| +-+- | 1 | 0 | 0 | 1 | 1 | 0 | 0 | 0 | 0 | 0 | 0 | 1 | 1 | 2 | 0 | 0 |
| --+- | 0 | 0 | 0 | 0 | 1 | 1 | 1 | 0 | 0 | 0 | 0 | 0 | 0 | 0 | 1 | 1 |
| ++-- | 3 | 1 | 0 | 0 | 1 | 0 | 1 | 0 | 1 | 0 | 0 | 0 | 0 | 0 | 0 | 0 |
| -+-- | 0 | 2 | 0 | 0 | 3 | 0 | 1 | 2 | 1 | 0 | 0 | 0 | 0 | 0 | 0 | 0 |
| +--- | 0 | 0 | 1 | 0 | 2 | 0 | 0 | 1 | 0 | 0 | 1 | 1 | 0 | 0 | 0 | 0 |
| ---- | 0 | 0 | 0 | 2 | 0 | 0 | 0 | 0 | 0 | 1 | 2 | 0 | 0 | 0 | 0 | 0 |

**TABLE S10.9: The experimentally measured raw data of a complete set of the 256 possible combinations of ten-photon coincidence events with each photon measured in the basis of** $(|H\rangle \pm e^{i7\pi/8}|V\rangle)/\sqrt{2}$ **(indicated by the $\pm$ label in the first column and the first line) to obtain** $\langle M_{7\pi/8}^{\otimes 8}\rangle$**. The four + or – label in the first column represent the measurement base choice for the four $o$ photons, while the four + or – label in the first line represent the measurement base choice for the four $e$ photons.**

| | ++ ++ | - ++ + | +- ++ | -- ++ | ++ -+ | -+- + | +-- + | --- + | ++ +- | - ++ - | +- +- | -- +- | ++ -- | -+- - | +-- - | ---- |
|---|---|---|---|---|---|---|---|---|---|---|---|---|---|---|---|---|
| ++++ | 1 | 1 | 1 | 1 | 0 | 0 | 0 | 0 | 0 | 0 | 0 | 0 | 1 | 0 | 1 | 0 |
| -+++ | 2 | 0 | 0 | 0 | 0 | 1 | 2 | 0 | 0 | 0 | 0 | 0 | 0 | 0 | 0 | 1 |
| +-++ | 0 | 0 | 0 | 0 | 1 | 1 | 0 | 0 | 0 | 2 | 2 | 1 | 1 | 0 | 0 | 2 |
| --++ | 1 | 0 | 0 | 0 | 0 | 0 | 0 | 3 | 1 | 1 | 0 | 0 | 0 | 1 | 1 | 0 |
| ++-+ | 2 | 0 | 0 | 1 | 1 | 0 | 0 | 0 | 0 | 0 | 0 | 1 | 0 | 0 | 0 | 1 |
| -+-+ | 0 | 0 | 0 | 1 | 0 | 0 | 0 | 0 | 2 | 0 | 0 | 0 | 0 | 0 | 0 | 0 |
| +--+ | 0 | 0 | 0 | 0 | 0 | 0 | 0 | 1 | 0 | 0 | 1 | 0 | 1 | 0 | 0 | 0 |
| ---+ | 0 | 0 | 0 | 0 | 0 | 2 | 1 | 2 | 0 | 1 | 0 | 0 | 0 | 0 | 0 | 0 |
| +++- | 0 | 0 | 0 | 0 | 0 | 0 | 0 | 0 | 0 | 0 | 1 | 0 | 2 | 1 | 1 | 1 |
| -++- | 0 | 1 | 1 | 0 | 1 | 1 | 0 | 0 | 0 | 1 | 0 | 0 | 0 | 0 | 1 | 0 |
| +-+- | 1 | 0 | 0 | 1 | 0 | 0 | 0 | 0 | 2 | 0 | 0 | 0 | 0 | 1 | 0 | 0 |
| --+- | 1 | 0 | 1 | 1 | 0 | 1 | 0 | 0 | 0 | 0 | 1 | 1 | 0 | 0 | 0 | 0 |
| ++-- | 0 | 3 | 2 | 0 | 0 | 1 | 1 | 1 | 1 | 0 | 0 | 0 | 0 | 2 | 1 | 0 |
| -+-- | 0 | 0 | 0 | 0 | 0 | 1 | 1 | 0 | 0 | 0 | 1 | 0 | 2 | 0 | 0 | 0 |
| +--- | 0 | 0 | 0 | 1 | 0 | 0 | 0 | 0 | 1 | 1 | 1 | 0 | 0 | 0 | 0 | 1 |
| ---- | 0 | 0 | 1 | 0 | 0 | 0 | 0 | 0 | 0 | 0 | 0 | 0 | 0 | 0 | 0 | 0 |

From table S10.1, we can summary that the counts for the desirred term of $(|H\rangle\langle H|)^{\otimes 8}+(|V\rangle\langle V|)^{\otimes 8}$ is 126 and the sum of the others is 42. Therfore, we can calculate the expectation of population is 0.750(33).

**TABLE S10.10: Summary of experimental results in table S10.2-S10.9 to calculate $\langle M_\theta^{\otimes 8}\rangle$**

| Phase $\theta$ (rad) | $0\pi/8$ | $1\pi/8$ | $2\pi/8$ | $3\pi/8$ | $4\pi/8$ | $5\pi/8$ | $6\pi/8$ | $7\pi/8$ |
|---|---|---|---|---|---|---|---|---|
| Eigen($M_\theta^{\otimes 8}$) = +1 | 74 | 30 | 84 | 18 | 72 | 22 | 87 | 23 |
| Eigen($M_\theta^{\otimes 8}$) = -1 | 17 | 80 | 29 | 77 | 27 | 84 | 24 | 71 |
| $\langle M_\theta^{\otimes 8}\rangle$ | 0.626(82) | -0.455(85) | 0.487(82) | -0.621(80) | 0.455(90) | -0.585(79) | 0.568(78) | -0.511(89) |

Then, through $F(\psi^8)=(\langle P^8\rangle+\langle C^8\rangle)/2=[\langle P^8\rangle+(1/8)\sum_{k=0}^{k=7}(-1)^k\langle M_{k\pi/8}^{\otimes 8}\rangle]/2$, we can obtain the fideliy of the ten-photon entanglement is 0.644(22)

## S11. Experimentally measured raw data to calculate the fidelity of eight-photon state

Similar to the calculation method in S8, we can calculate the fidelity of eighter-photon entanglement through the following tables S11.1-S11.7.

**TABLE S11.1: The experimentally measured raw data of a complete set of the 64 possible combinations of eight-photon coincidence events with each photon measured in the basis of $|H\rangle/|V\rangle$ (indicated by the H/V label in the first column and the first line) to obtain $\langle P^6\rangle$. The three H or V label in the first column represent the measurement base choice for the three $o$ photons, while the three H or V label in the first line represent the measurement base choice for the three $e$ photons.**

|  | HHH | VHH | HVH | VVH | HHV | VHV | HVV | VVV |
|---|---|---|---|---|---|---|---|---|
| HHH | 305 | 3 | 2 | 0 | 4 | 0 | 0 | 0 |
| VHH | 2 | 5 | 0 | 0 | 2 | 2 | 0 | 0 |
| HVH | 12 | 3 | 3 | 1 | 1 | 0 | 0 | 0 |
| VVH | 0 | 4 | 0 | 4 | 1 | 13 | 0 | 7 |
| HHV | 5 | 0 | 2 | 0 | 8 | 0 | 4 | 0 |
| VHV | 0 | 0 | 0 | 0 | 2 | 6 | 9 | 10 |
| HVV | 0 | 0 | 2 | 2 | 0 | 0 | 3 | 6 |
| VVV | 0 | 0 | 0 | 2 | 0 | 2 | 5 | 276 |

**TABLE S11.2: The experimentally measured raw data of a complete set of the 64 possible combinations of ten-photon coincidence events with each photon measured in the basis of** $(|H\rangle \pm e^{i0\pi/6}|V\rangle)/\sqrt{2}$ **(indicated by the $\pm$ label in the first column and the first line) to obtain** $\langle M_{0\pi/6}^{\otimes 6}\rangle$**. The three + or – label in the first column represent the measurement base choice for the three $o$ photons, while the three + or – label in the first line represent the measurement base choice for the three $e$ photons.**

|     | +++ | -++ | +-+ | --+ | ++- | -+- | +-- | --- |
| --- | --- | --- | --- | --- | --- | --- | --- | --- |
| +++ | 3 | 2 | 0 | 7 | 0 | 1 | 4 | 2 |
| -++ | 0 | 3 | 1 | 1 | 3 | 0 | 0 | 3 |
| +-+ | 2 | 4 | 3 | 0 | 5 | 2 | 1 | 0 |
| --+ | 3 | 1 | 1 | 3 | 1 | 3 | 3 | 1 |
| ++- | 0 | 5 | 3 | 1 | 4 | 0 | 0 | 2 |
| -+- | 8 | 1 | 0 | 1 | 2 | 3 | 8 | 2 |
| +-- | 3 | 1 | 3 | 4 | 0 | 2 | 3 | 1 |
| --- | 1 | 5 | 3 | 1 | 5 | 0 | 0 | 3 |

**TABLE S11.3: The experimentally measured raw data of a complete set of the 64 possible combinations of ten-photon coincidence events with each photon measured in the basis of** $(|H\rangle \pm e^{i1\pi/6}|V\rangle)/\sqrt{2}$ **(indicated by the $\pm$ label in the first column and the first line) to obtain** $\langle M_{1\pi/6}^{\otimes 6}\rangle$**. The three + or – label in the first column represent the measurement base choice for the three $o$ photons, while the three + or – label in the first line represent the measurement base choice for the three $e$ photons.**

|     | +++ | -++ | +-+ | --+ | ++- | -+- | +-- | --- |
| --- | --- | --- | --- | --- | --- | --- | --- | --- |
| +++ | 1 | 1 | 4 | 2 | 5 | 1 | 1 | 6 |
| -++ | 2 | 0 | 1 | 4 | 0 | 4 | 1 | 2 |
| +-+ | 6 | 0 | 0 | 1 | 1 | 5 | 2 | 0 |
| --+ | 3 | 7 | 0 | 0 | 2 | 1 | 1 | 3 |
| ++- | 7 | 3 | 1 | 2 | 0 | 2 | 3 | 1 |
| -+- | 0 | 4 | 3 | 2 | 4 | 0 | 1 | 4 |
| +-- | 1 | 3 | 11 | 0 | 3 | 0 | 1 | 5 |
| --- | 0 | 1 | 0 | 3 | 0 | 4 | 2 | 3 |

**TABLE S11.4: The experimentally measured raw data of a complete set of the 64 possible combinations of ten-photon coincidence events with each photon measured in the basis of** $(|H\rangle \pm e^{i2\pi/6}|V\rangle)/\sqrt{2}$ **(indicated by the $\pm$ label in the first column and the first line) to obtain** $\langle M_{2\pi/6}^{\otimes 6}\rangle$**. The three + or – label in the first column represent the measurement base**

choice for the three *o* photons, while the three + or − label in the first line represent the measurement base choice for the three *e* photons.

| | +++ | -++ | +-+ | --+ | ++- | -+- | +-- | --- |
|-----|-----|-----|-----|-----|-----|-----|-----|-----|
| +++ | 5 | 1 | 0 | 2 | 0 | 1 | 3 | 0 |
| -++ | 1 | 3 | 2 | 1 | 3 | 0 | 0 | 2 |
| +-+ | 1 | 3 | 3 | 2 | 2 | 1 | 0 | 8 |
| --+ | 0 | 0 | 2 | 2 | 0 | 4 | 6 | 0 |
| ++- | 0 | 10 | 5 | 2 | 2 | 0 | 1 | 3 |
| -+- | 2 | 0 | 0 | 0 | 1 | 2 | 0 | 1 |
| +-- | 2 | 2 | 0 | 2 | 1 | 6 | 4 | 0 |
| --- | 0 | 3 | 1 | 0 | 3 | 2 | 1 | 6 |

**TABLE S11.5: The experimentally measured raw data of a complete set of the 64 possible combinations of ten-photon coincidence events with each photon measured in the basis of** $(|H\rangle \pm e^{i3\pi/6}|V\rangle)/\sqrt{2}$ **(indicated by the** $\pm$ **label in the first column and the first line) to obtain** $\langle M_{3\pi/6}^{\otimes 6}\rangle$**.** The three + or − label in the first column represent the measurement base choice for the three *o* photons, while the three + or − label in the first line represent the measurement base choice for the three *e* photons.

| | +++ | -++ | +-+ | --+ | ++- | -+- | +-- | --- |
|-----|-----|-----|-----|-----|-----|-----|-----|-----|
| +++ | 0 | 5 | 3 | 1 | 2 | 2 | 1 | 4 |
| -++ | 4 | 2 | 2 | 3 | 3 | 5 | 4 | 1 |
| +-+ | 4 | 2 | 0 | 3 | 0 | 2 | 1 | 1 |
| --+ | 0 | 1 | 2 | 2 | 2 | 0 | 0 | 2 |
| ++- | 3 | 0 | 0 | 1 | 0 | 6 | 3 | 0 |
| -+- | 1 | 7 | 3 | 1 | 4 | 1 | 0 | 5 |
| +-- | 0 | 5 | 3 | 0 | 3 | 0 | 0 | 2 |
| --- | 6 | 3 | 2 | 5 | 1 | 2 | 5 | 1 |

**TABLE S11.6: The experimentally measured raw data of a complete set of the 64 possible combinations of ten-photon coincidence events with each photon measured in the basis of** $(|H\rangle \pm e^{i4\pi/6}|V\rangle)/\sqrt{2}$ **(indicated by the** $\pm$ **label in the first column and the first line) to obtain** $\langle M_{4\pi/6}^{\otimes 6}\rangle$**.** The three + or − label in the first column represent the measurement base choice for the three *o* photons, while the three + or − label in the first line represent the measurement base choice for the three *e* photons.

| | +++ | -++ | +-+ | --+ | ++- | -+- | +-- | --- |
|-----|-----|-----|-----|-----|-----|-----|-----|-----|
| +++ | 4 | 3 | 0 | 3 | 0 | 4 | 2 | 1 |
| -++ | 0 | 4 | 3 | 0 | 3 | 0 | 0 | 4 |
| +-+ | 1 | 4 | 5 | 1 | 3 | 2 | 1 | 2 |
| --+ | 2 | 3 | 0 | 2 | 1 | 6 | 3 | 1 |
| ++- | 1 | 3 | 5 | 3 | 3 | 0 | 1 | 1 |

| | +++ | -++ | +-+ | --+ | ++- | -+- | +-- | --- |
|---|---|---|---|---|---|---|---|---|
| -+- | 4 | 2 | 0 | 1 | 0 | 3 | 4 | 1 |
| +-- | 4 | 0 | 0 | 5 | 0 | 6 | 3 | 3 |
| --- | 0 | 3 | 2 | 1 | 1 | 1 | 1 | 6 |

**TABLE S11.7: The experimentally measured raw data of a complete set of the 64 possible combinations of ten-photon coincidence events with each photon measured in the basis of** $(|H\rangle \pm e^{i5\pi/6}|V\rangle)/\sqrt{2}$ **(indicated by the $\pm$ label in the first column and the first line) to obtain** $\langle M_{5\pi/6}^{\otimes 6}\rangle$**.** The three + or – label in the first column represent the measurement base choice for the three $o$ photons, while the three + or – label in the first line represent the measurement base choice for the three $e$ photons.

| | +++ | -++ | +-+ | --+ | ++- | -+- | +-- | --- |
|---|---|---|---|---|---|---|---|---|
| +++ | 0 | 6 | 1 | 1 | 4 | 3 | 0 | 0 |
| -++ | 2 | 1 | 1 | 4 | 1 | 6 | 3 | 0 |
| +-+ | 1 | 0 | 1 | 2 | 0 | 3 | 4 | 1 |
| --+ | 0 | 1 | 3 | 2 | 3 | 0 | 0 | 4 |
| ++- | 3 | 0 | 1 | 1 | 1 | 3 | 6 | 2 |
| -+- | 1 | 4 | 1 | 2 | 4 | 1 | 0 | 5 |
| +-- | 2 | 4 | 6 | 1 | 2 | 0 | 2 | 2 |
| --- | 2 | 0 | 0 | 1 | 1 | 4 | 8 | 1 |

From table S11.1, we can summary that the counts for the desirred term of $(|H\rangle\langle H|)^{\otimes 6} + (|V\rangle\langle V|)^{\otimes 6}$ is 581 and the sum of the others is 137, therfore, we can calculate the expectation of population is 0.809(15).

**TABLE S11.8: Summary of experimental results in table S11.2-S11.7 to calculate** $\langle M_\theta^{\otimes 6}\rangle$

| Phase $\theta$ (rad) | $0\pi/6$ | $1\pi/6$ | $2\pi/6$ | $3\pi/6$ | $4\pi/6$ | $5\pi/6$ |
|---|---|---|---|---|---|---|
| Eigen($M_\theta^{\otimes 6}$) = +1 | 111 | 28 | 100 | 27 | 108 | 26 |
| Eigen($M_\theta^{\otimes 6}$) = -1 | 27 | 113 | 20 | 110 | 28 | 103 |
| $\langle M_\theta^{\otimes 6}\rangle$ | 0.609(68) | -0.603(67) | 0.667(68) | -0.606(68) | 0.588(69) | -0.597(71) |

Then, through $F(\psi^6) = (\langle P^6\rangle + \langle C^6\rangle)/2 = [\langle P^6\rangle + (1/6)\sum_{k=0}^{k=5}(-1)^k \langle M_{k\pi/6}^{\otimes 6}\rangle]/2$, we can obtain the fideliy of the ten-photon entanglement is 0.710(16)

[S1]  Yao, X.- C. *et al*. Observation of eight-photon entanglement. *Nature Photon*. **6**, 225-228 (2012).

[S2]  Kim, Y.-H., Kulik, S. P., Chekhova, M. V., Grice, W. P. & Shih, Y. Experimental entanglement concentration and universal Bell-state synthesizer. *Phys. Rev. A* **67**, 010301 (2003).

[S3]  Dür, W. & Cirac, J. I. Multiparticle entanglement and its experimental detection. *J. Phys. A: Math. Gen.* **34** 6837–6850 (2001).